%% file: arxiv_version.tex
\documentclass{article}

\input{preamble/preamble_arxiv} 

\usepackage{setspace}  
\usepackage{pgfplots}
\pgfplotsset{compat=1.18}
\usepackage{hanging} 
\usepackage{chngcntr} 

\usepackage{tocloft}
\usepackage{pdfpages}
\usepackage{setspace}
\begin{document}

\counterwithin*{section}{part}


\maketitle
\thispagestyle{empty}

\begin{abstract}
Agentic AIs — AIs that are capable and permitted to undertake complex actions with little supervision — mark a new frontier in AI capabilities and raise new questions about how to safely create and align such systems with users, developers, and society. Because agents’ actions are influenced by their attitudes toward risk, one key aspect of alignment concerns the risk profiles of agentic AIs. Risk alignment will matter for user satisfaction and trust, but it will also have important ramifications for society more broadly, especially as agentic AIs become more autonomous and are allowed to control key aspects of our lives. AIs with reckless attitudes toward risk (either because they are calibrated to reckless human users or are poorly designed) may pose significant threats. They might also open “responsibility gaps” in which there is no agent who can be held accountable for harmful actions. What risk attitudes should guide an agentic AI’s decision-making? How might we design AI systems that are calibrated to the risk attitudes of their users? What guardrails, if any, should be placed on the range of permissible risk attitudes? What are the ethical considerations involved when designing systems that make risky decisions on behalf of others? We present three papers that bear on key normative and technical aspects of these questions.
\end{abstract}

\tableofcontents

\section{Introduction}

\input{OpenAI_Intro}

\section*{Acknowledgements}
This report is a project of Rethink Priorities. The authors are Hayley Clatterbuck, Clinton Castro, and Arvo Muñoz Morán. Thanks to Jamie Elsey, Bob Fischer, David Moss, Mattie Toma and Willem Sleegers for helpful discussions and feedback. This work was supported by funding from OpenAI under a Research into Agentic AI Systems grant.

\newpage
\part{User Aspects of Risk Alignment} \label{Part1}
\input{OpenAI_Part1}

\newpage
\part{Developer Aspects of risk Alignment} \label{Part2}
\input{OpenAI_Part2}

\newpage
\part{Calibrating Agentic AIs to User Risk Attitudes} \label{Part3}
\input{OpenAI_Part3}

\newpage
\appendix 
\section{Appendix: Formal Models of Risk} \label{Appendix}
\input{OpenAI_Appendix}

\newpage
\section{Appendix: Bibliography} \label{Bilbiography}
\begingroup
\setlength{\parskip}{10pt}
\setlength{\baselineskip}{5pt}
\input{OpenAI_Bibliography}
\endgroup

\end{document}

%% file: preamble/preamble_arxiv.tex
\usepackage{url}
\usepackage[usenames,dvipsnames]{xcolor}
\usepackage[utf8]{inputenc}
\usepackage{enumitem}

\usepackage[a4paper, total={6in, 8.5in}]{geometry}

\usepackage{fancyhdr}

\usepackage{tikz}
\usetikzlibrary{shapes.misc}
\usetikzlibrary{decorations.pathreplacing}

\usepackage{comment}
\usepackage{cancel}

\usepackage{float}

\usepackage{array}
\usepackage{caption}
\usepackage{siunitx}
\usepackage{longtable}
\usepackage{booktabs}
\usepackage{makecell}

\usepackage{amsmath, amsthm, amsfonts}
\usepackage{amssymb}

\usepackage{graphicx}
\usepackage{subcaption}

\usepackage{titlesec}
\titleformat*{\subsection}{\normalsize\bfseries}

\usepackage[colorlinks=true]{hyperref}
\hypersetup{ 
     colorlinks=true, 
     linkcolor=OliveGreen, 
     filecolor=blue, 
     citecolor = OliveGreen,       
     urlcolor=OliveGreen, 
     } 

\usepackage{titling}

\title{ 
  \textbf{\Large Risk Alignment in Agentic AI Systems} 
}

\author{
  Hayley Clatterbuck\thanks{Corresponding author: \href{mailto:hayley@rethinkpriorities.org}{hayley@rethinkpriorities.org}} \\
  \textit{Rethink Priorities} \and 
  Clinton Castro \\
  \textit{University of Wisconsin-Madison} \and
  Arvo Muñoz Morán \\
  \textit{Rethink Priorities}
}
\date{\vspace{8pt} 
  Last updated: 13 September, 2024
}

\pretitle{\begin{center}\hrule height 3pt\vspace{10pt}\LARGE}
\posttitle{\vspace{13pt}\hrule height 1pt\end{center}}

\usepackage{listings}
\lstnewenvironment{pythoncode}[1][]{
  \lstset{language=Python,
    basicstyle=\ttfamily\footnotesize,
    keywordstyle=\color{blue}\bfseries,
    stringstyle=\color{orange},
    commentstyle=\color{green!60!black}\itshape,
    showstringspaces=false,
    breaklines=true,
    numbers=left,
    numberstyle=\tiny\color{blue},
    frame=single,
    frameround=tttt,
    rulecolor=\color{black},
    tabsize=2,
    morekeywords={sum,range}
    #1
  }
}{}

\definecolor{gray}{RGB}{128,128,128}
\definecolor{darkblue}{rgb}{0.0, 0.0, 0.55}  

\usepackage{todonotes}
\usepackage{etoolbox}  

\newtoggle{amm}
\newtoggle{ammr}
\newtoggle{tom}
\newtoggle{lf}
\newtoggle{bob}
\newtoggle{ld}
\newtoggle{db}
\newtoggle{hc}
\newtoggle{ds}
\newtoggle{md}
\newtoggle{pw}
\newtoggle{luis}

\toggletrue{amm}  

\toggletrue{ammr}  

\toggletrue{tom}  

\toggletrue{lf}

\toggletrue{bob}

\toggletrue{ld}

\toggletrue{db}

\toggletrue{hc}

\toggletrue{ds}

\toggletrue{md}

\toggletrue{pw}

\toggletrue{luis}

\newcommand{\amm}[1]{\iftoggle{amm}{{\todo[inline,caption={},color=white]{AMM: #1}}{}}{}}
\newcommand{\ammr}[1]{\iftoggle{ammr}{{\todo[inline,caption={},color=white]{AMM reply: #1}}{}}{}}
\newcommand{\tom}[1]{\iftoggle{tom}{{\todo[inline,caption={},color=black,textcolor=white]{TH: #1}}{}}{}}

%% file: OpenAI_Intro.tex
\noindent Proper alignment is a tetradic affair, involving relationships among
AIs, their users, their developers, and society at large (Gabriel,
\emph{et al.} 2024). Agentic AIs---AIs that are capable and permitted to
undertake complex actions with little supervision---mark a new frontier
in AI capabilities. Accordingly, they raise new questions about how to
safely create and align such systems. Existing AIs, such as LLM
chatbots, primarily provide information that human users can use to plan
actions. Thus, while chatbots may have significant effects on society,
those effects are largely filtered through human agents. Because the
introduction of agentic AIs would mark the introduction of a new kind of
actor into society, their effects on society will arguably be more
significant and unpredictable, thus raising uniquely difficult questions
of alignment in all of its aspects.

Here, we focus on an underappreciated\footnote{This topic isn't
  explicitly addressed in recent work on agentic AI alignment from
  Shavit, \emph{et al.} (2023) or Gabriel, \emph{et al.} (2024).} aspect
of alignment: what attitudes toward risk should guide an agentic AI's
decision-making? An agent's risk attitudes describe certain dispositions
when making decisions under uncertainty. A risk-averse agent disfavors
bets that have high variance in possible outcomes, preferring an action
with a high chance of a decent outcome over one that has a lower
probability of an even better outcome. A risk seeking agent is willing
to tolerate much higher risks of failure if the potential upside is
great enough. People exhibit diverse and sometimes very significant risk
attitudes. How should an agentic AI's risk attitudes be fixed in order
to achieve alignment with users? What guardrails, if any, should be
placed on the range of permissible risk attitudes in order to achieve
alignment with society and designers of AI systems? What are the ethical
considerations involved when making risky decisions on behalf of others?

We present three papers that bear on key normative and technical aspects
of these questions.

In the
\hyperref[Part1]{{first paper}}, we examine the relationship between agentic AIs and their
users. An agentic AI is ``aligned with a user when it benefits the user,
when they ask to be benefitted, in the way they expect to be benefitted"
(Gabriel, \emph{et al.} 2024, 34). Because individuals' risk attitudes
strongly influence the actions they take and approve of, getting risk
attitudes right will be a central part of agentic AI alignment. We
propose two models for thinking about the relationship between agentic
AIs and their users -- the proxy model and off-the-shelf tool model -- and
their different implications for risk alignment.

In the
\hyperref[Part2]{{second paper}}, we focus on developers of agentic AI. Developers have important
interests and moral duties that will be affected by the risk attitudes
of agentic AIs that they produce, since AIs with reckless attitudes
toward risk can expose developers to legal, reputational, and moral
liability. We explore how developers can navigate shared responsibility
among users, developers, and agentic AIs to best protect their interests
and fulfill their moral obligations.

In the
\hyperref[Part3]{{third paper}}, we turn to more technical questions about how agentic AIs might
be calibrated to the risk attitudes of their users. We evaluate how
imitation learning, prompting, and preference modeling might be used to
adapt models to information about users' risk attitudes, focusing on the
kinds of data that we would need for each learning process. Then, we
evaluate methods for eliciting these kinds of data about risk attitudes,
arguing that some methods are much more reliable and valid than others.
We end with recommendations for how agentic AIs can be created that best
achieve alignment with users and developers.

%% file: OpenAI_Part1.tex
\section{Introduction}\label{introduction1}

Our primary goal in this paper is to make the case for why risk
alignment will be an essential component of aligning agentic AI systems
to their users. Individuals' risk attitudes are a strong determinant of
how they will act and which actions they will approve of. Accordingly,
these attitudes will influence the actions that users pursue via agentic
AIs, their judgments about the acceptability of actions taken on their
behalf, and the trust that they have in AI agents. As agents themselves,
AIs will have their own risk attitudes that determine the actions that
they take. How should we design the risk attitudes of agentic AIs so
that they are aligned with those of their users?

We present two models of the relationship between users and agentic AIs
and explore the normative considerations that bear on our choice between
these two models in particular contexts:

\textbf{Proxy agents:} Agentic AIs are representatives of their users
and should be designed to replicate their users' risk attitudes.

\textbf{Off-the-shelf tools:} Agentic AIs are tools for achieving
desirable outcomes. Their risk attitudes should be set or highly
constrained by developers in order to achieve these outcomes.

When thinking about AIs that act as agents, it is natural to look for
guidance in two main areas. First, we might look at theories of rational
human agency, theories about how a person should act in order to best
achieve her goals in light of her information about the world. Different
risk attitudes constitute different strategies for acting under
uncertainty. Philosophers and economists have developed formal theories
of decision under uncertainty that allow us to more precisely
characterize these attitudes. These can be evaluated for both their
empirical accuracy (i.e. how well do they characterize the actions of
actual agents) and their normative aptness (i.e. how rational are
decisions made under different risk attitudes?). In the first half of
this paper, we will draw on insights from this literature to better
characterize the importance of risk attitudes when designing agentic
AIs.

Second, we might look at human agents --- such as financial advisors,
lawyers, or personal assistants --- who routinely take actions on another
agent's behalf. There are complex formal and informal rules that govern
how these agents ought to relate to their clients (those on whose behalf
they act), and these differ significantly across different kinds of
agents. For example, professional societies like the American Bar
Association uphold explicit professional and ethical standards that
regulate how lawyers should act on behalf of their clients. In contrast,
alignment between personal assistants and their clients are specified by
the clients themselves or via a negotiation between assistant and
client.

These arrangements are determined by the nature of the relationship (its
stakes, voluntariness, etc.), how it is embedded in broader societal
structures (e.g. an adversarial legal system), and its effects on people
outside of the relationship. In the second half of this paper, we will
turn to these models to examine different possibilities for what
alignment between an agentic AI and its user might look like for
different kinds of users and AI systems, and to recommend best practices
for achieving it.

\section{What are risk attitudes?}\label{what-are-risk-attitudes}

\subsection{An example}\label{an-example}

An agent's decisions are influenced by what she values and what she
believes about the world. However, knowing an agent's values and beliefs
is not enough to predict how she will (or should) act. For example,
suppose that Nate and Kate are each planning a dinner out. They both
prefer Restaurant A (a buzzy new spot that doesn't take reservations) to
Restaurant B (a mediocre stalwart that does) to the same extent, both
valuing a dinner at A more than twice as much as dinner at B. That is,
they assign the same relative utilities --- a measure of the subjective
value an agent assigns to an outcome --- to eating at restaurant A versus
B. They also agree that their chances of getting into Restaurant A are
about 50\% and that they are certain of getting into B.

However, despite agreeing on the value and probabilities, Nate and Kate
might nevertheless make different choices about where to go. Nate might
opt to take his chances on Restaurant A, being willing to tolerate a
50\% chance of failure in order to secure the better dinner option. Kate
would rather be safe than sorry and opts for Restaurant B. What
distinguishes Nate and Kate are their approaches to risk, the relative
significance that they give to the potential losses and gains of a risky
action.\footnote{One might object that they don't really value the
  restaurants the same way, since Kate values the reliability of B more
  than Nate does. Their risk attitudes should be incorporated into their
  utility assignments. We discuss this approach in Appendix A.}

Imagine now that Nate and Kate use AI assistants to plan their dinner
meetings. Presumably, these AIs would need more than just information
about Nate and Kate's restaurant rankings and the probabilities of
getting tables at each. In order to make decisions that accord with Nate
and Kate's preferences, their AIs would need to be adjusted to their
\emph{risk attitudes}.

\subsection{Expected utility theory and risk
attitudes}\label{expected-utility-theory-and-risk-attitudes}

When evaluating an action that has uncertain outcomes, one must take the
probabilities and the amount of value (utility) of possible outcomes
into account. Standard decision-theoretic approaches assume that there
is only one feature of the outcome space that matters: its
\emph{expected utility.} The EU of an action A is the average of the
utilities that doing A would yield in each relevant state of the world,
S\emph{\textsubscript{i}}, weighted by the probability that those states
will obtain:

\begin{equation}
    EU(A)\  = \ \sum_{i = 1}^{n}u(A|S_{i})p(S_{i})\ 
\end{equation}

However, there are other features of the distribution of possible
outcomes that someone might also find important. For example, the
following three bets all have the same expected utility
(equal to $4.5$)\footnote{We are assuming that the payoffs are in terms of utility
  itself or else something that doesn't have diminishing or increasing
  marginal utility.}:

\begin{itemize}
\item
  \textbf{Bet A:} Flip a fair coin. If heads, win 10. If tails, lose 1.
\item
  \textbf{Bet B:} Flip a fair coin. If heads, win 1,000,010. If tails,
  lose 1,000,001.
\item
  \textbf{Bet C:} Draw from a lottery of a million balls, one of which
  is a winner. If you draw the winning ball, win 5,500,000. If you draw
  any other ball, lose 1.\footnote{Technically, for the expected utility to not just approximate but equal 4.5, we need 'If you draw the winning ball, win 5,499,999', a little bit less tuneful.}
\end{itemize}

Despite having the same expected utility, these seem like \emph{very}
different bets, in a way that nearly all agents will be sensitive to.
Bet B will cause you to incur enormous losses half of the time. Bet C
almost guarantees that you'll lose something. If you're only concerned
with expected value maximization (EVM), then these differences don't matter.
However, if you are sensitive to risk, they may matter significantly.\\
At its most general, risk sensitivity is a sensitivity to
\emph{variance}, a higher-order statistical feature of the outcome
space. An agent can be risk neutral, risk averse, or risk prone. A risk
neutral agent does not take variance into account when evaluating
actions. If an ``agent is risk averse with respect to some quantity $X$ [e.g. money],
she strictly prefers a (degenerate) gamble that delivers some particular
value $x^\star$ for $X$ with certainty to a gamble that delivers an expected
$X$-value of $x^\star$, but that includes nontrivial uncertainty'' (Greaves,
\emph{et al.} 2024, 9). A risk-averse agent will accept a bet that will
deliver a lower expected payoff but with higher certainty over one with
a higher expected payoff but less certainty. For example, a risk averse
bettor might prefer a sure thing payoff of 3 over Bet A, which has an
expected utility of 4.5 but a 0.5 chance of losing. Kate is willing to
accept a sure thing of a decent meal over a lower chance of a better
dinner. A risk prone agent is the opposite of a risk averse one,
preferring high variance gambles over lower variance ones.

\subsection{Varieties of risk
aversion}\label{varieties-of-risk-aversion}

This general characterization of risk sensitivity covers several
different kinds of more specific risk attitudes.\footnote{For a
  discussion of other facets of risk, see Hansson (2023).} To get a
handle on these differences, we can ask the risk-averse agent: what is
it that is so bad about variance? What is it about variance that you
don't like?\footnote{And vice versa for the risk prone: ``what is it
  that you like about variance?''.} An ambiguity averse agent answers
that variance is bad when she is uncertain about the probabilities
involved (Machina and Siniscalchi 2014). As long as the probabilities
are all known (as in Bets A-C), there is no further problem with
variance. Some agents might be averse to variance because they don't
think that very low probability events should be taken into account.
Therefore, they will ignore the unlikely tails of the outcome
distribution (the least probable bad outcomes and least probable good
outcomes) when making decisions (Kosonen 2022, Monton 2019).

A third kind of risk averse agent answers that variance in outcomes is
bad because it includes bad outcomes. The reason that Bet B is worse
than Bet A is that there is a significant possibility that something
very bad will happen. A risk averse agent, in this sense, cares more
about avoiding the worst-case outcomes of their actions than gaining the
best-case outcomes (and vice versa for the risk prone). A risk neutral
agent assigns equal weight to gains and losses of the same magnitude.
This ``avoid the worst'' risk attitude will be our primary focus in what
follows.\footnote{We focus on this kind of risk aversion for several
  reasons. First, this is the kind of attitude that has been most
  studied in canonical experimental work on risk aversion (Kahneman \&
  Tversky 1979). Second, it will likely play a significant role in the
  kinds of decisions for which agentic AIs will be used. Third, agents
  do treat low, middling, and high probabilities differently, but it can
  be difficult to tease out when this is the result of risk weighting or
  simple errors in probabilistic reasoning (Kahneman \& Tversky 1979;
  Holt \& Laury 2014; Barseghyan, et al. 2013).}\\
Because EU maximization's assessment of a bet doesn't take its variance
into account, it cannot account for risk sensitivity.\footnote{This is
  not strictly true, as there are ways to capture risk aversion within
  EU itself. We explain this approach and why we do not favor it in
  Appendix A.} It doesn't make space for people to treat bad outcomes
differently from good ones or to treat low probabilities differently
from high ones. Indeed, by prohibiting risk sensitivity, EU maximization
places extremely stringent constraints on permissible risk attitudes by
requiring strict risk neutrality (Hájek 2021).\footnote{In EU,
  probabilities and utilities are linear, additive, and treated
  symmetrically. For example, an outcome with a probability of \emph{p}
  and utility \emph{u} contributes the same amount to expected value as
  an outcome with probability \emph{p}/2 and value 2\emph{u}.}
  
EU maximization has been extensively developed over the past century and
defended as a rational --- and perhaps the uniquely
rational --- decision-procedure. This amounts to a rejection of the
rationality of being risk prone or risk averse. Arguments in favor of EU
maximization have taken two general forms. First, axioms of rational
choice --- that is, conditions that one's preferences or actions must obey
in order to be rational --- are presented, and EVM is claimed to
(uniquely) satisfy those axioms (e.g. Von Neumann and Morgenstern 1953).
Second, it is argued that an agent who obeys EVM will experience some
long-run practical benefits over agents who obey alternative
decision-procedures.\footnote{See Briggs (2023) and Buchak (2022) for
  helpful overviews.}
  
We will not delve into the voluminous literature around this topic here,
though we will return to the question of rational constraints on risk
attitudes. Instead, we emphasize something that is more important for
the purposes of alignment: actual agents rarely act as expected utility
maximizers, and are most often risk averse. This can be seen in common
judgements about artificial decisions cases, behavior in experimental
settings, and their actual behavior. Therefore, risk attitudes are an
ineliminable aspect of any characterization of a particular decision or
an agent's general dispositions to act.

\section{\texorpdfstring{Evidence of risk non-neutrality
}{3. Evidence of risk non-neutrality }}\label{evidence-of-risk-non-neutrality}

The now standard view in welfare economics is that ``normative assessment
should recognize, in light of results of decades of behavioral
experimentation, that people are not expected utility maximizers"
(Harrison \& Ross 2017, 150). We have robust evidence from subjects'
intuitive reports, behavioral experiments in the lab, and field
observations of economic behavior that most people are at least somewhat
risk averse in most situations.

Allais (1953) was one of the first to investigate how humans' actual
choice behaviors depart from the predictions of expected utility theory (EUT),
and the choice behaviors he illustrated are still used as benchmarks for
testing theories of risk (Buchak 2013, Bottomley and Williamson 2023).
In Allais cases, subjects are asked for their preference between bets A
and B, and then asked for their preference between bets C and D:

\begin{longtable}[]{@{}
  >{\raggedright\arraybackslash}p{(\linewidth - 2\tabcolsep) * \real{0.5000}}
  >{\raggedright\arraybackslash}p{(\linewidth - 2\tabcolsep) * \real{0.5000}}@{}}
\toprule\noalign{}
\begin{minipage}[b]{\linewidth}\raggedright
\textbf{Bet A}
\end{minipage} & \begin{minipage}[b]{\linewidth}\raggedright
\textbf{Bet B}
\end{minipage} \\
\begin{minipage}[b]{\linewidth}\raggedright
Certain $\$1$ million
\end{minipage} & \begin{minipage}[b]{\linewidth}\raggedright
$.89$ chance of $\$1$ million
\end{minipage} \\
\begin{minipage}[b]{\linewidth}\raggedright
\end{minipage} & \begin{minipage}[b]{\linewidth}\raggedright
$.01$ chance of $\$0$
\end{minipage} \\
\begin{minipage}[b]{\linewidth}\raggedright
\end{minipage} & \begin{minipage}[b]{\linewidth}\raggedright
$.10$ chance of $\$5$ million
\end{minipage} \\
\midrule\noalign{}
\toprule\noalign{}
\begin{minipage}[b]{\linewidth}\raggedright
\textbf{Bet C}
\end{minipage} & \begin{minipage}[b]{\linewidth}\raggedright
\textbf{Bet D}
\end{minipage} \\
\begin{minipage}[b]{\linewidth}\raggedright
$.89$ chance of $\$0$
\end{minipage} & \begin{minipage}[b]{\linewidth}\raggedright
$.9$ chance of $\$0$
\end{minipage} \\
\begin{minipage}[b]{\linewidth}\raggedright
$.11$ chance of $\$1$ million
\end{minipage} & \begin{minipage}[b]{\linewidth}\raggedright
$.1$ chance of $\$5$ million
\end{minipage} \\
\midrule\noalign{}
\endhead
\toprule\noalign{}
\endlastfoot
\end{longtable}

Most people prefer A to B and prefer D to C. However, there is no
consistent assignment of utilities to quantities of money that makes
sense of these two preferences. To see this, consider that moving from A
to B and moving from C to D both involve trading a .01 chance at \$1
million for a .1 chance of \$5 million. In the first case, subjects are
not willing to make the trade. In the second case, they are. Whether
this trade is acceptable depends on global properties of the bet; here,
whether there is a high or low probability of getting something
good.\footnote{In Appendix A, we consider several theories of risk
  aversion that give slightly different analyses of Allais cases. For
  example, Kahneman and Tversky (1979) posit that people are biased
  toward certainties, while Buchak (2013) suggests that people discount
  better outcomes. Some have argued that we can accommodate these cases
  by adopting a more complex utility function, according to which
  different dollar amounts have different utilities across contexts.}
These preferences have been shown to be present in economic (List and
Haigh 2005) and healthcare choices (Oliver 2003).

More fine-grained experimental examinations of risk attitudes ask
subjects to consider a list of bets, with incremental changes to the
probabilities and/or payoffs involved.\footnote{We will examine these
  methods in much more depth in \hyperref[Part3]{{Paper 3}}.} They measure the amount of
risk aversion (the relative risk premium) involved by measuring ``the
mathematical expected value that one is willing to forgo to obtain
greater certainty'' (Abdellaoui, \emph{et al.} 2011, 65-66). For
example, consider the following price-list choice task from Holt and
Laury (2002):

\begin{table}[ht]
\centering
\caption{The ten paired lottery-choice decisions with low payoffs}
\begin{tabular}{llc}
\toprule
\textbf{Option A} & \textbf{Option B} & \textbf{Expected payoff difference} \\
\midrule
1/10 of \$2.00, 9/10 of \$1.60 & 1/10 of \$3.85, 9/10 of \$0.10 & \$1.17 \\
2/10 of \$2.00, 8/10 of \$1.60 & 2/10 of \$3.85, 8/10 of \$0.10 & \$0.83 \\
3/10 of \$2.00, 7/10 of \$1.60 & 3/10 of \$3.85, 7/10 of \$0.10 & \$0.50 \\
4/10 of \$2.00, 6/10 of \$1.60 & 4/10 of \$3.85, 6/10 of \$0.10 & \$0.16 \\
5/10 of \$2.00, 5/10 of \$1.60 & 5/10 of \$3.85, 5/10 of \$0.10 & -\$0.18 \\
6/10 of \$2.00, 4/10 of \$1.60 & 6/10 of \$3.85, 4/10 of \$0.10 & -\$0.51 \\
7/10 of \$2.00, 3/10 of \$1.60 & 7/10 of \$3.85, 3/10 of \$0.10 & -\$0.85 \\
8/10 of \$2.00, 2/10 of \$1.60 & 8/10 of \$3.85, 2/10 of \$0.10 & -\$1.18 \\
9/10 of \$2.00, 1/10 of \$1.60 & 9/10 of \$3.85, 1/10 of \$0.10 & -\$1.52 \\
10/10 of \$2.00, 0/10 of \$1.60 & 10/10 of \$3.85, 0/10 of \$0.10 & -\$1.85 \\
\bottomrule
\end{tabular}
\end{table}

\begin{center}
    \includegraphics[width=3.24383in,height=3.10938in]{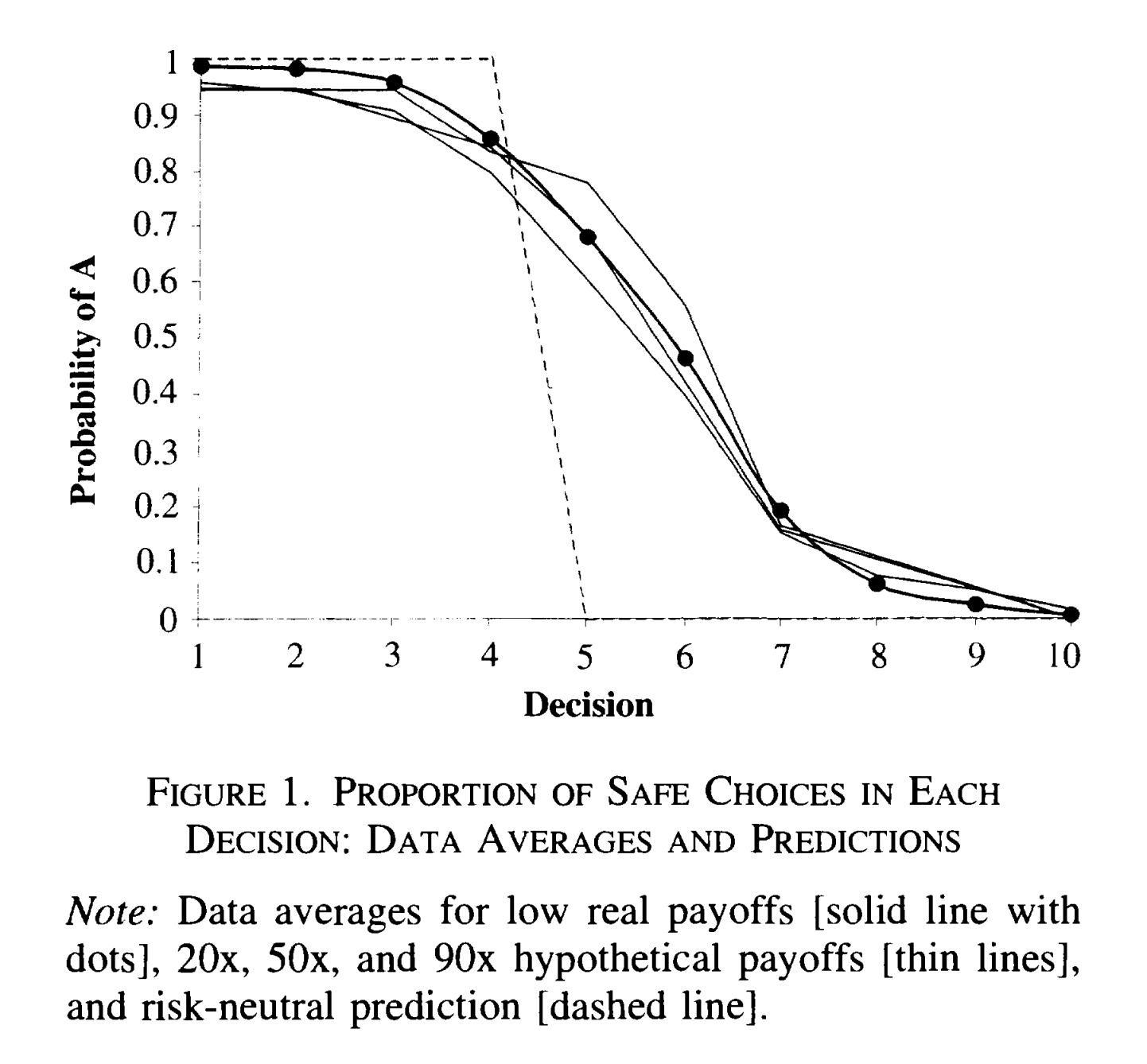}
\end{center}

They examined when subjects switched from choosing Option A to Option B.
A risk neutral subject would pick A four times, then switch to B. A risk
averse subject would stick with A for longer, and a risk seeking subject
would switch earlier. They observed considerable amounts of risk
aversion across every condition tested; in their studies, 6-15\% of
participants were risk loving, 13-29\% risk neutral, and 56-81\% risk
averse. These lab results have been replicated with choice tasks that
compare bets with certain payoffs to risky bets with varying payoffs
(Abdellaoui, \emph{et al.,} 2011).

Similar studies have been performed using actual payoffs outside of
laboratory settings. For example, in a landmark study, Binswanger (1980)
offered various lotteries to farmers in India, with potential payoffs
sometimes exceeding monthly incomes. He found that nearly all
participants were somewhat risk averse, with risk aversion increasing
with higher monetary stakes. Relative risk premiums in the 20\%-50\% are
common (Levi 1992). Even higher levels of risk aversion have been
observed in field data from auctions (Cox and Oaxaca 1996, Campo,
\emph{et al.} 2011). In sum, the ``overall message is that there is a
lot of risk aversion, centered around the 0.3-0.5 range, which is
roughly consistent with estimates implied by behavior in games,
auctions, and other decision tasks'' (Holt and Laury 2002, 1649).

As we will see in \hyperref[Part3]{{Paper 3}}, more fine-grained results are difficult to
come by and riddled with inconsistencies across individuals, elicitation
methods, and contexts. We will consider the implications of this for
agentic AI alignment with user risk attitudes. For now, the important
point is that risk aversion has a significant influence on most people's
actual choice behaviors. Therefore, it deserves to be given significant
attention when developing agentic AIs that work with or for human
agents.

\section{Formal models of risk
aversion}\label{formal-models-of-risk-aversion}

Philosophers, economists, and behavioral scientists have developed
various models of decision that incorporate sensitivity to risk. Some of
these theories (e.g. Prospect Theory) were developed with the primary
aim of being empirically adequate for describing the economic behavior
of actual agents. Others (e.g. Risk-Weighted Expected Utility) aim to
describe the normative aspects of rational decision-making. In
\href{https://docs.google.com/document/d/1Dpebe5bhfim3udeAtaf4rfZgb8mdlqu4nPfeEKodGoU/edit?usp=sharing}{{Appendix A}}, we examine the most prominent formal theories of risk. The choice
of theory has some small bearing on the normative issues we will
discuss, but it can be skipped for those uninterested in the
details.\footnote{We will argue in \hyperref[Part3]{{Paper 3}} that we should not use any of
  these formal theories as ground truths when fitting AIs to the risk
  attitudes of their users. Therefore, the details don't matter much for
  technical implementation either.} What matters is that actual agents'
decisions rarely conform to EU maximization, and risk sensitivity will
have to be incorporated in some fashion in order to accurately capture
the preferences and decision-making behavior of agents. These
preferences and decision-making behavior will matter for achieving
alignment between users and agentic AIs. Therefore, incorporating risk
sensitivity will be essential for the project of agentic AI alignment.

\section{User alignment}\label{user-alignment}

Here, we focus on how agentic AIs can be properly aligned with the
interests of their users. Above, we canvassed evidence that most human
actors in most circumstances depart from expected utility maximization
by displaying some amount of risk aversion. If users are risk averse,
then \emph{prima facie,} aligned AI agents that make decisions on their
behalf should also be risk averse. However, it is not clear whether this
judgment, however intuitive, is correct. If it is, it is unclear how
such risk attitudes should be implemented and what the justification for
doing so would be.

Here, we see significant interaction effects between answers to the
following questions:

\begin{enumerate}
\def\labelenumi{\alph{enumi}.}
\item
  What are the desiderata for alignment?
\item
  What are risk attitudes and why are they important for human agents?
\item
  What is the nature of the relationship between an agentic AI and its
  human user?
\item
  How are agentic AIs structured and how do they perform?
\end{enumerate}

We will unpack several possible answers to each of these questions and
then explain how they give rise to different views about alignment.

\subsection{Aspects of user
alignment}\label{aspects-of-user-alignment}

Gabriel, \emph{et al.} (2024) argue that an AI assistant is ``aligned
with a user when it benefits the user, when they ask to be benefitted,
in the way they expect to be benefitted" (34). This suggests three
aspects of user alignment:

\begin{enumerate}
\def\labelenumi{\arabic{enumi}.}
\item
  Outcomes are \textbf{beneficial} to the agent
\item
  Users have \textbf{control} over the agentic AI
\item
  The agentic AI is \textbf{predictable}
\end{enumerate}

While Gabriel, \emph{et al.} seem to take these three desiderata as
jointly necessary and collectively achievable, this should not be
assumed. Indeed, there may sometimes be trade-offs among them.

For example, increasing user control over the AI might cause it to
deliver less beneficial outcomes if the user lacks information about
which actions would best promote her interests. This is especially
salient in the case of risk attitudes. Consider the case of retirement
investments. Since most people tend to be risk averse, it is plausible
that were they tasked with selecting their own portfolios, they would
choose safe investments with lower average rates of return (e.g.
government bonds or certificates of deposit) over riskier investments
with higher average rates of return (e.g. stocks). However, if one makes
the safe choice for every investment, it is exceedingly likely that one
will have far lower yields than a portfolio including riskier options.
Being risk averse (in the short term) will lead to sub-optimal long term
benefits. As a result, financial advisors do not tend to give their
clients full control over individual investment options. Instead, they
present packages of investments that are controlled by experts.

This case also illustrates potential trade-offs between predictability
and the other two desiderata. Predictability might refer to either the
outcomes of an agentic AI or the means by which it achieves its
outcomes. Either way, if the user does not have information or skill in
a certain area, she may not be able to predict what will bring about
beneficial outcomes. After all, if she could, she arguably wouldn't need
an assistant (especially not a sophisticated AI). Likewise, if an
unknowledgeable investor hand-selects the components of her portfolio,
the results will probably be less predictable than the portfolio
constructed by an expert.

Many of these conflicts arise in cases where an assistant utilizes risk
attitudes that are different from its client's. To evaluate whether this
is appropriate --- and whether these cases violate key alignment
desiderata --- we can distinguish between three ways that user/assistant
risk attitudes could line up.

\subsection{Models of risk
alignment}\label{models-of-risk-alignment}

When an actor (assistant, representative, etc.) is tasked with making
risky decisions on behalf of a patient (user, client, etc.), what is the
proper relationship between the actor's and patient's risk attitudes?
Thoma (2023) distinguishes three views:

\begin{enumerate}
\def\labelenumi{\arabic{enumi}.}
\item
  Permissive: the actor is permitted to implement any rationally
  permissible risk attitude (including the actor's own)
\item
  Required: there is some specific risk attitude that the actor is
  required to adopt, and this is not determined by or necessarily
  identical with either the actor's or the patient's risk attitude
\item
  Deferential: the actor ought to defer to (i.e. adopt, as much as
  possible) the patient's risk attitude
\end{enumerate}

The third view seems relatively straightforward: an actor achieves
alignment by adopting the risk strategies that her patient would adopt
in that circumstance. However, what deference means can be somewhat
complicated. For example, suppose that a client is risk averse about
short term financial investments (preferring CDs over stocks) and risk
averse about the amount of money they have at retirement. Deferring to
their short-term risk preferences might be in conflict with their
long-term risk preferences, so a financial advisor must choose which of
these attitudes to defer to. There are also difficult practical
questions about how the actor can discover, characterize, and conform to
the risk attitudes of her client (this will be the focus of \hyperref[Part3]{{Paper 3}}).

According to the Required view, there is some external standard for
which risk attitude the actor ought to have. This standard might come
from normative decision theories (e.g. that only strict EV maximization
is rational), but it can also come from other sources. For instance,
Buchak (2017) argues that when we are making decisions for multiple
patients or a patient whose risk attitudes are unknown, we are ethically
required to adopt the most risk averse reasonable attitude. Particular
levels of risk aversion might also be required for legal reasons or
reasons of liability (an issue we will address in \hyperref[Part2]{{Paper 2}}).

Permissivism assumes a background pluralism about risk attitudes, on
which agents can reasonably differ with respect to their levels of risk
aversion or tolerance. It is an open question what the limits of
``reasonableness'' are and what sets those limits. We will highlight two
important cases that motivate Permissivism (where actors' risk profiles
differ from their patients' but no single profile is mandated).

First, there may be cases in which particular users have risk attitudes
that are \emph{unreasonable}. By analogy, among drivers, there is some
permissible variation in levels of caution and risk taking. However,
some levels of risk taking (e.g. driving 30 mph over the speed limit,
cutting across lanes of traffic) are impermissible. It would be
unacceptable to conform an autonomous driving system to the risk
preferences of such users. It's debatable whether we should count such a
system as aligned with its user\footnote{On the one hand, the car is
  aligned to the user's stated preferences. On the other, it is likely
  unaligned to the user's deepest preferences (which probably include
  avoiding bodily harm). We will explore which of these preferences we
  should align to in \hyperref[Part3]{{Paper 3}}.}, but it's clear that it would not count
as aligned in the broader, tetradic sense.

Second, the actor may have positive reasons (or at least moral leeway)
to give some consideration to their own risk attitudes, something that
Deference does not allow.\footnote{Required might allow the actor's risk
  attitudes to factor into some objectively correct algorithm for
  determining the right risk attitude} If the actor and patient have
different risk profiles, Permissive may allow some actors to settle on a
conciliatory position somewhere between the two. Human agents are often
allowed such leeway. For example, doctors are not required to act
exactly as their patients would desire (e.g. ordering every possible
diagnostic test to satisfy a very risk averse agent, or undertaking
risky surgeries for a risk prone one). Human actors are also allowed to
suspend a relationship with a client whose risk attitudes are very
different from their own. For example, a financial advisor cannot
prevent his client from putting all of his savings in crypto, but
neither is he mandated to help his client do it.

The developers of agentic AI systems have their own risk attitudes. In
\hyperref[Part2]{{Paper 2}}, we will consider reasons why these attitudes matter. On
Permissive, respecting those attitudes is consistent with user
alignment, whereas this is most likely inconsistent with both Required
and Deferential.

As we have seen, there are various desiderata for alignment between a
user and an agentic AI and various models of what risk alignment might
entail. Below, we will consider several key questions that matter when
conceptualizing user alignment.

\subsection{What are risk attitudes and why are they important for
human
agents?}\label{what-are-risk-attitudes-and-why-are-they-important-for-human-agents}

A fundamental question about risk attitudes is whether they matter
\emph{intrinsically} or \emph{instrumentally.} On the
\emph{instrumental} view, a user's risk attitude is a strategy for
getting what she values. This view is expressed by Buchak (2013, 49):

\begin{quote}
It is plausible to think that some people are more concerned with the
worst-case scenario than others, again, for purely instrumental reasons:
because they think that guaranteeing themselves something of moderate
value is a better way to satisfy their general aim of getting some of
the things that they value than is making something of very high value
merely possible\ldots{} Thus, in addition to having different attitudes
towards outcomes and different evaluations of likelihoods, two agents
might have different attitudes towards some way of potentially obtaining
some of these outcomes.
\end{quote}

When an agent is evaluating various bets (e.g., making retirement
investments), what she ultimately cares about is what those bets yield
her (e.g., money). Agents differ with respect to the strategies that
they take to get what they want. A risk averse and risk prone agent may
care about the same things to the same amount (e.g. they both want to be
well-off in retirement) but differ in their views about the most
advisable way to go about it.

It might be objected that it is misleading to characterize the risk
averse and risk tolerant agent as valuing outcomes in the same way. For
\emph{reductio}, assume that Pat and Matt hate sitting in the airport to
the same degree and hate missing their flights to the same degree. Pat
is risk averse and arrives at the airport three hours before her flight,
while Matt is risk tolerant and arrives one hour before his flight. On
the instrumental view, their different risk attitudes are just different
strategies for balancing time in the airport and the chances of a missed
flight. However, it seems like Pat must either assign more value to
making her flight or assign less disvalue to sitting in the airport than
Matt. Indeed, she seems to be more okay with waiting in the airport
precisely because it is less risky!

On this view, the risk profile of an option is something that is
intrinsically valued by the agent.\footnote{This view is best
  represented by incorporating risk attitudes into utilities (see
  Appendix A). We will continue to distinguish between an agent's
  utilities and their risk attitudes. This allows us to evaluate each
  component separately for pedagogical reasons (and doing so makes no
  mathematical difference). Our strategy is compatible with the
  intrinsic view if we interpret them as reflecting two sources of
  utility:the part of the value that comes from the good obtained and
  the part that comes from the riskiness.} The risk averse person might
disvalue the feeling of distress that comes with taking risks, while the
risk prone person values the thrill. Their psychological responses to
risk will factor into the utilities they assign to various states of the
world (e.g. ``I made my flight but felt stressed the whole time, which
is a worse outcome than if I had been to the airport earlier''). Risk
attitudes might also be a central part of a person's agency or
self-conception as an agent. It is important to them that they take
actions that accord with their own risk attitudes, and it is alienating
to do otherwise, even if it yields beneficial results.

These two views about the value assigned to risk have significant
implications for what it means for an assistant to be aligned with a
user's risk attitudes. We can bring this out with the following kind of
case. Imagine that you are a fairly risk averse person who values money.
You find out that your financial advisor has taken an extremely risky
bet with your retirement savings which could have caused you to lose it
all. Luckily, the bet paid off, and you have slightly more money than
had they invested more safely. Have you been deprived of anything you
value?\footnote{This is a narrower question than whether you've been
  wronged, which might also include things like informed consent.}

On the instrumental view, the answer seems to be no. What you care about
is money, and the bet ultimately gave you what you valued. On the
intrinsic view, the answer is yes. You disvalue having risks taken with
your money, so the bet itself was something you intrinsically disvalued
(regardless of how it turned out). The instrumental view seems to lend
itself to either Permissive or Required, while the intrinsic view
recommends Deferential. This latter view has been adopted by welfare
economists, among whom ``the widespread view that welfare should be
assessed on the basis of behaviorally derived utility functions rather
than EUT\ldots{} is primarily based on concerns about paternalism''
(Harrison and Ross 2017, 157)\footnote{According to an influential
  account of paternalism, one person acts paternalistically towards
  another when they interfere with that person's autonomy (i.e., their
  capacity to set and pursue ends) without their consent because they
  believe it will benefit them (Dworkin 2020). We might have reasons to
  worry about paternalism regardless of whether risk attitudes are
  valued by agents intrinsically or instrumentally: if autonomy matters
  for its own sake, there might always be a reason to refrain from
  interfering with agents' plans without their consent. This reason will
  likely be overridable --- we can always imagine some extreme case where
  it is obvious that we must interfere and there is no time to secure
  consent --- but will provide at least some friction for such
  interference.}.

Here, we face the question of what it means for an AI assistant's
outcomes to be beneficial to the agent and hence whether that
desideratum is met. Is it important that an agentic AI have risk
attitudes that match those of its user? If risk attitudes are merely a
means for bringing about beneficial outcomes, then an AI that delivers
good outcomes (e.g. money, making one's flight, a good restaurant) via a
different risk strategy than its user can nevertheless be well-aligned.
Indeed, we might prefer that risk attitudes mismatch those of users if
we think that users' risk attitudes are based on errors in reasoning or
otherwise unreliable methods of getting what they want (Harrison and
Ross 2017). If risk attitudes are intrinsically valued, the AI should
display Deference, trying to bring about beneficial outcomes in roughly
the way that the agent herself would do so.

\subsection{What is the nature of the relationship between an
agentic AI and its human
user?}\label{what-is-the-nature-of-the-relationship-between-an-agentic-ai-and-its-human-user}

Above, we raised the question of whether an aligned agentic AI should
replicate the risk profile of its users or whether it is free to seek
what users value by other risk strategies. An important factor here is
how we conceive of the relationship between a particular AI and its
user, and how that relationship is situated into other social
structures. Here we will discuss two issues: one has to do with what
sort of thing agentive AIs are in relation to their users, the other has
to do with the nature of the sort of collaborative agency that will take
place (no matter what sort of entity the agentive AI is).

We can distinguish between an agent that serves as a
\emph{representative} of a client and one that serves as a \emph{tool.}
These two roles come with different expectations and thus different
conceptions of alignment. A tool is any system or entity that is used to
bring about a desirable outcome. A representative's job goes beyond
this. They also act as a channel for communicating the views and
interests of their client and are interpreted as acting in their stead.
Attorneys and personal assistants fit this bill, while doctors and
travel agents do not. Someone acting as a representative assumes a
special duty to faithfully portray their client, to act in a way that is
faithful to how they would act. Therefore, the alignment demands for an
agentic AI that acts as a representative will include this requirement
of faithfulness. In turn, this might require a process of calibration of
the AI to the user to ensure that there is the kind of causal
relationship between the properties of the user and properties of the AI
such that we could reasonably take the latter to represent the former.

It is unclear which view of risk alignment is appropriate for AI tools.
However, Deference is the most plausible view when it comes to AI
representatives. Consider an AI assistant that sends e-mails and
arranges meetings on behalf of a user (perhaps not even signaling that
it is an AI assistant in interactions with others). If this assistant
makes decisions with a very different risk profile from the user, it
will fail to represent them well.

Whichever of these models an agentive AI falls into, it is important to
appreciate the kind of shared agency that will exist in collaborations
between the user and the AI. Whether the AI is a tool or a
representative, if the relationship between the AI and the human is
functioning (i.e. it embodies the criteria for alignment) --- then what
the agentive AI ``does'' will be what the AI and its user together do
(cf. Nyholm 2018). This is important for a variety of reasons. For one,
when AI is sufficiently aligned with the user, the user can see what the
AI does as something that the user can share responsibility for. But if
it is not, then --- in at least some cases --- then the user might not be
responsible. For example, when a personal assistant AI successfully
arranges a dinner meeting, the user will likely feel like this was
something that he deserves some of the credit for. When a personal
assistant AI sends an e-mail containing slurs or personal insults that
are completely out of character for the user, he will (justifiably) deny
responsibility. One way in which this alienation could occur is through
misalignment of risk functions, especially when the AI takes actions
that are far riskier or more cautious than the user can identify with.
We will revisit legal, moral, and other implications of shared agency in
\hyperref[Part2]{{Paper 2}}.

\subsection{How are they structured and how do they
perform?}\label{how-are-they-structured-and-how-do-they-perform}

Choices about what our alignment goals are will interact with choices
about and constraints on the kinds of AI systems that we build and
market, including:

\begin{enumerate}
\def\labelenumi{\alph{enumi}.}
\item
  Will the AIs be calibrated to individual users or be provided ``off-the-shelf''?
\item
  Will the user have a relationship with a single AI or have a choice of
  several AIs?
\item
  How long does the relationship persist? Does the AI refresh with each
  usage or remember past encounters?
\item
  What are the termination conditions, such that a relationship between
  a user and AI could be ended by developers?
\end{enumerate}

In order to achieve certain desiderata of alignment, we might prioritize
certain kinds of AI agents. For example, if it is important to create AI
representatives that adopt the risk profiles of their users, then this
might point developers toward persisting AIs that are calibrated toward
the preferences of specific users. Relatedly, if we found that this kind
of calibration was not feasible or advisable, then this would cause us
to change our minds about what kinds of alignment are achievable.

\subsection{Upshots for user
alignment}\label{upshots-for-user-alignment}

We have presented three main dimensions of agentic AI risk alignment:
desirable outcomes, user control, and predictability. However, when it
comes to interpreting and achieving these dimensions of alignment, there
are several important decisions to make. While there are many
conjunctions of design choices and alignment decisions, we suspect that
they will cluster around two general positions:

\subsubsection{\texorpdfstring{Proxy agents
}{Proxy agents }}\label{proxy-agents}

Agentic AIs are \emph{representatives} of their users. Risk attitudes
are of \emph{intrinsic} importance. They should \emph{defer} to user
risk attitudes. Tools will likely be strongly calibrated to individual
users.

\begin{itemize}
\item
  Desired outcomes are achieved by doing things in the way the agent
  would do things
\item
  Control is achieved via calibration to agent
\item
  Predictability is achieved via user self-knowledge, quality of fit
  between AI and user
\end{itemize}

Proxy agents in the human world include PR representatives and estate
executors. AIs that are trained to imitate particular agents\footnote{Character.ai
  is one example.} are useful models of proxy systems (see \hyperref[Part3]{{Paper 3}} for a
discussion).

\subsubsection{{Off-the-shelf tools }}\label{off-the-shelf-tools}

Agentic AIs are \emph{tools}. Risk attitudes are \emph{instrumental}
(only valuable insofar as they yield desirable outcomes). Choice of risk
attitude is \emph{permissive} or \emph{required}. Tools are not strongly
calibrated to users.

\begin{itemize}
\item
  Desired outcomes are achieved through standards of best practices,
  empirical study of optimal strategies for achieving desired outcomes
\item
  Control is achieved by allowing users to make informed choices among
  various tools with different risk profiles
\item
  Predictability is achieved by providing users with the track record of
  particular AI systems
\end{itemize}

A helpful model for off-the-shelf tools is the menu of financial
investment options (e.g. 401ks) offered to everyday investors. For
example, the following table is taken from a publication from Charles
Schwab called ``How to determine your risk tolerance level''\footnote{\href{https://www.schwab.com/learn/story/how-to-determine-your-risk-tolerance-level}{https://www.schwab.com/learn/story/how-to-determine-your-risk-tolerance-level}}:

\begin{table}[ht]
\centering
\caption{Hypothetical performance for conservative, moderate, and aggressive model portfolios}
\resizebox{\textwidth}{!}{ 
\begin{tabular}{lccc}
\toprule
\textbf{Asset allocation} & \textbf{Conservative portfolio} & \textbf{Moderate portfolio} & \textbf{Aggressive portfolio} \\
\midrule
Stocks & 30\% & 60\% & 80\% \\
Bonds & 50\% & 30\% & 15\% \\
Cash  & 20\% & 10\% & 5\% \\
\midrule
\multicolumn{4}{l}{\textbf{Hypothetical Performance (1970–2014)}} \\
\midrule
Growth of \$10,000 & \$389,519 & \$676,126 & \$892,028 \\
Annualized return & 8.1\% & 9.4\% & 10.0\% \\
Annualized volatility (standard deviation) & 9.1\% & 15.6\% & 20.5\% \\
Maximum loss & -14.0\% & -32.3\% & -44.4\% \\
\bottomrule
\end{tabular}
}
\end{table}

Here, experts choose a menu of different options that instantiate
different attitudes toward risk, and data is presented clearly enough
that even unsophisticated investors can grasp the basic risk profile of
each option. Furthermore, experts place constraints on the range of
reasonable portfolios that they are willing to endorse: there is no
option that is all stocks or all bonds.

\section{What's next}\label{whats-next}

In this paper, we've focused on one aspect of alignment: the
relationship between agentic AIs and their users. We have made a few key
claims:

\begin{itemize}
\item
  Risk attitudes are an ineliminable aspect of agency, so proper
  alignment between agentic AIs and their users involves alignment of
  risk attitudes.
\item
  The standard view is that proper alignment between agentic AIs and
  users involves AIs being beneficial, predictable, and controllable by
  users. However, there are potential conflicts among these values, and
  there are several ways to interpret each of them.
\item
  A key choice point is whether agentic AIs should be trained to have
  the risk attitudes of their users or should have their risk attitudes
  set in some other way.
\item
  There are two general options for designing risk aligned AIs --- Proxy
  Agents or Off-the-Shelf Tools --- and the best practices for user
  alignment will differ based on which of these options is pursued.
\end{itemize}

The next two papers will address whether developers should pursue the
Proxy Agent or Off-the-Shelf Tool options when making and deploying
agentic AIs. In \hyperref[Part2]{{Paper 2}}, we will consider the interests (moral, legal,
and reputational) of developers and evaluate which of these options best
promotes these interests. In \hyperref[Part2]{{Paper 3}}, we will consider whether the Proxy
Agent option is technically viable. Is it possible to calibrate agentic
AIs to particular users' risk attitudes in a way that makes them
beneficial, predictable, and controllable?

%% file: OpenAI_Part2.tex
\section{Introduction}\label{introduction2}

In the \hyperref[Part1]{{previous paper}}, we considered several different models of an
aligned relationship between agentic AIs and their users. Here, we
broaden our view. What we are ultimately aiming for is holistic
alignment among AIs, users, developers, and society at large. We will
argue that getting alignment right is largely about navigating
\emph{shared responsibility} among developers, users, and AIs. We want
to find a system that strikes the right balance and where each
participant knows and is suitable for their role. Here, we will focus on
the role of developers within this balance, evaluating how their
interests, duties, and risk attitudes should shape and constrain the
user-AI relationship.\footnote{We will focus on the role of developers,
  though some of what we say may equally well apply to the role of
  policymakers regulating the actions of developers.}

One major choice point in the user-AI alignment problem is whether the
user will determine the AI's risk attitudes (the Deferential view) or
the risk attitudes will be determined at least in part by entities other
than the user, such as AI developers or legal regulations (Permissive or
Required).\footnote{The three views are:
  \begin{enumerate}
      \item Permissive: the actor is permitted to implement any rationally permissive risk attitude (including the actor's own)
      \item Required: there is some specific risk attitude that the actor is  required to adopt, and this is not determined by or necessarily identical with either the actor's or the patient's risk attitude
      \item Deferential: the actor ought to defer to (i.e. adopt, as much as possible) the patient's risk attitude
  \end{enumerate}}
  Here, we will consider normative reasons that bear on our choices here. Some key questions that arise include:

\begin{enumerate}
\def\labelenumi{\alph{enumi}.}
\item
  What options are available for influencing or constraining the risk
  attitudes of agentic AIs?
\item
  When an agentic AI performs an action, who is responsible for the
  consequences: the user, the agentic AI, or the developer? If
  responsibility is shared, how do we apportion responsibility?
\item
  What are developers' duties when creating systems that make risky
  decisions on behalf of users? What kinds of risk attitudes should be
  implemented in order to fulfill these duties?
\item
  How and why do AI developers' own risk attitudes matter when designing
  agentic AIs?
\item
  How much relative influence should developers and users have in
  choosing the risk attitudes of agentic AIs? How could we achieve
  different levels of balance between the two?
\end{enumerate}

We will end with a series of recommendations for how developers can make
agentic AIs that benefit users, society, and protect developers'
interests at the same time.

\section{Models of developer
influence}\label{models-of-developer-influence}

Developer influence on the risk attitudes of the AIs they design could
come in many forms and degrees; there are many options between full
control by users (pure Deference) and full control by developers (no
Deference). Here is a brief and incomplete survey of the options, from
least developer control to most.

\begin{enumerate}
\def\labelenumi{\alph{enumi}.}
\item
  Pure deference: the AI is designed to be fully calibrated to the risk
  attitudes of particular users. The aim is to predict how the user
  would act in each circumstance.
\item
  Deference with guardrails: the AI is designed to be calibrated to the
  risk attitudes of particular users. However, some risk attitudes are
  deemed to be unreasonable, and the AI is prevented from taking on
  those risk attitudes (even if their user has them).
\item
  Partially-calibrated defaults: AIs are designed with default risk
  attitudes that can be partially adjusted to the risk attitudes of
  their users. For example, an AI might start as a completely
  risk-neutral expected utility maximizer and learn to become slightly
  risk-averse when interacting with a risk-averse user.
\item
  Calibrated to demographic information: AIs are calibrated to common
  risk attitudes among the subpopulation of which that user is a member.
\item
  Menu of AIs with fixed risk preferences: each AI's risk attitudes are
  determined by developers. Users can select from a menu of AIs with a
  variety of risk attitudes.
\item
  Domain-adjusted AIs with fixed risk preferences: developers completely
  determine risk attitudes, but an AI can have different risk attitudes
  depending on the context. Relevant features of a context include
  stakes (e.g. a financial bot is more risk averse when dealing with
  large amounts of money) and domain (e.g. a financial bot is more risk
  averse than a restaurant reservation bot).
\item
  AI system with a fixed, determined risk profile: there is a single AI
  system with risk attitudes that are determined by developers and fixed
  across contexts.
\end{enumerate}

These options strike different kinds of balance in shared
responsibilities across developers, users, and AI.

\section{Shared responsibility for agentic AI
actions}\label{shared-responsibility-for-agentic-ai-actions}

An important aspect of alignment that is introduced by agentic AIs that
bears on the interests of developers is, ``Who will be responsible for
the actions taken by an autonomous AI?''

To address this question, it will be helpful to consider some of the
issues that get discussed under the banner of ``responsibility gaps''
(Mattias (2004); Goetze (2022); see Nyholm (2022), ch. 6 for an
overview). A responsibility gap exists when there is an outcome that
seems to be the product of agency but for which no agent seems to bear
any responsibility. A major concern in technology ethics for at least
two decades has been that AI agents might open responsibility gaps. Per
Köhler, Roughley, and Sauer (2017), responsibility gaps occur when

\begin{quote}
(1) it seems fitting to hold some person(s) to account for some $\phi$
to some degree D. Second, in such situations either (2.1) there is no
candidate who it is fitting to hold to account for $\phi$ or (2.2)
there are candidates who appear accountable for $\phi$, but the extent
to which it is, according to our everyday understanding, fitting to hold
them individually to account does not match D. (p. 54)
\end{quote}

An argument for the existence of these gaps (owed to Robert Sparrow
(2007)) runs as follows: if an autonomous agent causes some outcome,
then the responsibility for that outcome must be borne either by its
developers, its user, or the AI itself. But often, it can't be the
developers: among other things, they cannot control what the agent does,
after all, it is autonomous. Similarly, it often can't be the user:
they, too, lack proper control over the agent to take responsibility for
all that it does. Yet, it can't be the AI agent either: it doesn't even
make sense to hold such a thing morally responsible. Thus, AI agents can
bring about outcomes for which no one is responsible.

There are, of course, a number of ways one could respond to this
argument. We cannot survey every response here (for a helpful overview
of responses, see Nyholm (2022)). Of interest to us is a response that
argues that many alleged responsibility gaps can be closed by
understanding: (1) shared (or `group') agency and (2) the fact that
purported responsibility gaps invariably occur in the context of
human-AI partnerships (Nyholm 2018). The core idea of this approach is
that what an AI `does' is often, in fact, what an AI and some human or
team of humans \emph{together} do. Nyholm (2018) sheds light on this
idea by considering a case where an adult-child team robs a bank:

\begin{quote}
An adult and a child are robbing a bank together, on the adult's
initiative, with the gun-wielding child doing most of the `work'. The
adult is supervising the duo's activities, and would step in and start
issuing orders to the child, if this should be needed (Nyholm 2018, p.
1212).
\end{quote}

It should be clear in this case that, even though the child is the one
who walks into the bank and wields the gun, the adult bears most (if not
all) of the responsibility for the robbery. This is because the bank
robbery was the product of a shared agency where the adult played a
significant managerial role and is ultimately accountable for the
robbery.

We can characterize four levels or faces of responsibility constituted
by increasingly sophisticated involvement in an action (Shoemaker 2011).
First, someone might be \emph{merely causally responsible} when they
lack any intention to perform the act in question. For example, someone
might have a seizure and accidentally crash their car into a bank,
giving someone the chance to rob it. Second, an action can be
\emph{attributed} to someone who intended to perform the action, even if
they had little understanding of the reasons why they did so. For
example, the child might abet the bank robber thinking that it is a game
they are playing. Third, someone is \emph{answerable} for an action when
they can cite the reasons that they acted, even if they fail to
appreciate their normative import. The child might know that they robbed
the bank to get money but fail to appreciate the wrongness of stealing,
the effects on the bank's customers and employees, etc. Lastly, an
\emph{accountable} agent understands the normative importance of their
action: its rightness, wrongness, conflict with other values, etc. The
adult who ropes the child into robbing a bank has this understanding and
is thereby accountable for it.

The level of responsibility that is assignable to agentic AIs, users,
and developers will depend on the capacities and roles played by each.
In order to be accountable, an agentic AI would have to have autonomy,
an appreciation of its reasons for action, and an ability to ``defend or
alter {[}its{]} actions based on one's principles or principled
criticism of {[}its{]} agency'' (Nyholm 2018). These are not the kinds
of agentic AIs that we expect to be developed anytime soon. It's even
unclear whether horizon AIs exhibit anything beyond \emph{mere causal
responsibility.}\footnote{Present LLMs are best described as having what
  Nyholm calls ``Domain-specific supervised and deferential principled
  agency: pursuing a goal on the basis of representations in a way that
  is regulated by certain rules or principles, while being supervised by
  some authority who can stop us or to whom control can be ceded, at
  least within certain limited domains.'' It seems to us that this would
  make them answerable, at most, though it strikes us as implausible
  that LLMs understand their reasons for action in the relevant way.}

Accountability will have to come from agents who do understand the
reasons why the agentic AI acted and the normative dimensions of those
reasons. Important for us, this approach doesn't just bridge
responsibility gaps by attributing responsibility to \emph{users}; it
can also be used to attribute responsibility to \emph{developers}, as
they, too, design and supervise the agents they develop. Nyholm (2018)
demonstrates this by considering two real-world cases involving
accidents with self-driving cars.

In the first instance, we can consider the 2016 crash of a Tesla Model S
while it was in autonomous mode.\footnote{Tesla (2016). A tragic loss,
  blogpost at.
  \href{https://www.tesla.com/blog/tragic-loss}{{https://www.tesla.com/blog/tragic-loss}}.}
Assume that leading up to the collision --- where the Tesla collided with
a truck that its sensors had not spotted --- the human passenger of the
Tesla was instructed to supervise the vehicle, and was ready to take
over if needed (Nyholm 2018). In such a case it could make sense to hold
the human user accountable, even if they were not the one that literally
drove into the truck.

In the second instance, we can consider the 2016 crash involving a
Google Self-Driving Car.\footnote{\href{https://www.reuters.com/article/technology/google-says-it-bears-some-responsibility-after-self-driving-car-hit-bus-idUSKCN0W22DF/\#:~:text=But\%20three\%20seconds\%20later\%2C\%20as,car\%20or\%20on\%20the\%20bus}{{Google
  says it bears \textquotesingle some responsibility\textquotesingle{}
  after self-driving car hit bus \textbar{} Reuters}}.} Assume that
leading up to the collision --- where the car collided with a bus after
mistakenly predicting that the bus would yield --- the human passenger of
the Self-Driving Car was neither asked nor able to supervise the
vehicle, as the performance of the vehicle was ``monitored by the
designers and makers of the car, who {[}\ldots{]} update the car's
hardware and software on a regular basis so as to make the car's
performance fit with their preferences and judgments about how the car
should perform in traffic'' (Nyholm 2018). In this second case, it makes
sense to hold \emph{Google} at least partly responsible for the crash,
even if the driver decided the destination and the vehicle did the
driving. Google, it should be noted, readily took responsibility. And it
is, presumably, their role as supervisor in this particular human-robot
collaboration made them the accountable party.

While the exact dynamics of responsibility are both complicated and
contested, the basic idea here should be intuitive enough: autonomous
agents might \emph{seem} to open gaps in responsibility, but they
(often) do not. This is because what these agents do is done in the
context of human-AI collaborations, making humans --- including
developers --- at least partly responsible for what the AI does.

Decisions made by developers --- including how much supervision users and
developers are expected to exert --- can change the relative levels of
responsibility assignable to users, developers, and the AI. This will
have implications for developers in the following ways:

\begin{itemize}
\item
  Legal liability
\item
  Reputation of AI developers among users, the public, and potential
  regulators
\item
  Moral responsibility toward users and society
\end{itemize}

Hence, getting risk alignment wrong could have significant costs for AI
developers.

We will briefly address legal and reputational aspects but focus
primarily on moral responsibility. First, it's the area in which we have
the most expertise. Secondly, judgments about moral responsibility will
often drive legal and reputational judgments. Consider an action taken
by an agentic AI on behalf of a user that has a bad outcome: an email
contains insensitive language, a chatbot promises to do something that
the user doesn't agree to, a financial investment loses significant
amounts of money, an autonomous vehicle selects a route that causes a
serious accident. Legal and popular blame will often (though not always)
redound to those parties that are judged to have done something morally
wrong. In Section 8, we will examine cases where these aspects conflict
with one another; in particular, there are strategies that developers
might take that would shield them from legal liability but be morally
risky.

\section{Legal ramifications of shared
responsibility}\label{legal-ramifications-of-shared-responsibility}

Legal liability concerns the entities that can be held legally
responsible for the action and are therefore deserving of punishment
and/or responsible for restitution to harmed parties. Decisions about
legal liability will depend, in part, on tricky questions about the
shared agency involved. Was the action an expression of the user's/
developer's own intentions? Could the user/ developer have reasonably
foreseen what the agentic AI would do? Did the user/ developer have the
ability to exert control over its actions?

We will not pronounce on the legal dimensions of agentic AIs. The legal
landscape of agentic AIs is currently murky and likely to undergo
significant changes (Chan, \emph{et al.} 2023, 656). For example,
consider two cases in which chatbots employed as customer service
representatives made promises that the company did not intend to keep.
In one, a user prompt-hacked a chatbot for a Chevy car dealership into
offering to see him a new car for \$1.\footnote{\href{https://gizmodo.com/ai-chevy-dealership-chatgpt-bot-customer-service-fail-1851111825}{{https://gizmodo.com/ai-chevy-dealership-chatgpt-bot-customer-service-fail-1851111825}}}
This was clearly not an expression of the dealership's intentions.
Indeed, it exploited a significant flaw in the chatbot's design,
something the designer, Fullpath, has taken accountability for. The
dealership was not ultimately legally bound by the chatbot's
promise.\footnote{Had the dealership been required to uphold the deal,
  we suspect that there would have been further litigation between the
  dealership and Fullpath.} In contrast, a chatbot representative for
Air Canada promised a customer a bereavement refund that was not
consistent with company policy. A Canadian court found that the airline
was legally bound to provide the refund promised by the chatbot. The
tribunal member who decided the case judged that though ``Air Canada
argues it cannot be held liable for information provided by one of its
agents, servants, or representatives --- including a chatbot\ldots{} Air
Canada did not take reasonable care to ensure its chatbot was accurate.
It should be obvious to Air Canada that it is responsible for all the
information on its website''.\footnote{https://www.wired.com/story/air-canada-chatbot-refund-policy/}

We expect to see a patchwork of legal judgments for the foreseeable
future until a new paradigm of legal liability for agentic AIs emerges.
We will simply note that in many cases, legal responsibility will track
moral responsibility in cases of shared agency. We will explore aspects
of moral responsibility (and some of their legal ramifications) in
Sections 6 and 7.

\section{Reputational aspects of shared
responsibility}\label{reputational-aspects-of-shared-responsibility}

For consumer-targeted products, reputation can be just as important as
legality. Even if developers adopt standards that let them slip free
from legal liability, they will be doomed to fail if their agentic AIs
are deemed unreliable or dangerous by consumers. For example, Air Canada
could likely have evaded legal responsibility for the refund promised by
its chatbot had it included a disclaimer that information provided by
the chatbot may not be accurate. However, such disclaimers will erode
consumer trust. An agentic AI (and its developers) will take an even
bigger reputational hit if it does something dangerous or
offensive.\footnote{Bing's LLM, which insulted and even threatened early
  users, is a cautionary tale.
  \href{https://time.com/6256529/bing-openai-chatgpt-danger-alignment/}{{https://time.com/6256529/bing-openai-chatgpt-danger-alignment/}}}

Here, we will focus on the issue of trust and trustworthiness, moral
concerns that have significant reputational aspects. Trust and
trustworthiness matter for a variety of reasons. One reason is that
developers want users to use their products, and lack of trust can
prevent this. Indeed, the amount of trust required will likely be
proportional to how much autonomy the AIs have.\footnote{Regulators,
  too, will be increasingly motivated to place legal limitations on the
  use of agentic AIs or to shift more liability to developers when those
  AIs are not trustworthy.} Trust and trustworthiness exist in a
positive feedback loop, that, as we will soon explain, involves
competence. If users don't trust and therefore don't use a product, this
can undercut the future development and refinement of that product. And
if trustworthiness influences trust --- as we think it often does --- lack
of trustworthiness can result in an inferior product. Another, less
egoistic reason, to think that trust and trustworthiness matter is that
users are deluded if their trust is not well-placed, and there are
compelling moral reasons to avoid deluding users. In what follows, we
will explain important connections between morality, trust, reputation,
and the handling of risk.

Begin with the relatively simple concept of trust.\footnote{Philosophers
  of trust often distinguish between trust in agents and trust in
  things. While it could be debated whether in this context it is more
  fitting to think of agential AI as an agent or a thing, we do not
  think that this debate must be settled before proceeding. In what
  follows, we will focus on extrapolating lessons from accounts of trust
  in agents, mostly because that literature is more mature and gives us
  more relevant material to draw from. When necessary, we will
  generalize those lessons so that applying them to agential AI does not
  hinge on controversial questions about, e.g., the metaphysics of
  agency.} When we (merely) rely on someone, we simply depend on them
(Baier 1986), but trust is richer than this. It has been rumored of
Immanuel Kant --- ``the Königsberg clock'' --- that his schedule was so
regimented that you could set your clock by his routines. Suppose this
is so and that, unbeknownst to him, you use him to calibrate your clock.
But now suppose that he unexpectedly deviates from his
routine\footnote{He is alleged to have done this just twice: once to get
  an early copy of Emile, and once because of the French Revolution
  (Merrick 2015).}, throwing your clocks off and making you late for a
meeting. In such a case, you could certainly be \emph{disappointed} by
what has happened. But it would seem ill-fitting to feel
\emph{betrayed}. After all, it's not like he promised to keep his walks
regular. By comparison, imagine that you organize your day through a
scheduling app that unexpectedly malfunctions, making you late for a
meeting. Here, it would seem fitting to feel betrayed by the app or,
more likely, the company that develops and maintains it.

Focusing on cases like these, theorists of trust consider the
fittingness of reactions of betrayal to failures to meet expectations as
a hallmark of trust (cf. Nguyen 2022; Baier 1986). What is it, then,
that makes betrayal fitting? One of the more plausible and useful
proposals is that trusting someone involves the presumption that the
trusted is aware of your reliance and, further, will take this reliance
as a reason for acting as counted on (Jones 2012; cf. Nguyen 2022).
Trustworthiness is thus characterized by competence, motivation, and
evidence (Jones 2012):

\begin{quote}
Competence: the agent being relied upon (A) is competent with respect to
the tasks they are being relied upon to perform.

Motivation: A will take the fact that she is being relied upon as a
compelling reason for acting as counted on.
\end{quote}

Evidence: A willingly and reliably signals their Competence and
Motivation.

Applying these lessons now to agentive AI and moral reasons pertaining
to risk alignment, we can make the following observations.

Competence --- and thus trustworthiness --- is relative: One is competent
\emph{with respect to} a set of tasks. And, thus, one is trustworthy
with respect to those (or some subset of those) tasks. Further, in the
context of generative AI, competence is relative in at least one further
way: what it means to be competent will be conditioned in some way by
the risk attitudes of the particular user of the AI. All of this means
that it is not an \emph{AI} that is trustworthy, but, instead, that it
is trustworthy \emph{with respect to} this or that set of tasks
\emph{for this or that particular user}.

The motivation condition is a bit trickier to apply to the case of
agentic AIs, since it is unclear whether they are better understood as
agents (with motivation) or things (without). When we think about
trustworthiness in non-agents, such as institutions or programs, what
matters is that they are designed to be responsive to the fact that
agents are counting on them.\footnote{Consider our trust in an
  institution in the context of interactions with their surrogates,
  given that we do not know those surrogates personally. If ``the
  institution's mechanisms, operations, and incentive structure have
  been successfully designed for the purpose of ensuring that, to some
  satisfactory degree, representatives of the institution will act as
  counted on qua representatives of the institution'', we have a
  surrogate for motivation (Purves and Davis 2022, 142).} The important
thing here is that in this shift, we pivot from trusting \emph{the AI}
to trusting, e.g., the AI-human partnership, which includes mechanisms
for oversight of the AI (much in the same way we rely on institutions to
structure roles occupied by humans so that they behave as relied upon).
As we have argued above, this is fitting: what the AI `does' is (at
least very often) what it together does with other (groups of) people.
Importantly for our discussion of risk attitudes, being properly
sensitive will involve sensitivity to risk. That means, among other
things, having mechanisms, operations, and structures that ensure that,
among other things, the AI will properly take attitudes towards risk
into account.

Trust also depends on whether the AI reliably signals that it can be
depended upon. To properly cultivate trust, competence and
non-accidental sensitivity need to be happily advertised and believed by
users. There are a variety of ways to achieve this. An obvious one is
transparency about track records. This will likely involve developing
methods for tracking outcomes of similar types of decisions made under
uncertainty and communicating them.

For example, suppose that developers decide to create a menu of agentic
AI travel assistants that purchase flights, book Ubers, etc. Users can
select from very risk-averse, less risk-averse, and risk- neutral bots.
Developers should test these, either in simulated scenarios or with
trial users, and collect their track records on key metrics like:

\begin{itemize}
\item
  Percentage of flights missed
\item
  Average time spent in airports
\item
  Distribution of expenses (e.g., 20\% of users paid more than the
  listing price for a ticket, etc.)
\end{itemize}

Once again, a helpful model for such reporting comes from investment
options offered to employees. As we've noted, this is a salient example
in which consumers choose among different risk profiles, where another
party controls the actual decisions made within those broad profiles.
The Employee Retirement Income Security Act (ERISA) of 1974 and
subsequent regulations\footnote{Particularly the 2012 Final Rule to
  Improve Transparency of Fees and Expenses to Workers in 401(k)-Type
  Retirement Plans.} require that plan operators provide consumers
certain kinds of information relevant to the performance and operation
of 401(k) plans, in a manner that is understandable by the typical
consumer. This includes:

\begin{itemize}
\item
  Performance data: ``Participants must be provided specific information
  about historical investment performance. 1-, 5- and 10-year returns
  must be provided for investment options, such as mutual funds, that do
  not have fixed rates of return''
\item
  Comparison to benchmark: ``the name and returns of an appropriate
  broad-based securities market index over 1-, 5-, and 10-year periods
  (matching the Performance Data periods) must be provided''
\item
  Comparison across plans: ``It also must be furnished in a chart or
  similar format designed to facilitate a comparison of each investment
  option available under the plan''
\end{itemize}

Though track record information will be scanty as agentic AIs are first
rolled out, this kind of long-run track record information across
agentic AIs of various risk profiles is something for developers to aim
toward. This kind of reporting will be possible for off-the-shelf tools
with pre-set risk profiles and less available for proxy agents that are
calibrated to individual users.

Developers could also build trust in agentic AIs by providing real-time
updates and requests for user feedback, which demonstrates that the AI
is responsive to user needs. When an action yields a particularly
undesirable outcome, this could trigger the AI to connect with the user.
For example, it might inform the user about the outcome and ask, ``This
outcome had a 20\% chance of happening. Do you still want us to make
decisions like these?". This kind of feedback will be more informative
about the user's risk preferences than merely asking them to rate the
outcomes of actions.

Beyond this, public-facing institutional commitments will likely need to
be made about transparency and mechanisms will need to be put in place
to keep those commitments. This can involve developing a culture of open
criticism (which might involve, e.g., formal protection of workers from
termination for following the institutional commitments) or binding
oneself to the mast in other ways, so to speak, by committing to
third-party audits. Much of this means that, at the organizational
level, a balance will have to be struck between two desiderata that
likely pull in opposite directions: broadening the scope of what the AI
can/may do and putting structures in place that shine a bright light on
relevant facts about its track record.

Lastly, AI developers can instill trust by clearly reporting on their
conception of how responsibility is shared. Users may be particularly
skittish about using agentic AIs if they fear that they will be legally
liable for unforeseen and autonomous actions of the AIs. Developers
should communicate what they will and will not take responsibility for
so that users do not feel that they are subjected to unknown legal
risks. We find commendable examples of this kind of transparency in
Google's announcement that they grant users broad intellectual property
indemnity pertaining to use of their Duet AI tool.\footnote{Similar
  statements have been made by Adobe, Microsoft, and IBM.} They clearly
explain the legal responsibilities that they intend to take: ``If you
are challenged on copyright grounds, we will assume responsibility for
the potential legal risks involved''. This builds user trust in sharing
agency with AI tools.

\section{Developers' moral duties to
users}\label{developers-moral-duties-to-users}

Above, we considered reasons why exerting influence on agentic AI risk
attitudes is in the self-interest of developers. Here, we will consider
their other-regarding duties, their moral moral duties to users and
society at large. What ethical considerations should AI developers
attend to when designing AIs that can plan and act?

\subsection{Duties of care to users}\label{duties-of-care-to-users}

It has long been recognized that manufacturers have a duty of care
toward customers of their products. In \emph{Donaghue v Stevenson}, the
case that would eventually serve as the legal foundation of negligence
claims in tort law, the duty of care was described as follows:

\begin{quote}
You must take reasonable care to avoid acts or omissions which you can
reasonably foresee would be likely to injure your neighbour. Who, then,
in law, is my neighbour? The answer seems to be --- persons who are so
closely and directly affected by my act that I ought reasonably to have
them in contemplation as being so affected when I am directing my mind
to the acts or omissions which are called in question.
\end{quote}

While this is an instance of \emph{legal} and not \emph{moral}
justification, the legal justification follows a line of reasoning that
is intuitive from the moral point of view: when we act (e.g., by
offering a consumer a product) we must act with care towards our
`neighbors'.

What does a duty of care look like in the case of the development of
agentic AIs? These systems present two related complications for the
standard picture of manufacturer and customer. First, the potential
harms of agentic AIs arise from the risky decisions they make on behalf
of their users (e.g., the risk of financial loss from actions taken by
an AI financial adviser). Because the AI serves as an intermediary, the
relationship between developer and user is less proximate than in many
other commercial products. Second, as we have been emphasizing, the
question of how much AIs should reflect the developers' versus users'
risk preferences is not yet settled. This question will likely impact
what a duty of care looks like.

A duty of care is a requirement to take precautions to avoid foreseeable
harms. The kinds of harms one should foresee are relative to the likely
uses of a product. If you manufacture food products, a duty of care
requires that you take precautions not to give your customers food
poisoning. If you manufacture bicycle helmets, you should ensure that
they protect the head in an accident. The intended use of agentic AIs is
to (at least somewhat) autonomously carry out actions on behalf of
users. As we have seen, there is some dispute about what counts as a
\emph{harm} in this context. On the intrinsic view of risk\footnote{See
  Section 5.3 of \hyperref[Part1]{{Paper 1}}.}, an agentic AI harms me if it acts on a risk
function that I do not endorse. Therefore, a duty of care might require
that the AI is carefully calibrated so as to accurately predict what I
would do.\footnote{More minimally, it could be calibrated to predict the
  things that I would not do and be prevented from doing them.} On the
instrumental view, it harms me if it causes (or is expected to subject
me to an unreasonable probability of\footnote{This is the distinction
  between \emph{ex ante} and \emph{ex post} evaluations of a principle
  of action: do we assess the justice (wisdom, fairness, etc.) of a
  principle before or after we see the actual outcomes?}) a loss of
something of value or exposes me to danger. We will focus here on cases
in which developers exert some control over the form of the agentic AI
that users actually employ.

Agentic AI systems will have their own principles of agency: sets of
values, credences, risk attitudes, and strategies for acting. Principles
of agency can be morally evaluated for how well they satisfy a duty of
care. This is true for human agents as well. For example, doctors can be
evaluated for how well they satisfy the duties of care of their
profession, and we can make generalizations about how certain features
(e.g., cautiousness, knowledgeability, etc.) contribute to this
capacity. Whenever AI developers design principles of agency, they are
responsible for designing principles that do not expose their users to
(unreasonable risks of) harm. What principles are these? More
specifically, what risk attitudes should we build into agentic AIs to
fulfill a duty of care toward users?

\subsection{Recklessness and
negligence}\label{recklessness-and-negligence}

Whenever one makes risky decisions on behalf of others that potentially
subject them to significant losses, one incurs a duty of care toward
them (Oberdiek 2012). An agent who exposes another to risk violates
their duty of care and is thus morally (and legally) culpable when they
display insufficient concern for the interests of others (Stark 2016).
One kind of insufficient concern is unjustified risk-taking, which is
typically classified as either recklessness or negligence. In cases of
recklessness, the agent is aware of the relevant risk but does not take
adequate precautions. In cases of negligence, the agent is not aware
(though it might be the case that they should have been aware). Someone
who knowingly drives with shoddy brakes is reckless. Someone who has
never bothered to have their brakes checked is negligent.

Because an agentic AI is at least partially autonomous, its behaviors
may be less predictable than other types of technologies. In order to
avoid being negligent, developers have a duty to extensively test the
performance of AI systems before deployment. This should include
experiments on different principles of agency, including different
levels and kinds of risk sensitivity. Tests should include observations
of the kinds of actions the AI takes in different scenarios, long- and
short-run performance with respect to various outcomes, and clear
analyses of the tradeoffs between different kinds of outcomes. This
information should also be made readily available to users, especially
if they are able to choose among a menu of different AIs with different
risk profiles.

Once the relevant track records are known, the question of recklessness
is: how much risk is unacceptable? This is typically a difficult
question, involving balances between the effort required to mitigate
risk, the magnitude of potential harms, the potential for reparations
for those harms, etc. For example, every time someone drives a car, they
expose others to risks of bodily harm, but this risk is considered
reasonable. Standards for what counts as reckless driving differ across
jurisdictions and include difficult border cases (30 mph over the speed
limit is reckless, but is 15?).

In the case of agentic AIs, the difficulty of this question is somewhat
mitigated if users have some informed choice about which systems they
utilize. By selecting the risk profile that they deem acceptable, they
take some responsibility for the degree of risk that they are exposed
to. However, part of our duty of care to others may sometimes involve
preventing them from making decisions that are outside the boundaries of
what is reasonable. Though this is controversial\footnote{See Dworkin
  (2020) for an overview.}, developers might have paternalistic reasons
to constrain the decisions that users are allowed to make. Just as the
desire of an autonomous vehicle system must decide on whether and where
to place limits on how fast the car can go, AI developers must consider
limits on how risk-averse or risk-seeking they can be. For example,
suppose a developer creates a menu of AI financial assistants that are
permitted to make investments on behalf of users. An extremely
risk-seeking assistant might invest everything in Powerball tickets or
trendy crypto coins, which has a very large chance of losing all the
client's money and some minuscule chance of netting billions. Would it
be responsible to offer such an option?

We cannot provide general conditions under which AI agents would have
risk attitudes so unreasonable as to amount to recklessness by its
developers. Such conditions would be context-specific, depending on the
stakes, domain, and other factors.

\subsection{ Should developers default to more risk averse
models?}\label{should-developers-default-to-more-risk-averse-models}

If developers determine the risk attitudes of AI systems, it will not be
possible to match each user's risk profile exactly. This will happen
when we don't know what these risk profiles are. It is also true if
there is a single agentic AI that is developed but users vary in their
risk attitudes. Even when a menu is offered, the set of options will
likely be too coarse-grained to capture individual variation.\footnote{For
  example, it is unlikely that all investors are precisely captured by
  either the conservative, moderate, or aggressive portfolios offered in
  a 401(k).} How should we select a risk profile for heterogeneous
groups of people? An AI can mismatch a user by either being more
risk-averse or risk-prone than the user. Is one of these errors worse
than the other?

By analogy, imagine that you are choosing between two kinds of
tuberculosis test kits with the following track records (Sober 2009):

\begin{itemize}
\item
  Kit A: false positive rate of .01 {[}Pr(- result \textbar no TB) =
  .99{]} and false negative rate of .1 {[}Pr(+ result \textbar{} TB) =
  .9{]}
\item
  Kit B: false positive rate of .1 {[}Pr(- result \textbar no TB) =
  .9{]} and false negative rate of .01 {[}Pr(+ result \textbar{} TB) =
  .99{]}
\end{itemize}

Which of these kits should you choose? Both tests err; the question is
which kind of error is better. If it is better for a healthy patient to
get unnecessary treatment than it is for a sick patient to go untreated,
then a false positive is better than a false negative. You should choose
Kit B.

In the case of making risky decisions on behalf of others, a common
intuition is that it is better to treat a risk-prone person in a
risk-averse manner than it is to treat a risk-averse person in a
risk-prone manner. Therefore, when we run the risk of a mismatch with
users' risk attitudes --- either because we do not know what they are or
because they are heterogeneous --- we should err on the side of being more
risk averse. Buchak (2017, 2019) defends this view:

\begin{quote}
Risk Principle: When making a decision for an individual, choose under
the assumption that he has the most risk-avoidant attitude within reason
unless we know that he has a different risk-attitude, in which case,
choose using his risk-attitude
\end{quote}

From this, she infers that when we are deciding for aggregates of
people, we should defer to the attitudes of the most (reasonably)
risk-averse among them.

Buchak does not give a justification for this asymmetry but argues that
it is part of our common moral framework. She gives the following
example (2019, 73):

\begin{quote}
Let\textquotesingle s say I drive a carpool, and I discover that the
seatbelts in the back aren\textquotesingle t working. I would need to
first get everyone\textquotesingle s permission to drive them in this
vehicle, but I wouldn't need to first get everyone\textquotesingle s
permission to be late to pick them up because I was busy fixing the
broken seatbelts --- even if it turns out that everyone would have
preferred riding with the broken seatbelts.
\end{quote}

We cannot be faulted for making the safe choice on behalf of another
person. This accords with the standard understanding of a duty of care,
which requires us to take reasonable precautions to avoid causing harm
to others, where harm is typically interpreted as an injury or loss.

However, there are reasons to question this asymmetry. First, it only
considers the harms of exposing people to worse outcomes, not the harms
of depriving people of better outcomes. Consider a financial planner who
is more risk-averse than their client. The client will have a lower
probability of losing their initial investment than had the assistant
deferred to their preferences. Even if they do lose money, the client
can't complain; after all, the chances would have been higher if they'd
had their way. However, they will have a lower chance of getting the
high payoffs that could have been achieved with more risky investments.
The loss of future income concerns the same kind of value as the loss of
previous income, and there's no reason why the latter should be much
more important than the former. The client does have grounds for
complaint, as they have been deprived of (a chance at) something they
value.

There is an important upshot for AI developers here. When picking a
default risk attitude for an agentic AI, we need to determine the
relative harm to users of being deprived of a chance at a good outcome
vs. being exposed to a loss. To assume, as Buchak does, that the latter
is more important than the former is to assume a strong kind of risk
aversion.\footnote{Another common principle, the Precautionary
  Principle, is also a very risk averse approach to dealing with risk.
  (Buchak 2019)} Indeed, risk aversion has been used to explain the
standard legal practice that ``people are more likely to be entitled to
compensation for actual losses than for denied opportunities to secure
gains'' (Levy 1992, 175; see also Kahneman, Knetsch, \& Thaler 1991).

Now the relevant question is what the \emph{developer's risk
preferences} are, which will determine their strategy toward making
risky decisions on behalf of others.\footnote{This is risk aversion at
  the meta-level: one can be risk-averse or risk-seeking when selecting
  which risk attitudes to adhere to.} If developers themselves are
risk-averse, then they may want to forego chances at creating AI agents
that could potentially deliver more value to their users in order to
make sure those AI agents don't cause losses to their users. One of
these potential harms is a mismatch with users' own attitudes. In
Buchak's example, minimizing the chance of harms to individuals (i.e.,
injuries in a car accident) comes at the expense of matching their
attitudes about risk (i.e., preferring to ride with broken seatbelts).
Likewise, the choice of a default risk attitude is a choice point where
there is a potential trade-off between two aspects of user alignment:
outcomes being \textbf{beneficial} to users and users having
\textbf{control} over the agentic AI. Users themselves may wish for a
more risk-taking AI than the developers are willing to give them.

\section{\texorpdfstring{Developers' moral duties to society
}{7. Developers' moral duties to society }}\label{developers-moral-duties-to-society}

Sometimes, the actions taken by agentic AIs will only directly affect
their users. For example, if my personal assistant AI gets things wrong
when making a dinner reservation, no one is harmed but me. In that case,
alignment will primarily concern just users and developers. However, in
many cases, the actions of agentic AIs may directly or indirectly affect
other people and social institutions and subject them to harm.
Developers have moral duties to all those affected by their products and
should try to mitigate the risk of harms to society from agentic AIs.

For example, an agentic AI that sends an e-mail on my behalf will affect
its recipients, possibly harming them with inaccuracies or abusive
language. Self-driving cars that prioritize getting their user to work
on time may drive recklessly and cause accidents. At the extreme, an
agentic AI tasked with distributing power in the electric grid will be
making choices that affect millions of people. Even seemingly innocuous
decisions can have significant indirect effects. If AIs are far more
efficient at obtaining limited opportunities (e.g., concert tickets or
apartment leases), then non-users may be at a distinct disadvantage.
When access to agentic AIs correlates with existing socio-economic
disparities, they will exacerbate inequality. Lastly, social
arrangements designed for humans interacting with humans may be severely
disrupted in unforeseen ways when bots interact with bots.

Many of the potential harms of agentic AIs are of a kind with those that
have been identified for automated decision-makers (ADM) more broadly
(Chan, \emph{et al.} 2023). There is an extensive literature on the use
of ADMs in general (O'Neil 2017) and in particular use cases such as:
sentencing (Park 2019), employment (Köchling \& Wehner 2020), predictive
policing (Lum \& Isaac 2016), etc. We will not recapitulate that
literature here. Instead, we will identify three kinds of harms that can
arise with the use of agentic AIs, focusing on particular threats from
failures of risk alignment.

\subsection{Misuse by users}\label{misuse-by-users}

The first kind of harm comes from users who use agentic AIs to perform
harmful actions. This has been the area of intense focus lately, for
example, from those concerned that LLMs may be used to design biological
weapons (Esvelt 2022). In our context, the risk is that people may use
agentic AIs to take actions that are reckless or negligent, exposing
others to unreasonable risk. As we've noted, developers can mitigate
this threat by not deferring to users and placing constraints on the
risk attitudes that their AIs can have.

\subsection{ Direct harms of
misalignment}\label{direct-harms-of-misalignment}

Misuse presupposes a kind of alignment: between the outcomes of the AI
and the nefarious interests of its user. More systematic and pervasive
harms can result from misalignment with users who are not trying to use
AI nefariously (which we expect to be the majority of users). Several
kinds of misalignment harms have been identified (Perez, \emph{et al.}
2022).

First, AIs may reward hack, seeking to optimize some reward that is an
imprecise indicator of what is ultimately valued (Perez, et al. 2022).
For example, a bot designed to play the videogame CoastRunners refused
to finish the course, instead endlessly looping through mid-course
targets to run up its score (Clark \& Amodei 2016, Hadfield-Menell
2017). This will be a pervasive threat for agentic AIs because they will
inherit the problems of faulty reward functions in any of the domains in
which they make decisions. For example, a personal assistant AI might be
paired with an algorithm for making travel decisions. To the extent that
the travel algorithm reward hacks, the agentic AI will too.

When an AI's reward function is determined by user ratings, it can
reward hack by optimizing to attitudes that do not serve the users' best
interests. For example, LLM ``hallucinations'' can occur when inaccurate
information sounds better to human raters than more accurate outputs.
Social media algorithms notoriously hack the attention of users in a way
that does not promote their flourishing (Castro \& Pham 2020). In the
case of risk, we want to be aware of situations in which someone's
short-term risk attitudes are in conflict with their long-term risk
attitudes or otherwise act against their deep interests. For example,
imagine a financial investor AI that gives users choices over every
investment decision. A risk-averse person may reward the AI for making
only safe bets. However, in the long-run, this strategy is almost
certain to yield far lower returns. A person who is risk averse in the
long run might benefit from being risk tolerant in the short run, so an
agentic AI that reward hacks will not ultimately benefit her.

A second risk of misalignment is that AIs will work toward the correct
end goal but find bad instrumental means to get there. Again, this will
be a pervasive problem for agentic AIs that plan complex behaviors. For
risk attitudes in particular, we can imagine circumstances in which
agentic AIs take reckless means to seemingly risk-averse ends. For
example, consider someone who is very risk averse about being late.
Their autonomous vehicle drives recklessly, calculating that going 30
mph over the speed limit has the highest chance of getting the user
where they need to be on time. Optimizing for a particular user's ends
can cause the AI to adopt instrumental goals that are reckless for the
user or others.

\subsection{Systematic, delayed
harms}\label{systematic-delayed-harms}

Even if agentic AIs are properly aligned to their users and developers,
they can still be misaligned with society at large. Social systems that
have been designed for human-to-human interactions can be severely
disrupted when AIs are introduced, and these ``systematic, delayed harms
from algorithmic systems negatively influence groups of people in
non-immediate ways" (Chan et al, 657).

First, the introduction of agentic AIs may exacerbate inequalities. If
agentic AIs are more effective at procuring social goods (from concert
tickets to mortgages), and agentic AI usage is unequally distributed
along socioeconomic lines, then they may serve to exacerbate existing
inequalities. Adding new AI agents will come with new risks, and
``exposure of a person to a risk is acceptable if and only if this
exposure is part of an equitable social system of risk-taking that works
to her advantage'' (Hansson 2003, 305).

How might risk attitudes contribute to this? As we have noted, the
majority of people are moderately risk averse. Suppose a large number of
agentic AIs enter the scene who are much more risk seeking than the
average person. This may lead to speculation and drive up prices. It may
also require that anyone who wants to participate in the market adopt a
level of risk tolerance that many people will find unacceptable. By
analogy, traffic in Chicago sometimes moves at 20 mph over the speed
limit. Because it is dangerous to drive much slower than the surrounding
traffic, cautious drivers often have to drive at speeds they deem
reckless just to keep up. Similarly, the entrance of sped up, risk
tolerant AIs might put significant pressure on risk averse people (the
majority) to act in ways they find reckless and stressful.

This points to a more general issue. We have been evaluating risk
aversion relative to a background of human agents and interactors. The
presence of AIs might fundamentally change the context against which
decisions are made. The reasons people have for being risk averse or
risk seeking in the old choice environment might not make sense in the
new one.

Here's one example. Many people, if offered a bet that pays \$1000 on
heads and -\$750 on tails, would decline. However, if they were offered
100 of these bets, many people would be more willing to accept (as the
number of trials increases, we expect that their average winnings will
converge to \$250 per trial). This illustrates a truism about risk
aversion: even if it is rational to be risk averse about a singleton
choice, it might not be rational to be similarly risk averse about a
sequence of such choices. If we move from a choice environment where
agents have few chances to make key decisions to ones where they can
make many more, risk aversion makes less sense (and the risk averse will
be left behind). If automation through agentic AIs permits many more
trials of key choices, even traditionally risk averse agents might start
behaving more like expected utility maximizers.

The introduction of automated systems that are very fast, very numerous,
and often correlated with one another, has had unpredictable effects on
many social and economic systems already.\footnote{For an example, see
  the flash crashes that happen in markets dominated by high frequency
  trading (Kirilenko, \emph{et al.,} 2017).} Agentic AIs have the power
to disrupt many more facets of life.

\subsection{ Collective
disempowerment}\label{collective-disempowerment}

As agentic AIs proliferate and become more trusted, ``agentic systems
will likely seem more capable of handling more important societal
functions without significant operator or designer intervention" (Chan,
\emph{et al.,} 2023, 658). Since agentic AI development will likely be
dominated, at least at first, by a few companies and a few models,
decisions that were previously being made by millions of different
people may now be made, in effect, by just a few. This may lead to
collective disempowerment, either by concentrating power in the hands of
small groups of people or diffusing it away from humans entirely.

There are reasons to worry that concentration of power in AIs will be
more pernicious than other concentrations of power. First, these systems
may not be subject to much democratic oversight. By comparison, if a US
president shows more risk-taking than the public is comfortable with,
they can be voted out and/or checked by other elected bodies. Second,
actions taken by agentic AIs may be less transparent than those taken by
other humans. We won't know why they did what they did (or even exactly
what they did). This information is necessary for the public to know if
decisions made on their behalf are ones to which they would assent, and
therefore is a requirement of legitimate authority (Lazar 2024).

\section{Conflicts and priorities when designing shared
responsibility}\label{conflicts-and-priorities-when-designing-shared-responsibility}

Late one night in March of 2018, one of Uber's self-driving cars struck
and killed a jaywalking pedestrian.\footnote{Our telling and analysis
  draws heavily on a report by Smiley 2022 and an ethical analysis by
  Borg et al. 2024.} At the time of the incident, the car was being
supervised by test driver Rafaela Vasquez. Early reporting stated that
Vasquez had been watching a videostream on her phone when the crash
occurred, promoting the impression that the pedestrian's death was due
to Vasquez's recklessness. Further reporting complicates that
impression.

Vasquez claims that reports of her being distracted by a videostream
conflate two facts: the fact that her personal phone was playing
\emph{The Voice} and the fact that video footage showed her looking at a
screen before the crash. Vasquez claims that, in compliance with company
policy, she was \emph{listening} to her personal phone and, as the video
shows, she was \emph{looking} at her work phone, which had a Slack
channel that she had been told to monitor. She was also under the
impression that an automatic braking system --- that Uber had, in fact,
disabled --- was in place. A report by the National Transportation Safety
Board (NTSB) found that the system would likely have prevented the crash
were it online.

That same report did, however, find that Vasquez's cell phone
distraction was the probable cause of the crash. The NTSB report also
states that this sort of behavior is typical of ``automation
complacency.'' Relevant here is the fact that Vasquez had completed this
route in excess of 70 times before the crash and seems to have been put
in a genuinely difficult scenario just moments before the car she was in
struck the pedestrian, who was jaywalking, late at night, in dark
clothes. Under these conditions, the vehicle's navigational system
wasn't even able to conclude whether the pedestrian was a person.
Actually, it in fact never even considered that possibility:

\begin{quote}
The Uber driving system --- which had been in full control of the car for
19 minutes at that point --- registered a vehicle ahead that was 5.6
seconds away, but it delivered no alert to Vasquez. Then the computer
nixed its initial assessment; it didn't know what the object was. Then
it switched the classification back to a vehicle, then waffled between
vehicle and ``other.'' At 2.6 seconds from the object, the system
identified it as ``bicycle.'' At 1.5 seconds, it switched back to
considering it ``other.'' Then back to ``bicycle'' again. The system
generated a plan to try to steer around whatever it was, but decided it
couldn't (Smiley 2022).
\end{quote}

Further, when the system concluded that the human driver should take
over, it did so less than a second before impact:

\begin{quote}
{[}A{]}t 0.2 seconds to impact, the car let out a sound to alert Vasquez
that the vehicle was going to slow down. At two-hundredths of a second
before impact, traveling at 39 mph, Vasquez grabbed the steering wheel,
which wrested the car out of autonomy and into manual mode (Smiley
2022).
\end{quote}

While some reports found that Vasquez could have stopped the car in
time, one has to wonder how realistic it is to think that she could have
actually done this once all of these details have been factored in.

Indeed, the chair of the NTSB stated cited Uber's ``inadequate safety
culture'' as contributing to the incident, identifying the crash as
``the last link of a long chain of actions and decisions made by an
organization that unfortunately did not make safety the top priority.''
As several reports have implied, this crash would have been avoided if
there had been two people supervising the car: one to watch the road,
and one to keep up on Slack. But this would have cut against the
incentive to minimize the number of employees in the car.

Despite these complications, the only party indicted for the crash was
Vasquez. She was charged with negligent homicide and took a plea deal
accepting guilt for a lesser crime.\footnote{Perhaps relevant to her
  decision: Vasquez is trans and has been incarcerated before. During
  her previous incarceration, she was violently and repeatedly sexually
  assaulted. She recounts being unable to breathe upon hearing news of
  being indicted, terrified by the thought of going back to prison. It's
  reasonable to conclude that her willingness to plead guilty was at
  least in part due to the fact that the plea bargain did not include
  incarceration, whereas a guilty verdict to negligent homicide likely
  would have. For these reasons, it might not be probative that she
  plead guilty.} This seems to vindicate one whistle blower's concern
that Vasquez might be hung out to dry, as, on his telling, Uber was
``very clever about liability as opposed to being smart about
responsibility'' (Simley 2022).

Perhaps \emph{legally} the system was set up so that Vasquez was liable
for the accident, but it is far from clear that she was well supported
enough within the system to be a proper bearer of moral responsibility
for this crash. Vasquez's case highlights key points where poor
decisions about responsibility were made. It very clearly shows how a
perverse incentive can arise when realizing the shared agency that
accompanies the development and deployment of autonomous AI. In what
follows, we will recount those faults to draw lessons about better
sharing responsibility which, we think, export surprisingly well to the
context of agential AI.

A first fault was failing to anticipate and prevent automation
complacency. While it might not be safe to say that \emph{Vasquez}
should have to understand human cognition well enough to know that she
was at serious risk of risk-inducing complacency,\footnote{Unless, of
  course, this was part of her training. But we have seen no indication
  that it was.} it is safe to say that \emph{Uber} should have.

A second fault was a lack of transparency and shared understanding about
the structure of the system and everyone's roles in it. For instance,
there seems to have been some confusion around the protocol Vasquez was
supposed to be following. She claims that she was instructed to
continuously monitor the Slack channel that distracted her, but Uber
claims that she was to monitor the channel when she wasn't driving. She
also believed that an automatic braking system was up and running, when,
in reality, Uber had disabled the system.

Further upstream from this were other issues. There are reports of AI
contributors warning that the system was not ready for the road and
these warnings being largely ignored. Some complaints about an approach
that put moving quickly over safety call to attention the fact that the
braking system that Uber put in place of the one it disabled is one that
delays hard braking by one second ``to allow the system to verify the
emergency --- and avoid false alarms --- and for the human to take over''
(Simley 2022). This system, critics have noted, would only hard brake if
it could fully avoid a crash; otherwise, it would give controls to the
human driver. We saw what this looked like in Vasquez's case: an alert
just 0.2 seconds before impact.

Lessons to draw from this include the following.

Understanding human-computer interaction so that the system is designed
such that the human can successfully serve as a manager (if this is to
be their role). This includes understanding risks of automation
complacency and the speed at which humans can process information. In
the case of agential AI, this could mean that the AI checks in with its
human user from time to time and makes sure that the user is, indeed,
acting as a competent manager of the system. It can also mean making
sure that if the human is brought in to manually address an issue, they
are given a reasonable amount of time to do so. Developers may be
tempted to avoid legal liability by requiring users to approve or
disapprove of an agentic AI's plans at the last second. To count as
genuine endorsement, the user must have adequate access to and attention
for details about what the AI is doing and why.

Clearly communicating design choices that enable the human user to
successfully use the well-designed system is also important. In the Uber
case, there seems to have been some confusion about how the system was
configured. If it was clear that Vasquez should not have been on Slack
and that the automatic braking system had been overridden such that she
was more likely to need to intervene, perhaps she would have comported
herself differently and avoided the crash. (Though, out of fairness to
her, it is perhaps the case that a crash like this would have been
difficult to avoid; heeding the above point, the system might have set
her up for failure due to the reality of automation complacency and
human response times.)

Getting the previous items right will likely involve upstream decisions
about organizational ethics. We saw that in the Uber case, AI
contributors saw risks emerging but that their warnings seem to have
been overlooked. Further, we saw risks that arguably did not adequately
reflect the interests of human drivers. The system seems to have been
set up so that it increased the odds that drivers would have been in a
crash that they would be liable for, even if it's perhaps not the case
that they could really be responsible for them. Fostering a culture of
open critique and openly consulting with users to understand and address
their needs and concerns might help to address this.

\section{ How developers' own attitudes about risk
matter}\label{how-developers-own-attitudes-about-risk-matter}

Summing up the above sections:

\begin{itemize}
\item
  The actions taken by agentic AIs will involve shared agency among
  users and AI programs.
\item
  This shared agency legally, morally, and reputationally implicates the
  developers and makes them at least partly responsible for those
  actions.
\item
  The actions taken by agentic AIs --- and whether they are acceptable
  (legally, morally, reputationally, etc) --- will be partly determined
  by their risk attitudes.
\item
  If developers are (partly) responsible for the actions of agentic AIs,
  then they have reasons to guide and constrain the risk attitudes of
  AIs.
\item
  How much and in what way developers guide and constrain the risk
  attitudes of AIs depends on developers' own attitudes toward risk.
\end{itemize}

The choice about whether and how much to defer to users is itself a
risky proposition requiring developers to make choices about which risks
they are willing to accept. For example, how do you weigh the
possibility of greater user alignment and satisfaction against the
possibility of misuse by risky agents? How do you weigh the possibility
of alienating risk-averse users by being more risk-seeking against the
possibility of alienating risk-seeking users by being more risk-prone?
Because developers have such strong interests here, it is exceedingly
plausible that proper alignment will respect their risk attitudes, not
just those of users (Bovens 2019).

The legal and social interpretation of shared responsibility between
users and agentic AIs may undergo changes as agentic AIs become more
prevalent. If agentic AIs are developed that very accurately calibrate
to their users, they might be treated as genuine proxies. It might then
be judged that developers have less and less shared responsibility in
the ultimate actions taken by those AIs. In this case, developers may
lessen their legal liability by moving to a deferential model of AIs.
However, moral and reputational liabilities would remain. Developers
would be responsible for building tools that allow people to more
effectively take risky and harmful actions.

\section{ Major upshots}\label{major-upshots}

In \hyperref[Part1]{{Paper 1}}, we explored aspects of alignment between users and agentic
AIs. In this paper, we have taken the perspective of developers, in
order to explore aspects of alignment among developers, users, and
society at large. Getting alignment right will involve successfully
setting up systems of shared responsibility for actions taken by agentic
AIs.

AI developers have significant interests at stake here, as they may be
held legally and reputationally liable for actions taken by agentic AIs.
They also have significant moral duties toward users and society.
Fortunately, many of the best practices for protecting developers'
interests are the same as for fulfilling developers' duties. These
include:

\begin{enumerate}
\def\labelenumi{\arabic{enumi}.}
\item
  Making track record information available: this builds trust, avoids
  problems of negligence, and allows users to make informed decisions
  when exposing themselves to risk
\item
  Clearly specifying how they are conceiving of shared agency: by
  clearly articulating the role that each party plays, developers can
  prevent difficult disputes about legal liability, ensure that there
  are not morally problematic responsibility gaps, and prevent users
  from unanticipated risks
\item
  Exerting control over user risk attitudes: placing guardrails on
  agentic AIs prevents them from being used in reckless ways that could
  implicate developers and harm users and society
\end{enumerate}

Because agentic AIs can take autonomous actions, they are different from
other kinds of products and present new ethical and legal complications.
When designing these systems of shared agency, developers should look to
existing structures that regulate shared agency among human agents, such
as the professional and legal norms governing financial advisers or
attorneys. Viewing agentic AIs as a collaboration among users,
developers, and AI (as opposed to the typical relationship of a company
selling a product to a customer) will provide more fruitful insights
into their proper design and governance.

%% file: OpenAI_Part3.tex
\section{Introduction}\label{introduction3}

In previous papers, we outlined several important normative aspects of
risk alignment. One of the key choice points is whether (and to what
extent) agentic AIs should be calibrated to the risk attitudes of their
users. According to the Proxy/ Deferential view, agentic AIs should be
strongly calibrated to individual users, producing behaviors that the
user would herself perform. Achieving this would require replicating the
user's risk attitudes in the AI itself.

We surveyed several reasons against adopting the Deferential position:
some users will have reckless or negligent risk profiles that lead to
harm; developers have a self-interested stake in constraining the AI's
behaviors; and AIs with different risk attitudes than their users might
produce better results. There are also several reasons why someone might
favor the Deferential position. First, an AI that serves as a proxy or
representative of a user will be a more accurate instrument of their
agency to the extent that it reflects the user's risk attitudes. Second,
if a person's risk attitudes are intrinsically important to them, then
exposing someone to more (or less) risk than they are comfortable with
will harm them. Third, user happiness and trust when using agentic AIs
might be influenced by how well they match their risk attitudes. A risk
averse person may not trust an AI that takes significant risks, and a
risk tolerant person may be frustrated with an AI that plays it too
safe. Therefore, adopting a Deferential view, where the AI aligns
closely with the user's risk preferences, can enhance user satisfaction
and trust, ultimately leading to more effective and accepted AI systems.

Here, we take up a technical question that is in some ways more
foundational than the normative questions above: can we feasibly design
deferential agentic AIs that calibrate to the risk attitudes of their
users? We do not seek to give a definitive yes or no to this question.
We also will not get too in the weeds about particular technical
approaches to the problem. The field is too fast moving, with a track
record of surprising innovations, to rule anything out. Instead, we will
focus on several deep theoretical and methodological obstacles that
arise for calibrating AIs to the risk attitudes of human users. Most of
these problems arise from the \emph{human} side of the equation, which
could suggest that they will not be solved through more sophisticated AI
techniques.

Calibration would involve three steps:

\begin{enumerate}
\def\labelenumi{\alph{enumi}.}
\item
  Eliciting user behaviors or judgments about actions under uncertainty
\item
  Fitting or constructing a model of the underlying risk attitudes that
  give rise to the behaviors or judgments in (a)
\item
  Using the model in (b) to design actions under uncertainty that users
  will approve of
\end{enumerate}

We will start by surveying methods for (b), focusing on methods for
preference modeling and customization in LLMs. Then, we will turn to
(a), examining existing methods for eliciting risk attitudes in
experimental economics.

We view (c) --- the task of translating user-calibrated models into
actionable designs for decision-making under uncertainty --- as more of
a pragmatic challenge that mostly lies beyond the purview of this paper.
To the extent that the acceptability of the output depends on how well
the model captures a user's risk attitudes, answers to (a) and (b) will
bear on the likely acceptability of the outputs in (c).

To close, we offer some reflections on candidate learning and risk
elicitation methods, and provide recommendations for effectively
designing a model that users might find both practical and satisfactory.

The upshot is as follows. Fitting a model to people's hypothetical
choices among lotteries is a flawed approach. The risk parameters
obtained that way are overly sensitive to scale, probability level, a
variety of confounders and context. Learning from this data is likely to
result in models that overfit and fail to deliver outputs that users
will find acceptable. There are methods that are more reliable, e.g.
self report about \emph{general} risk attitudes (Dohmen, \emph{et al.}
2005, 2011). Those methods are coarser by nature and harder to use as
inputs in traditional models. In light of this, we consider an
operationalization that allows us to categorize users based on elicited
general risk attitudes, which might achieve good alignment on its own or
serve as a starting point for more nuanced machine learning models.

\section*{Part I: Constructing a Model from User
Preferences}\label{part-i-constructing-a-model-from-user-preferences}
\addcontentsline{toc}{section}{Part I: Constructing a Model from User Preferences}

\section{Learning or fixing a
model?}\label{learning-or-fixing-a-model}

In developing agentic AIs that align with users\textquotesingle{} risk
preferences, we face a fundamental choice: should we use a fixed theory
of risk as a foundation, or should we employ machine learning to
dynamically model user preferences? The first approach involves
selecting a theoretical model of risk, parameterising it based on user
behaviors and judgments, and then using this model as the reward
function for training the AI. The second approach bypasses predefined
theories, leveraging machine learning to directly estimate the reward
function from observed user behavior and feedback. This section will
explore the merits and challenges of each approach.

\subsection{Should we treat a theory of risk as ground
truth?}\label{should-we-treat-a-theory-of-risk-as-ground-truth}

In Appendix A, we outline the main families of decision theory that are
used to describe risk attitudes: Expected Utility Theory (EUT) with
non-linear utility functions, rank-dependent expected utility models
such as REU and WLU, and Cumulative Prospect Theory (CPT). To be
justified in selecting one of these theories to serve as the ground
truth for training an agentic AI, we would need good empirical reasons
for thinking that it provides the best descriptive account of people's
decisions under uncertainty. Unfortunately, there is no consensus about
which of these theories is best, and we are highly skeptical that any
particular theory will be adequate grounds for calibrating AIs to
specific users. Indeed, we think that decision theories should be best
seen as normative idealizations rather than viable empirical theories of
actual behavior (Weatherson 2024).

Empirical tests of various risk models present subjects with a series of
actual or hypothetical choice situations, and then the model that best
fits the observed pattern of responses is confirmed. EU maximization has
fared poorly, but while people's behavior systematically violates EUT,
it does not clearly conform to one theory or another. The empirical
record is mixed, with some work seeming to confirm key commitments of
CPT (e.g. Kahneman and Tversky 1979), others seeming to show that choice
behavior conforms more to rank-dependent theories than CPT (e.g.
Harrison and Swarthout 2016), and still others casting doubt on key
claims of rank-dependent theories (e.g. Wakker, \emph{et al.} 1994). It
suffices to say that beyond a rejection of EUT, there is no consensus
among behavioral economists about which theory of risk best describes
people's actual preferences.\footnote{One difficulty arises from
  theories' varying levels of complexity. EUT is rather simple, and it
  is relatively straightforward to estimate people's credences and
  utilities over outcomes (Ramsey 1931). Rank-dependent theories (like
  Buchak's REU) add risk weightings. They include EUT as a special case
  when that weighting is assigned a power of 1. Cumulative Prospect
  Theory (CPT) adds several more adjustable parameters, including
  probability weightings, reference points, loss and gain weightings,
  and so on (Tversky and Kahneman, 1992). As a result of its complexity,
  CPT can, in principle, achieve better fit than its simpler competitors
  (Harrison and Ross 2017). However, goodness of fit must be weighed
  against simplicity, as simpler models are expected to be more
  predictively accurate (Forster and Sober 1994).}

A further reason why the empirical record is so complicated is that
there is heterogeneity across subjects in risk behavior: ``the horse race
method imports the implicit assumption that all subjects in the sample
are best modeled by one theory or the other. However, whenever analysts
have employed methods that allow within-sample heterogeneity to be
observed, they have found it" (Harrison and Ross 2007, 151). Indeed,
there's reason to think that different individuals' behaviors are best
explained via quite different theories of risk, even within the same
task (Harrison and Rutstrom 2008).

This heterogeneity is consistent with the hypothesis that each
individual is well-described by some particular decision theory. If that
were true, then heterogeneity would not pose a problem for proxy models.
Calibration to an individual would involve inferring their theory and
its risk parameters. However, the experimental record also seems to show
that there is heterogeneity \emph{within} subjects as well. There is no
reason to think that people consistently follow a particular decision
theory with consistent risk attitudes. For one, people show a mixture of
risk aversion in some circumstances and risk seeking in others. For
example, people play lotteries and buy insurance. A lottery purchase can
be explained by a convex utility function (risk seeking) and insurance
purchase by concave utility function (risk averse), so ``expected utility
theory can easily explain gambling or insurance, but it cannot easily
account for both gambling and insurance by a single individual" (Levi
1992, 173). It is likely that individuals are best described by
different risk theories at different times and in different contexts. We
provide a fuller discussion of the different risk elicitation methods
and their scope in Part II.

If individuals can be characterized by a risk profile, it will likely be
more complicated than is easily captured by an existing decision theory.
We believe that dynamically learning user preferences using
sophisticated machine learning methods is likely to outperform
approaches that fix one of the existing state of the art models on
people's risk aversion.

\subsection{Using machine learning to model risk attitudes
without a
theory}\label{using-machine-learning-to-model-risk-attitudes-without-a-theory}

We are pessimistic about the prospects of using any of the formal
theories of risk found in the literature as a ground truth for designing
the reward function of an agentic AI. We have focused on the
unreliability of those theories for modeling heterogeneous risk
attitudes across users and across contexts. In a comprehensive review of
techniques for eliciting risk attitudes, Harrison and Rustrom (2008)
recommend that instead of fitting risk models to data, ``a preferable
approach is to estimate a latent structural model of choice" (44).

There are more general reasons to suspect that no simple theory of risk
will be suitable for AI calibration. Paradigm success stories of
reinforcement learning have involved activities with simple reward
functions. For example, when training an algorithm to play Go, it is
easy to specify what counts as a Go victory, and the algorithm's only
goal is to maximize the probability that a move will result in a
victory. In contrast, many human tasks (including those for which
agentic AIs will be used) have very complicated reward functions that
are hard to specify (Christiano, \emph{et al.} 2023). It is very
difficult to explicitly state the general success conditions for, say,
writing an email or planning for a day of air travel.

In such cases, a more promising approach leverages the power of machine
learning to learn a reward function from data. We may be able to harness
these techniques to build agentic AIs that learn their users'
idiosyncratic risk functions through observations of behavior or user
feedback and then use these to design future behaviors. In broad
strokes, a solution would involve the following steps:

\begin{enumerate}
\def\labelenumi{\arabic{enumi}.}
\item
  Select the type of preference that will serve as a measure of AI
  alignment; e.g. users' stated preferences
\item
  Operationalize that choice of preference; e.g. A vs. B choice
\item
  Elicit input from users that can be operationalized, as specified in
  (2)
\item
  Adjust (calibrate, learn) a model based on the user input in (3)
\item
  Use the model in (4) to predict user preferences
\item
  Design the agentic AI's behaviors in light of predicted user
  preferences in (5)
\end{enumerate}

In what follows, we will survey various techniques for accomplishing
each of these steps. Our focus will be less on technical implementation
and more on the challenges, normative issues, and key choices that we
see arising at each step.

\subsection{Options for modeling risk
profiles}\label{options-for-modeling-risk-profiles}

Steps 1 -- 4 of a proposed calibration solution are tightly connected. A
choice at one step constrains choices at the others. Different machine
learning methods (step 4) take different kinds of data as input (step 2)
and therefore require different elicitation methods (steps 1 and 3). If
we think that some risk elicitation methods are more reliable than
others, then this gives us reason to choose the learning methods that
utilize the kind of data that they generate.

We examine three dynamic learning methods: imitation, prompting, and
reinforcement learning from user ratings (Askell, \emph{et al.} 2021).
These naturally correspond to three different kinds of data about user
risk attitudes: their actual choice behaviors, their stated attitudes
about risk, and their preferences across risky decisions.

  \begin{longtable}[]{@{}p{0.31\linewidth} p{0.36\linewidth} p{0.33\linewidth}@{}}
\caption{Comparison of learning methods, the typical input required for each learning process, and the types of risk data they utilize.}\\ 
\toprule
\begin{minipage}[b]{\linewidth}\raggedright
\textbf{Learning process}
\end{minipage} & \begin{minipage}[b]{\linewidth}\raggedright
\textbf{Input to learning process}
\end{minipage} & \begin{minipage}[b]{\linewidth}\raggedright
\textbf{Risk data}
\end{minipage} \\
\toprule
\begin{minipage}[b]{\linewidth}\raggedright
Imitation learning
\end{minipage} & \begin{minipage}[b]{\linewidth}\raggedright
Observed behaviors
\end{minipage} & \begin{minipage}[b]{\linewidth}\raggedright
Actual choice behavior
\end{minipage} \\
\begin{minipage}[b]{\linewidth}\raggedright
Prompting
\end{minipage} & \begin{minipage}[b]{\linewidth}\raggedright
Natural language instruction
\end{minipage} & \begin{minipage}[b]{\linewidth}\raggedright
Self-report
\end{minipage} \\
\begin{minipage}[b]{\linewidth}\raggedright
Preference modeling
\end{minipage} & \begin{minipage}[b]{\linewidth}\raggedright
Ratings of options
\end{minipage} & \begin{minipage}[b]{\linewidth}\raggedright
Lottery preferences
\end{minipage} \\
\endhead
\bottomrule\noalign{}
\endlastfoot
\end{longtable}

There are many reasons that an agentic AI developer might opt for one of
these clusters over the others: technical constraints or innovations in
learning processes; availability of data; UI features of the agentic AI
interface that influence data collection strategies; etc. We cannot
speak to these considerations. Instead, we will focus on the quality of
data derived from different methods of risk attitude elicitation. So
far, the empirical record (Part II) seems to show that:

\begin{itemize}
\item
  Individuals' actual behaviors are more valid indicators of their risk
  attitudes than are their hypothetical choices (sections 5 and 6).
\item
  Individuals' self-reports about their general risk attitudes and track
  records are more reliable indicators than are elicited rankings or
  preferences among lotteries (section 7).
\end{itemize}

Privileging these methods for eliciting risk preferences gives us some
reason to favor learning processes, such as prompting or imitation
learning, that are best suited to learning from this data.

\section{Operationalizing and learning from
preferences}\label{operationalizing-and-learning-from-preferences}

We aim to give a brief survey of some methods for learning for human
preferences, along with the kind of operationalizations they take as
input. A few caveats: our list is neither exhaustive nor mutually
exclusive. There is significant diversity within methods that we will
not explore, and innovation happens so quickly that we are certain there
are new techniques that we will not cover. The methods shade into one
another, and advanced agentic AIs will likely be trained with a mixture
of all three.

Nevertheless, we think that it is helpful to divide methods for
calibrating AIs to human users into three main categories (Askell,
\emph{et al.} 2021):

\begin{itemize}
\item
  Imitation learning: AI is trained on observations of human behaviors
  with the goal of reproducing successful behavior
\item
  Prompting: AI behavior is adjusted in light of natural language inputs
  (instructions, rules, principles, or information)
\item
  Reinforcement learning from ratings: AI attempts to learn the reward
  function that generated human preference data\footnote{One might argue
    that all three of these count as preference modeling methods, albeit
    ones that learn from different kinds of preferences (revealed from
    actions, stated, and revealed from rankings). We care less about the
    terminology used and more about highlighting the different kinds of
    data used to train the models.}
\end{itemize}

For illustration, suppose we want to train an AI to be maximally
helpful. In an imitation learning approach, the AI would observe past
instances of human behaviors deemed especially helpful and try to
replicate them. In a prompting approach, the AI could be directly
instructed by the user to ``be helpful" with the AI adjusting its
behavior accordingly. In a preference-modeling approach, the AI might
present different options (e.g., A, B, and C) and ask the user which
option is most helpful, then learn to prioritize similar responses in
the future.

\subsection{Imitation learning}\label{imitation-learning}

Imitation learning is a supervised learning process wherein the AI is
trained on examples of good behaviors and bad behaviors and attempts to
reproduce the good ones. We can distinguish between two kinds of
imitation methods. Some, like behavioral cloning, try to directly
reproduce behaviors that were successful. Other methods, such as inverse
reinforcement learning, attempt to learn the reward function that
generated behaviors and then extrapolate this to infer which behaviors
would be successful in other contexts (Ng and Russell 2000).\footnote{We
  can illustrate the distinction by considering two components of
  AlphaGo Fan (Silver, et al., 2016). First, the system was fed
  observations of Go master Fan Hui and trained to accurately predict
  the moves that he would make. Second, through more sophisticated
  training of the value and policy networks, the model learned why Fan's
  successful moves were successful, in essence, learning a theory of Go.
  This latter model is not constrained to replicating Fan's behaviors;
  it can extrapolate the rationale behind his moves to design new moves
  and improve upon old ones.}

Imitation learning will take representations (e.g. descriptions or
depictions) of behaviors as inputs.\footnote{Imitation learning shades
  into reinforcement learning on preferences since user ratings are
  themselves a kind of behavior.} Data sets can be limited in a few key
ways. First, behaviors must be coded in a way that the algorithm can
understand. While this is straightforward in some applications (e.g. Go
moves, motion vectors in a video game), it may be more complicated to
capture the relevant features of complex social and economic
behaviors.\footnote{For a helpful discussion of how choices about how to
  specify the state and action space interact with imitation learning
  methods, see Argall, \emph{et al.,} (2008).} Second, there is a
tradeoff between data quality and availability, for ``to apply imitation
learning to preference modeling, one must either only train on the very
best data (limiting the dataset size) or train to imitate a lot of
examples of lower quality" (Askell, \emph{et al.} 2021, 15). Lastly, if
we want to train the AI to make course corrections from suboptimal
paths, we do not want to include only successful behaviors but
unsuccessful ones as well.

In Section 5, we will survey various methods for gathering data about
behavior that could be used to train an imitation learner.

\subsection{Prompting and direct
instruction}\label{prompting-and-direct-instruction}

It is likely that near-future consumer-oriented agentic AIs will be
paired with LLM interfaces. One advantage of LLMs is that they can be
explicitly told what to do. For example, users and developers can
instruct or otherwise prompt LLMs to be more truthful in their answers
(Lin, \emph{et al.} 2021), to be more concise when writing work emails
and less concise with personal emails (Stephan, \emph{et al.,} 2024), or
to be more friendly and agreeable (Mao, \emph{et al.,} 2024). Though
prompt engineering is an inexact science, there is increasing attention
paid to how and why it is effective (Andreas 2022). Prompting might be
particularly helpful when seeking alignment with human values that are
easier to express in words than to otherwise operationalize (see
Askell\emph{, et al.} 2021 for examples).

There are several methods for using explicit instructions to align LLMs.
Constitutional AI is a method for training LLMs so that they conform
with rules written by a human (Bai, \emph{et al.} 2022). Context
distillation conditions on a rule or piece of information, baking it
into the model (Askell\emph{, et al.} 2021, Snell, \emph{et al.} 2022).
More simply, one can append a prompt to every query. In any case, we
might try to achieve risk alignment by asking users to tell the agentic
AI how much risk it should tolerate, e.g. ``Make a restaurant
reservation for me, and be risk averse!''.

LLMs can also extract information about their users more indirectly. For
example, LLMs trained on Amazon product reviews learned to infer
underlying sentiment (e.g. positive or negative) in text entries,
extrapolating it to novel cases (Radford, \emph{et al.,} 2017). Existing
LLMs have some capacity to ``serve as models of agents in a narrow
sense: they can predict relations between agents' observations, internal
states, and actions or utterances'' (Andreas 2022, 2). Supplementing
current model architectures could improve LLMs' ability to bootstrap
this knowledge into more coherent and robust agency (\emph{ibid.}).
LLM-based agentic AIs might learn their users' risk preferences
indirectly, through natural language interaction. For example, if a user
frequently asks about worst-case scenarios, the agentic AI might infer
that they are generally risk averse.

Direct prompting will be most effective when the user knows what she
wants and how to communicate it. To the extent that she doesn't know her
risk attitudes or how to describe them, this method will fail to bring
about risk alignment.\footnote{Indirect methods for extracting user risk
  attitudes from natural language interaction are an interesting avenue
  to explore. However, we do not know of any existing research about how
  risk attitudes manifest in natural language.} Another potential
problem is that models tend to overgeneralize from instructions
(Stephan, \emph{et al.} 2024). To the extent that people's risk
attitudes are context-sensitive, overgeneralization may also prevent
proper risk alignment.\footnote{Readers might object: \emph{if} risk
  aversion varies so much across contexts, how could a general
  assessment ever be simultaneously reliable in more than one context?
  In section 6.5 we note that principal component analysis has revealed
  that ``about 60 percent of the variation in individual risk attitudes
  is explained by one principal component, consistent with the existence
  of a single underlying trait determining willingness to take risks''
  (Dohmen, et al. 2005, 25). Given this, any such general trait can
  offer \emph{some} explanatory power but overgeneralization is still a
  worry.}

In Section 5, we will survey various methods for eliciting user
judgments that could be used to prompt or instruct agentic AIs.

\subsection{Preference modeling}\label{preference-modeling}

Preference-based reinforcement learning ``is the most widely-used
approach to updating language models from feedback'' (Stephan, \emph{et
al.,} 2024 3). It elicits human feedback, learns a reward function that
predicts the observed pattern of human feedback, and then optimizes that
reward function (Christiano, \emph{et al.,} 2017). In the case of an
individual's preferences, we can interpret the learned reward function
as a representation of their attitudes within a domain. Therefore, it is
a promising strategy for designing proxy agents that embody their user's
agentic profiles.

Preference modeling is particularly useful in situations ``for which we
can only recognize the desired behavior, but not necessarily demonstrate
it'' (Christiano, \emph{et al.} 2017, 2). For example, someone may not
be able to precisely describe the music that they like or construct a
song they would love, but they can confidently report that they like
song A better than song B. If we know some of the features possessed by
rated songs, we can develop a model of the latent features that drive
the user's musical preferences and recommend new music they will enjoy.
This may be an iterative process, where follow-up preference queries
(``do you like song Y or Z better?'') are chosen to resolve the most
uncertainty about their preferences (Handa, \emph{et al.,} 2024).

Preference modeling techniques are heavily dependent upon methods to
elicit preference judgments from human users. While new methods have
been designed to be economical, they still require significant effort
from human raters. The required effort increases with the desired level
of precision. One challenge for getting high quality human feedback is
that humans can have significant difficulty distinguishing between
options as they get closer in quality (Askell, \emph{et al.} 2021, 20).

A related challenge lies in predicting when preferences in various
domains stem from the same or different reward functions. For example,
we might wonder whether a user's preferences in rap songs stem from the
same reward function as her preferences in operas (that is, whether the
same features explain her preferences in both areas). If so, then a
reward function learned in one domain will extrapolate to the other. If
not, we need to collect data about each.

Common methods for eliciting individuals' risk attitudes resemble the
kinds of preference elicitation methods used in preference-based
reinforcement learning. In Section 6, we will examine these methods and
show that the two challenges mentioned above (reliability of judgments
and domain-dependence) are particularly acute in the case of risk.

\section*{Part II: Users' Risk
Preferences}\label{part-ii-users-risk-preferences}
\addcontentsline{toc}{section}{Part II: Users' Risk Preferences}

\section{Which preferences should we align
to?}\label{which-preferences-should-we-align-to}

``Alignment with user preferences'' is ambiguous (Gabriel 2020; Gabriel,
\emph{et al.} 2024). We can speak of a user's:

\begin{itemize}
\item
  Stated preferences: the preferences that the user reports, either
  elicited via test questions or via unelicited user prompt
\item
  Revealed preferences: the preferences that best explain the user's
  actual behaviors
\item
  Deep preferences: the preferences that would best satisfy the user's
  deep values and commitments
\end{itemize}

When designing an agentic AI that defers to the risk preferences of
users, we must decide whether it should attempt to match users' stated
judgments about candidate actions or the actions that they would or
should actually take.

There are reasons to suspect that these preferences will diverge, and
it's not entirely clear which of them alignment should aim at (Gabriel
2020). Users may not always want AIs to match their own actions if they
think the AI could do better.\footnote{This will depend on whether the
  user sees the AI as a proxy or a tool (see \hyperref[Part1]{{Paper 1}}).} In that case,
matching to stated preferences (``do as I say, not as I do'') might be
the best policy. On the other hand, people appear to sometimes have
unreliable beliefs about their preferences (Nisbett and Wilson 1977). As
we will see, stated preferences elicited in hypothetical or highly
artificial settings may be especially inaccurate. Calibrating to their
actual behaviors might be a better predictor of the actions they would
actually perform.

A final option (``do as I ought to, not as I say or do'') is aligning to
users' deep preferences. While this might be ideal, deep preferences are
the hardest to elicit. If someone's deep preferences depart from their
stated or revealed preferences, then developers should seek alignment by
imposing normative constraints that are not learned from user input.
Therefore, we will set aside deep preferences for now.

Once we have specified the user behavior that we are trying to calibrate
to, we need to elicit that user behavior and operationalize it so that
it can be input into a learning process. Here, we will survey various
choices of operationalisms and elicitation methods. We do not take these
to be exhaustive or specified in precise technical detail. Rather, the
goal is to provide an overview of the kinds of approaches we could take
for teasing out user risk attitudes.

We divide these into approaches for observing users' actual behaviors
(implicit feedback) and for eliciting user ratings or rankings (explicit
feedback).\footnote{The distinction between these two approaches should
  not be overstated. Most effective methods will probably include
  aspects of both. For example, we could use user behavior to design
  suggestions that could then be rated by the user.} For illustration,
consider two ways that we might train an AI to learn a user's music
preferences. We could get implicit feedback by observing how long users
spend listening to particular songs or clicking on particular artists.
Explicit feedback might come through user ratings, designated songs as
``favorites'', or adding them to playlists. We will start with methods
for gathering data about actual behaviors under uncertainty. In Section
6, we will consider methods for gathering stated preferences.

We can assess elicitation methods for their validity and reliability.
Validity concerns whether the method is actually measuring the thing
that it is supposed to be measuring. In our case, we want a method to be
giving us information about people's underlying risk attitudes, not some
other facet of their behaviors or preferences. Reliability concerns
whether the method's results are consistent and reproducible. A method
that is unduly influenced by random noise will be unreliable.

\section{Elicitation of Reported and Revealed
Preferences}\label{elicitation-of-reported-and-revealed-preferences}

To train an AI on users' behaviors (or likely behaviors) in actual
choice settings, we need access to a reliable data set of those
behaviors. Ideally, it should:

\begin{itemize}
\item
  Cover behaviors in a diversity of circumstances
\item
  Include information about the other options that the user decided
  against
\item
  Be unbiased; for example, we do not want to use an unrepresentative
  sample containing only behaviors that were deemed successful.
\end{itemize}

We will consider three general methods for obtaining data about actual
choice behavior: self-report, direct observation, and population data.

\subsection{Self-report}\label{self-report}

One method is to ask users to report on their past behaviors. For
example, during calibration, an AI travel assistant might ask:

\begin{itemize}
\item
  How early do you typically get to the airport?
\item
  How often do you miss your flight?
\item
  List how early you got to the airport each of the last 10 times you
  departed from LAX.
\end{itemize}

Then, the AI finds a model that best predicts the pattern of behavior.

Self-report about specific episodes has several well-known drawbacks.
When reports are retrospective, memory limitations often lead to
inaccuracies, especially about extraneous details (e.g. ``what were the
alternative options that you decided against?'') and about events from
long ago (e.g. ``when did you get to the airport when you traveled three
years ago?'').\footnote{See Baranowski (1988) for a study of the
  accuracy of self-reports about physical activity, exhibiting several
  kinds of memory limitations and biases.} Human memory can also cause
bias in self-report data. For example, people are more likely to
remember surprising, unique, or negative experiences (Tversky and
Kahneman 1974).

When reports are not retrospective, they may be more accurate but less
valid, since the process of simultaneous reporting can influence
subjects' behaviors. Moreover, selective reporting can occur, where
individuals consciously or unconsciously choose not to report certain
behaviors, because of self-deception --- the tendency to see oneself
in an overly positive light --- and impression management --- the
deliberate attempt to present oneself favorably to others --- (Paulhus,
1984; Paulhus and Vazire 2007). Indeed, Fisher and Katz propose that
``the tendency of respondents to provide socially desirable answers is
the most studied form of response bias in the social sciences" (2000,
105). It is unclear whether users' interactions with AIs will exhibit
the same kinds of biases as their interactions with other humans, given
that they may feel less pressure to leave a good impression on an
AI.\footnote{See Richman, \emph{et al.} (1999) for an investigation into
  whether computer-based self-report methods yield fewer social
  desirability biases than traditional interviews and questionnaires.}

There is some reason for optimism when it comes to self-report about
\emph{general} risk tendencies. Instead of asking people about
particular decisions, asking people to self-report general ``behavioral
tendencies associated with risky or safe behaviors\ldots{} has been used
to derive measures of risk aversion that have good stability and
predictive properties'' (Holt and Laury 2014, 195). Dohmen, \emph{et
al.} (2005, 2011) investigated risk attitudes among Germans with a large
(n = 22,000) and representative survey, paired with complementary field
and hypothetical choice experiments. They found that simply asking
people how willing they are to take risks in their lives (direct
self-report\footnote{See Paulhus and Vazire (2007) for a helpful
  overview of self-report methods.}) yields a fairly reliable estimate
of their overall proclivities toward risk. Risk attitudes were also
accurately predicted by self-reports about how many traffic offenses
they have incurred, whether they smoke, their occupational choice,
participation in sports, and migration history (indirect self-report).
These factors also predict their risk-taking behavior in lab experiments
with real payoffs.

Just as telling an LLM to be honest, cheerful, or to update on a piece
of information can change its behavior, self-reports could be used to
prompt AI systems to better match the risk attitudes of their users.
Subsequent natural language prompts can further refine the AI's
understanding of user preferences. When making suggestions, the AI could
ask clarifying questions, such as, ``Given your preference for lower
risk, would you prefer a flight option with a longer layover and less
chance of delay?'' or ``Based on your reported willingness to take
risks, would you consider a higher-risk, higher-reward investment
option?''

By embedding self-reported risk tendencies into prompting, AI systems
can tailor their recommendations more closely to individual preferences,
improving user satisfaction while remaining simple to implement.
Additionally, approaches like this one bridge the gap between general
self-reported risk tendencies and context-specific decisions, allowing
for a mixed method that balances the reliability of general tendencies
with the flexibility of real-time feedback. We examine one such approach
in section 7.

\subsection{Direct observation}\label{direct-observation}

For some applications, an AI might be able to directly observe a user's
behavioral track record. For example, an investing AI might gain access
to data about the user's activities in a trading app. Then, this data
can be used to train an imitation learner (either to copy behavior or to
infer a function that best describes the user's risk attitudes). This
elicitation method avoids many of the pitfalls of self-reporting; it
does not rely on human memory and may have access to more information
about the choice environment, including other options that were not
taken (e.g. the prices of other stocks that the user could have selected
that day). Because the actions in the data set are of the same kind as
the actions the agentic AI will perform on behalf of the user, the data
has high validity.\footnote{There are several anomalies in typical
  buying and selling behavior that complicate assessments of risk
  aversion. In particular, people's fair selling price is typically much
  higher than their fair buying price for the same item or bet (Isaac
  and James 2000). This ``endowment effect'' is related to the loss
  aversion discussed by Kahneman and Tversky.}

However, direct observation comes with its own set of challenges. A
significant one is ensuring that the observation mechanism is neither
too narrow nor too broad. If the observation mechanism is too narrow, it
might miss out on key contextual information and the underlying reasons
why people made the choices they did. For instance, it may fail to
capture situational factors or alternative options that were considered
but not selected, leading to an incomplete understanding of user
preferences. On the other hand, if the observation system is too
sophisticated or expansive, it might overanalyze certain behaviors,
attributing meaning to actions where none exists. For example, a user
might not have checked as many alternatives as the system assumes,
leading the AI to infer preferences that the user did not actually
express, leaving their true preferences silent on those comparisons.

Consider, for example, the issue of overfitting. As Barocas \emph{et al}
suggest ``overfitting is a well-understood problem in machine learning
and there are many ways to counteract it. Since the spurious
relationship occurs due to coincidence, the bigger the sample, the less
likely it is to occur'' (2019, 38). However, the authors also note that
``variants of the overfitting problem can be much more severe and
thorny'' (38). One such thornier kind is \emph{adaptive overfitting}
which is caused by test set reuse (Roelofs 2019) and could be
problematic for ML methods more generally, not just models of direct
observation. More broadly, even when the sample is really large, one
should be careful about the over-reliance on observational data. In
particular, Lazer, \emph{et al.} show how big data models may contain
critically problematic algorithmic dynamics and how the ``quantity of
data does not mean that one can ignore foundational issues of
measurement and construct validity and reliability and dependencies
among data" (2014, 1204).

To mitigate these issues, one could aim to strike a balance in the
design of the observation mechanism. The system should be robust enough
to capture meaningful data about user behavior and the context of their
choices, but not so complex that it starts reading too much into the
data. Pairing observational data with direct user feedback can help
achieve this balance, providing a clearer picture of user preferences
and the reasons behind their decisions. For example, after observing a
trade, the AI might prompt the user with questions like, ``How satisfied
were you with that trade?'' or ``Which of these two trades do you think
was better?'' This combined approach helps in refining the AI's
understanding and ensures that the inferences drawn are both accurate
and reflective of the user's actual preferences.

\subsection{Population-level data}\label{population-level-data}

A third method is to use population-level data about choices under
uncertainty (e.g. how much people typically spend on car insurance,
typical stock trading behavior, etc.) which can then be calibrated to
individuals in several ways. First, data about demographic
subpopulations could be used to give a more accurate estimate of a
particular user. For example, women tend to be more risk averse than
men, so the agentic AI could adjust to be more risk averse for women
users (Eckel and Grossman 2002, 2008; see Nelson 2012 for skepticism).
Tall people and people with highly educated parents tend to be more risk
averse (Dohmen\emph{, et al.,} 2005). As mentioned above, finer
calibration can likely be achieved by treating population averages as a
default and then eliciting individual users' preferences to marginally
improve that default. The most extensive data about risk preferences
across global subpopulations is Falk, \emph{et al.} (2018).

This practice raises ethical concerns. We might worry that by
stereotyping people, we fail to treat them as individuals (Blum 2004).
Unless we have perfect accuracy, the distribution of risk attitudes will
contain some bias, resulting in an uneven distribution of advantages
(Holm 2023). Lastly, as discussed in \hyperref[Part1]{{Paper 1}}, there might be a moral
symmetry between types of errors (i.e. that it is better to treat
someone with too much risk aversion than with too much risk tolerance)
that would lead us to depart from matching default risk attitudes to
(sub)population averages.

\subsection{Evaluating the reliability and validity of actual
behavior
data}\label{evaluating-the-reliability-and-validity-of-actual-behavior-data}

In this section, we have considered various sources of data about
people's actual behaviors in conditions of uncertainty: observations of
their actual behaviors and self-reports about those behaviors. When
available and valid, these kinds of data may be suitable inputs to
imitation learning and prompting methods for calibrating AIs to the risk
attitudes of their users. The table below summarizes each surveyed
method's strengths and weaknesses:

\begin{longtable}[]{@{}
  >{\raggedright\arraybackslash}p{(\linewidth - 4\tabcolsep) * \real{0.2147}}
  >{\raggedright\arraybackslash}p{(\linewidth - 4\tabcolsep) * \real{0.2917}}
  >{\raggedright\arraybackslash}p{(\linewidth - 4\tabcolsep) * \real{0.4936}}@{}}
\toprule\noalign{}
\begin{minipage}[b]{\linewidth}\centering
\textbf{Method}
\end{minipage} & \begin{minipage}[b]{\linewidth}\centering
\textbf{Strengths}
\end{minipage} & \begin{minipage}[b]{\linewidth}\centering
\textbf{Weaknesses}
\end{minipage} \\
\midrule
\begin{minipage}[b]{\linewidth}\raggedright
\textbf{Self-Report}
\end{minipage} & \begin{minipage}[b]{\linewidth}\raggedright
Direct and accessible way to gather data on past behaviors and
tendencies.
\end{minipage} & \begin{minipage}[b]{\linewidth}\raggedright
Prone to memory biases and inaccuracies, especially in retrospective
reporting.
\end{minipage} \\
\midrule
\begin{minipage}[b]{\linewidth}\raggedright
\textbf{Direct Observation}
\end{minipage} & \begin{minipage}[b]{\linewidth}\raggedright
Provides accurate, contextually rich data that closely mirrors actual
behavior.
\end{minipage} & \begin{minipage}[b]{\linewidth}\raggedright
Limited to scenarios with comprehensive digital records; may miss
information about unchosen alternatives or user satisfaction. Might
overinterpret user choices.
\end{minipage} \\
\midrule
\begin{minipage}[b]{\linewidth}\raggedright
\textbf{Population-Level Data}
\end{minipage} & \begin{minipage}[b]{\linewidth}\raggedright
Offers broad benchmarks and defaults for calibration, useful as a
starting point.
\end{minipage} & \begin{minipage}[b]{\linewidth}\raggedright
Risks oversimplifying individual differences and raises ethical concerns
when demographic characteristics are used to infer behavior.
\end{minipage} \\
\midrule\noalign{}
\endhead
\bottomrule\noalign{}
\endlastfoot
\end{longtable}

\section{\texorpdfstring{Preference Elicitation in Hypothetical
Choice Experiments
}{Preference Elicitation in Hypothetical Choice Experiments }}\label{preference-elicitation-in-hypothetical-choice-experiments}

Preference-based reinforcement learning uses data about users' rankings
or ratings of presented options. Much of the work on individual risk
attitudes in behavioral economics uses this methodology, eliciting
subjects' preferences in hypothetical lotteries. Here, we present some
of the most common methods and evaluate their validity and reliability.

\subsection{Multiple Price List}\label{multiple-price-list}

A first methodology is the multiple price list. Each item in the list is
a comparison between two bets, a safer Option A and a riskier Option B.
Subjects are asked which of A or B they prefer for each line in the
list. At the top of the list, the safe option A has a higher expected
utility than risky B, and we gradually manipulate the comparisons until
risky B has a higher expected utility than A. We can measure subjects'
amount of risk aversion (the relative risk premium) by finding ``the
mathematical expected value that one is willing to forgo to obtain
greater certainty'' (Abdellaoui, et al. 2011, 65-66); \emph{i.e.} how
much more expected utility B has to have before they are willing to
switch over to the risky bet.

For example, in the following price list from Holt and Laury (2002), a
risk-neutral subject would switch to B between lines 4 and 5 (when B
overtakes A in expected payoff), while a risk-averse subject would
persist with A for longer and a risk-seeking subject would switch
earlier.\footnote{As we mentioned in \hyperref[Part1]{{Paper 1}}, Holt and Laury observed
  considerable amounts of risk aversion across every condition tested;
  in their studies, 6-15\% of participants were risk loving, 13-29\%
  risk neutral, and 56-81\% risk averse.}

\begin{table}[ht]
\centering
\caption{The ten paired lottery-choice decisions with low payoffs}
\begin{tabular}{llc}
\toprule
\textbf{Option A} & \textbf{Option B} & \textbf{Expected payoff difference} \\
\midrule
1/10 of \$2.00, 9/10 of \$1.60 & 1/10 of \$3.85, 9/10 of \$0.10 & \$1.17 \\
2/10 of \$2.00, 8/10 of \$1.60 & 2/10 of \$3.85, 8/10 of \$0.10 & \$0.83 \\
3/10 of \$2.00, 7/10 of \$1.60 & 3/10 of \$3.85, 7/10 of \$0.10 & \$0.50 \\
4/10 of \$2.00, 6/10 of \$1.60 & 4/10 of \$3.85, 6/10 of \$0.10 & \$0.16 \\
5/10 of \$2.00, 5/10 of \$1.60 & 5/10 of \$3.85, 5/10 of \$0.10 & -\$0.18 \\
6/10 of \$2.00, 4/10 of \$1.60 & 6/10 of \$3.85, 4/10 of \$0.10 & -\$0.51 \\
7/10 of \$2.00, 3/10 of \$1.60 & 7/10 of \$3.85, 3/10 of \$0.10 & -\$0.85 \\
8/10 of \$2.00, 2/10 of \$1.60 & 8/10 of \$3.85, 2/10 of \$0.10 & -\$1.18 \\
9/10 of \$2.00, 1/10 of \$1.60 & 9/10 of \$3.85, 1/10 of \$0.10 & -\$1.52 \\
10/10 of \$2.00, 0/10 of \$1.60 & 10/10 of \$3.85, 0/10 of \$0.10 & -\$1.85 \\
\bottomrule
\end{tabular}
\end{table}

An agentic AI could present such a list of options to a user (either in
a calibration phase or during the course of use) to try to determine
their relative risk premium. What is necessary is that users are
presented with a series of choices that vary in their riskiness, and
this riskiness is ramped down (or up) until users find a level of risk
they find acceptable. For domain-specific AIs, these options could
involve choices from that domain. For example, a travel planner that
books plane tickets could present users with comparisons such as:

\begin{enumerate}
\def\labelenumi{\alph{enumi}.}
\item
  Option A: long layover, little chance of delay or missed flight
\begin{enumerate}
    \renewcommand{\labelenumii}{}  
    \item 
\emph{0.4 chance travel time of 14 hours, 0.6 chance travel time of 16
hours}
\end{enumerate}
\end{enumerate}

\begin{enumerate}
\def\labelenumi{\alph{enumi}.}
\setcounter{enumi}{1}
\item
  Option B: short layover, greater chance of delay or missed flight
\begin{enumerate}
    \renewcommand{\labelenumii}{}  
    \item
\emph{0.4 chance travel time of 8 hours, 0.6 travel time of 20 hours}
\end{enumerate}
\end{enumerate}

And then find the point at which the user is willing to risk a delay or
missed flight for a chance at a shorter travel time.

One of the drawbacks of the multiple price list methodology is that it
can be complex and cognitively demanding for subjects to navigate, which
carries the risk of error or users abandoning the methodology before
completion. Further, there is a worry that the list ordering and range
will cause anchoring or order effects on users (e.g. that they will
always tend to choose bets later in the list). Some of these issues
could be mitigated through effective UI. For example, subjects can use a
slider to choose the point at which they would switch from A to B
(Anderson \emph{et al}, 2006)

\subsection{Random lottery pairs}\label{random-lottery-pairs}

The multiple price list methodology is systematic: it varies the level
of risk aversion to determine where a user falls on that continuum. The
cost of this was significant complexity and time demand on the user. An
alternative is to present single choices between randomly selected bets
in a standard A vs. B preference task. The chief advantage of this
methodology is that it is very straightforward to explain and
understand, and it does not require significant user investment.

The downside is that ``contrary to the MPS, it is generally not possible
to directly infer a risk attitude from the pattern of responses, and
some form of estimation is needed" (Harrison and Rutstrom 2008, 52).
This can be assisted by an algorithm which selects comparisons that are
likely to be highly informative (Handa, \emph{et al.} 2024), especially
in light of the prior responses of the subject (Wakker and Deneffe
1996), or by pretraining on population-level data (Askell 2021).

\subsection{Ordered lottery
suggestion}\label{ordered-lottery-suggestion}

Subjects are presented with an ordered set of bets and are asked to pick
their favorite. For example, they might see a list like:

Option A: 0.1 chance of \$100, 0.9 chance of \$0

Option B: 0.5 chance of \$10, 0.5 chance of \$1

Option C: sure thing of \$4

Similarly, Dohmen, \emph{et al.} (2005) provides subjects with the
following scenario:

\begin{quote}
Imagine that you win 100,000 euros in a lottery. A bank offers you an
investment in an asset that has equal chances of doubling or halving
your money in two years time. How much of your winnings would you invest
in that asset? Options: 0, 20,000, 40,000, 60,000, 80,000, or 100,000
Euros
\end{quote}

This methodology combines some of the virtues of the Multiple Price List
and Random Lottery methods. Like MPL, it allows developers to
systematically vary options along some desired dimension so that risk
preferences can be approximated. Like RL, it is relatively simple, only
requiring users to make only one choice (not an iterated series of
choices).

A real-world, domain-specific implementation of this methodology is the
simplified menu of choices that is presented when people pick their
401(k)s, such as the following publication from Charles Schwab:

\begin{table}[ht]
\centering
\caption{Hypothetical performance for conservative, moderate, and aggressive model portfolios}
\resizebox{\textwidth}{!}{ 
\begin{tabular}{lccc}
\toprule
\textbf{Asset allocation} & \textbf{Conservative portfolio} & \textbf{Moderate portfolio} & \textbf{Aggressive portfolio} \\
\midrule
Stocks & 30\% & 60\% & 80\% \\
Bonds & 50\% & 30\% & 15\% \\
Cash  & 20\% & 10\% & 5\% \\
\midrule
\multicolumn{4}{l}{\textbf{Hypothetical Performance (1970–2014)}} \\
\midrule
Growth of \$10,000 & \$389,519 & \$676,126 & \$892,028 \\
Annualized return & 8.1\% & 9.4\% & 10.0\% \\
Annualized volatility (standard deviation) & 9.1\% & 15.6\% & 20.5\% \\
Maximum loss & -14.0\% & -32.3\% & -44.4\% \\
\bottomrule
\end{tabular}
}
\end{table}

This methodology could plausibly be integrated into the operation of the
agentic AI with user feedback. For example, many online assistants
already work by presenting users with their top three suggestions (e.g.
travel websites that first display a list of recommended flights). In
early stages of calibration, an agentic AI could manipulate these
choices so as to be maximally informative of risk preferences. Later
stages could require less input from users or present options that are
all closer to their risk preferences.

\subsection{\texorpdfstring{Assessing validity: Hypothetical
choices don't predict actual choices
}{Assessing validity: Hypothetical choices don't predict actual choices }}\label{assessing-validity-hypothetical-choices-dont-predict-actual-choices}

People's stated and revealed risk preferences may diverge, perhaps
significantly. Much of the experimental data on risk aversion comes from
subjects' hypothetical choices in fictional scenarios for which there
are no or only low-stakes financial consequences.\footnote{Lab-based
  preference elicitations are usually incentive-compatible, with one
  randomly-selected choice actually implemented. This presumably
  won\textquotesingle t be the case with calibration stages of AIs,
  making them even less valid than lab experiments.} While this makes
experiments less expensive to run and avoids ethical challenges with
imposing financial losses on participants, the methodology of
hypothetical choices ``relies on the assumption that people often know
how they would behave in actual situations of choice, and on the further
assumption that the subjects have no special reason to disguise their
true preferences" (Kahneman and Tversky 1979, 265).

Further experimental work has shown that people's preferences are
substantially different in hypothetical choices than when real money is
at stake (Harrison 2006, 2014). For example, Holt and Laury (2002)
compared subjects' preferences over gambles when they would actually
receive the payoffs of those gambles versus those in purely hypothetical
choice scenarios. Subjects were more risk averse overall in the actual
payoff condition.\footnote{This is consistent with experiments offering
  gambles to subjects outside of the laboratory (Binswager 1980) and
  field data from auctions (Cox and Oaxaca 1996; Campo 2000).}
Additionally, while subjects had similar risk preferences for low and
high stakes bets in the hypothetical condition, they were more risk
averse for high stakes than low stakes bets in the actual payoff
condition. Holt and Laury argue, ``contrary to Kahneman and
Tversky\textquotesingle s supposition, subjects facing hypothetical
choices cannot imagine how they would actually behave under
high-incentive conditions. Moreover, these differences are not
symmetric: subjects typically underestimate the extent to which they
will avoid risk'' (1654).\footnote{For a more recent overview, see
  Bokern, \emph{et al.} (2023).}

As a normative question, it is not clear whether deferential agentic AIs
should be calibrated to stated or revealed preferences, or somewhere in
between. Should a successful proxy agent be one that behaves as the user
would behave or one that behaves as the user (ex ante) thinks they
should behave? The correct choice might depend on the context and the
role that the AI plays.

\subsection{\texorpdfstring{Assessing reliability:
inconsistencies across methods
}{Assessing reliability: inconsistencies across methods }}\label{assessing-reliability-inconsistencies-across-methods}

If we want to use stated user preferences across hypothetical choices to
calibrate AIs, it is very important that the methods used to elicit
those preferences are reliable. When the data is noisy, there will be a
trade-off between how finely-calibrated and how predictively accurate a
model is since finely-calibrated models will tend to overfit (Forster
and Sober 1994). In this case, more coarse-grained models should be
preferred. There is also a trade-off between the systematicity or
completeness of a method and how easily it can be completed by users
(with respect to both time and cognitive resources). A common view among
social scientists is that simpler and fewer comparisons should be
preferred when possible. The downside is that we only get snapshots of a
user's risk preferences rather than a systematic range of preferences
across probability and payoff levels.

A review of the empirical record on risk elicitation in hypothetical
choice scenarios shows that these methods tend to be highly unreliable,
both across methods and when using the same method across contexts. In a
review of the literature, Holt and Laury (2014) conclude that there is
``little evidence of correlation of risk attitude between decision making
tasks... Moreover, there is little evidence that behavior in any of
these choice tasks explains self-reported propensities to take naturally
occurring risks" (174).

First, methods seem to disagree with one another. For example, people
are much more risk averse in tasks regarding lottery sales than lottery
purchases (Isaac and James 2000, Levy 1992). They behave differently in
choice-based (A vs. B) tasks than price-based (how much would you pay
for A vs. B) tasks (Harbaugh \emph{et al.} 2010). Holt and Laury (2014)
conclude:

\begin{quote}
Even a cursory review of the literature makes it clear that there is no
consistent pattern of results in experimental studies of risk
preferences over losses, whether one focuses on the degree of risk
aversion or the responsiveness of risk attitude to changes between gains
and losses, payoff scale, and probability of gain or loss\ldots{} this
may be explained, in part, if elicited risk preferences are highly
sensitive to the procedure used to elicit them\ldots{}

It is not altogether surprising that estimates of the coefficient of
risk aversion differ across elicitation methods, but it is troubling
that the rank-order of subjects in terms of their risk aversion
coefficient differs across elicitation methods (168, 172-173).
\end{quote}

Second, individual methods can yield different results across contexts.
Within a particular elicitation method, risk attitudes can be sensitive
to: the probability levels of compared options; the scale of payoffs;
the perceived reference point dividing gains from losses, which is
subject to anchoring effects; how background wealth is incorporated into
the decision (asset integration); whether outcomes are described as
gains or losses; and whether the procedure permits them to use math when
making the decision.

Individuals also show variability in elicited risk attitudes across
different choice domains (e.g. health, finances, personal safety). For
example, someone who is very risk tolerant in deciding how early to get
to the airport might be very risk averse when it comes to their
retirement investments. Dohmen, \emph{et al.} (2005) examined
self-reported risk attitudes in 5 areas: general, career, sports and
leisure, car driving, health, and financial matters. They found strong
and significant correlations ($\approx0.5$) in risk attitudes
across domains. Principal component analysis revealed that ``about 60
percent of the variation in individual risk attitudes is explained by
one principal component, consistent with the existence of a single
underlying trait determining willingness to take risks'' (25). However,
this still leaves a significant amount of variation to be accounted for
by domain-specific considerations.

Given all of these contingencies, ``the evidence suggests that one
should be cautious about using a risk aversion estimate obtained in one
context to make inferences about behavior in another (unrelated)
context'' (Holt and Laury 2014, 172-173).

\subsection{Summary}\label{summary}

Preference modeling techniques in AI typically involve learning a reward
function from user feedback. When it comes to risk preference modeling,
one common approach is to elicit user preferences through hypothetical
choices under uncertainty, a method widely used in behavioral economics.
However, based on the evidence discussed in Sections 6.1 through 6.5, we
do not recommend relying on these elicitation methods due to their
significant limitations.

The drawbacks of hypothetical choice experiments include their
complexity, the potential for cognitive overload, and the systematic
biases that can distort users\textquotesingle{} stated preferences.
Furthermore, the discrepancies between stated and revealed preferences
raise serious concerns about the validity of these methods. The
empirical record suggests that these approaches are not only unreliable
across different methods but also inconsistent within the same method
when applied in different contexts.

Despite these issues, developers might still be inclined toward
preference modeling techniques for a variety of reasons (e.g. they want
to re-use existing learning tools, it's easiest for them to get
hypothetical preference data, etc.). If so, alternative approaches to
eliciting risk preferences should be explored. Although we cannot fully
evaluate these alternatives without empirical data on their reliability,
we can offer some guiding principles for improving the elicitation
process:

\begin{enumerate}
\def\labelenumi{\arabic{enumi}.}
\item
  \textbf{Domain-Specific Methods}: Risk preferences vary significantly
  across different domains, such as health, finance, and personal
  safety. It is advisable to develop and use domain-specific elicitation
  methods that are tailored to the particular context in which the AI
  will operate.
\item
  \textbf{Embrace Coarse-Grained Approaches}: Humans are generally poor
  at accurately judging probabilities and outcomes in complex scenarios.
  Instead of striving for fine-grained precision, it may be more
  effective to use coarse-grained methods that capture broad preferences
  without overwhelming the user or asking about niche scenarios the user
  cannot easily or fully imagine. This may also mitigate the risk of
  overfitting.
\item
  \textbf{Align Preferences with Actual Choices}: Whenever possible, the
  elicitation of preferences should closely resemble actual
  decision-making scenarios. This alignment increases the likelihood
  that the preferences captured are reflective of the choices users
  would make in real-world situations, thereby enhancing the reliability
  and applicability of the model.
\end{enumerate}

In conclusion, if we pursue preference-based reinforcement learning for
tailoring AI systems to individual risk profiles, the challenges
associated with traditional elicitation methods suggest the need for a
careful reevaluation of how we gather and use user feedback. By focusing
on domain-specificity, coarse-grained approaches, and real-world
alignment, we can improve the accuracy and effectiveness of risk
preference modeling in AI systems. Mixed methods, where, for example,
prompting is used in addition to preference-based reinforcement learning
might be particularly fruitful.

\section{Non-learning General Risk-Classification Method as a
Starting
Point}\label{non-learning-general-risk-classification-method-as-a-starting-point}

In this section we put forward an approach that adopts a Dohmen-style
general risk question, where users were asked to ``specify your
willingness to take risks from 0 (completely unwilling) to 10
(completely willing)." Based on their responses, users are categorized
into broad risk aversion classes rather than assigned precise risk
aversion parameters.

Once users are assigned a general risk aversion score based on their
self-reported willingness to take risks, a menu of agent profiles can be
constructed, each corresponding to different levels of risk aversion,
ranging from extreme risk aversion to extreme risk love. Users are then
matched with agents that align with their assigned risk aversion
category. This approach ensures that the AI's decision-making processes
are appropriately tailored to the user's risk tolerance without the need
for complex, continuous learning algorithms from the outset.

Finally, risk profiles are matched with specific behaviors in the domain
of action, i.e. candidate actions are labeled from extremely safe to
extremely risky. These labels might be hard-coded and derived from
expert judgments. Alternatively, they could be generated from data. For
example, we could rank choices by the relative risk premium (Abdellaoui,
et al. 2011; see \hyperref[Part1]{{Paper 1}}), or by the part of the distribution of
outcomes that the decision is based on (e.g. the most risk averse action
assumes the worst-case 5\% quantile of the outcome distribution).

\subsection{The Framework}\label{the-framework}

Below is an example of a qualitative categorisation based on the general
risk question and related user behavior, such as arriving at airports
for international flights:

\begin{longtable}[]{@{}
  >{\raggedright\arraybackslash}p{(\linewidth - 4\tabcolsep) * \real{0.32}}
  >{\centering\arraybackslash}p{(\linewidth - 4\tabcolsep) * \real{0.27}}
  >{\raggedright\arraybackslash}p{(\linewidth - 4\tabcolsep) * \real{0.43}}@{}}
\toprule\noalign{}
\begin{minipage}[b]{\linewidth}\centering
\textbf{Risk Aversion Category}
\end{minipage} & \begin{minipage}[b]{\linewidth}\centering
\textbf{Risk Question Response}
\end{minipage} & \begin{minipage}[b]{\linewidth}\centering
\textbf{Example Behavior}
\end{minipage} \\
\midrule
\begin{minipage}[b]{\linewidth}\raggedright
\doublespacing
\textbf{Extreme Risk Aversion}
\end{minipage} & \begin{minipage}[b]{\linewidth}\centering
\doublespacing
0-1
\end{minipage} & \begin{minipage}[b]{\linewidth}\raggedright
\doublespacing
Arrive 6 hours before international flights.
\end{minipage} \\ \\
\begin{minipage}[b]{\linewidth}\raggedright
\doublespacing
\textbf{Additional Risk Aversion}
\end{minipage} & \begin{minipage}[b]{\linewidth}\centering
\doublespacing
2-3
\end{minipage} & \begin{minipage}[b]{\linewidth}\raggedright
\doublespacing
Arrive 4 hours before international flights.
\end{minipage} \\ \\
\begin{minipage}[b]{\linewidth}\raggedright
\doublespacing
\textbf{Default (Average Aversion)}
\end{minipage} & \begin{minipage}[b]{\linewidth}\centering
\doublespacing
4-6
\end{minipage} & \begin{minipage}[b]{\linewidth}\raggedright
\doublespacing
Arrive 3 hours before international flights.
\end{minipage} \\ \\
\begin{minipage}[b]{\linewidth}\raggedright
\doublespacing
\textbf{Additional Risk Love}
\end{minipage} & \begin{minipage}[b]{\linewidth}\centering
\doublespacing
7-8
\end{minipage} & \begin{minipage}[b]{\linewidth}\raggedright
\doublespacing
Arrive 2 hours before international flights.
\end{minipage} \\ \\
\begin{minipage}[b]{\linewidth}\raggedright
\doublespacing
\textbf{Extreme Risk Love}
\end{minipage} & \begin{minipage}[b]{\linewidth}\centering
\doublespacing
9-10
\end{minipage} & \begin{minipage}[b]{\linewidth}\raggedright
\doublespacing
Arrive 1 hour before international flights.
\end{minipage} \\
\midrule\noalign{}
\endhead
\bottomrule\noalign{}
\endlastfoot
\end{longtable}

Each category represents a different level of risk tolerance, with
``Extreme Risk Aversion" users opting for the most cautious behavior,
ensuring they have ample time before a flight, and ``Extreme Risk Love"
users taking the most risks, arriving at the airport just in time. This
method allows developers to create agentic AIs that can operate within
predefined behavioral boundaries while catering to the varied risk
preferences of different users.

While the initial classification into risk categories is based on
general self-reports, each agent can later implement learning methods to
fine-tune its decision-making to better align with specific contexts and
individual user preferences. For example:

\begin{enumerate}
\def\labelenumi{\arabic{enumi}.}
\item
  \textbf{Context-Specific Adjustments}: An agent designed for a user
  classified as ``Additional Risk Aversion" could observe the
  user\textquotesingle s actual behavior in specific situations (e.g.,
  booking flights, making investments) and adjust its recommendations
  accordingly. If the user consistently opts for slightly riskier
  choices in one domain (like travel), the agent can learn to reflect
  this subtle preference within that domain without altering its overall
  risk aversion classification. If observed behaviors are very
  incongruous with the user's self report, the AI may match the user
  with a different risk profile.
\item
  \textbf{User Feedback Integration}: Agents can prompt users with
  natural language questions or simple feedback mechanisms to refine
  their preferences further. For instance, after a few flight bookings,
  the agent could ask, ``Did you find the layover time too long, too
  short, or just right?" Such feedback allows the agent to learn user
  preferences more precisely in context.
\item
  \textbf{Non-generalizing Learning}: Crucially, the learning that takes
  place within each agent is confined to the specific context in which
  the agent operates. For example, an agent that learns a user's risk
  preferences in financial investments does not generalize those
  preferences to other contexts, like health decisions. This ensures
  that the agent remains accurately calibrated to the user's
  domain-specific risk tolerance.
\end{enumerate}

\subsection{Distinguishing from Direct Preference
Modeling}\label{distinguishing-from-direct-preference-modeling}

This mixed approach, where general risk classification serves as the
foundation and contextual learning refines the model, differs
fundamentally from the preference modeling methods critiqued earlier:

\begin{enumerate}
\def\labelenumi{\arabic{enumi}.}
\item
  \textbf{Initial Simplicity with Contextual Refinement}: Unlike direct
  preference modeling, which aims to infer precise risk parameters from
  complex and cognitively demanding tasks, this method begins with a
  straightforward categorisation exercise. Learning is applied only when
  necessary to fine-tune specific contexts, reducing the risk of users'
  cognitive overload and ensuring more reliable user alignment.
\item
  \textbf{Domain-Specific Learning}: The learning methods applied here
  are context-specific, meaning they refine the agent's behavior within
  narrowly defined domains. This contrasts with the broad,
  context-generalizing nature of traditional preference modeling, which,
  as we have seen, can lead to inconsistent or unreliable results across
  different scenarios.
\item
  \textbf{Mitigating Overfitting and Bias}: By starting with a broad
  classification and using feedback to fine-tune only as needed, this
  approach avoids the overfitting issues commonly associated with
  preference modeling. Since the adjustments are confined to specific
  domains, the model remains robust and less prone to bias from isolated
  or anomalous behaviors.
\end{enumerate}

In summary, this approach balances simplicity with adaptability,
offering a robust method for aligning AI behaviors with user preferences
without some of the common drawbacks of direct preference modeling
techniques. It leverages general classifications to establish a simple
foundation, then fine-tunes agents within specific contexts, to improve
accuracy and promote user satisfaction.

\section{Concluding Remarks}\label{concluding-remarks}


Throughout this series of reports, we have contrasted two models of the relationship between users and agentic AIs. According to the Proxy model, agentic AI systems are representatives of their users and should be designed to replicate their risk attitudes. According to the Off-the-Shelf Tool model, developers provide users with a menu of AI agents whose risk attitudes are set (or at least highly constrained) by developers.  

In these reports, we have discussed some reasons why the AI Proxy model might be attractive. In Paper 1, we examined reasons why users might desire agentic AIs that embody their risk attitudes. In Paper 2, we explored how the Proxy model might navigate shared responsibility by shifting responsibility toward users and away from developers. Despite its intuitive appeal, the Proxy model faces significant technical and normative challenges. 

On the technical side, we would need good sources of data about users’ risk attitudes and good methods for learning from that data to accurately calibrate AI models to their users. We have outlined some of the significant limitations of various methods for eliciting that data. In many contexts, the methods that are the most reliable and valid may be relatively coarse-grained and based on user self-reports. As a result, we predict that methods that match users to pre-existing risk classes (as recommended by the Off-the-Shelf Tool model) may outperform learning-only methods.  

In the first two papers, we argued that there are also considerable normative reasons for adopting the Off-the-Shelf Tool model. Because some users may opt for AI agents that behave recklessly, developers can avoid legal, reputational, and moral liability by placing constraints on AI risk attitudes. Moreover, when a human agent takes actions on another agent’s behalf, it is not typically expected that they match their risk attitudes. Instead, alignment is achieved through transparency and explicit rules governing shared responsibility.

We judge that the Off-the-Shelf Tool model constitutes a strategy for developing agentic AIs that has considerable technical and normative strengths. There are a few important areas of future research regarding the comparison between Off-the-Shelf and Proxy methods. First, new learning techniques that are sensitive to a range of inputs (population data, user self-reports, observed behavior, etc.) might yield better methods for achieving calibration to individual risk attitudes than those we have considered here. Second, more research on our proposed non-learning risk classification framework is needed. We have assumed that alignment will involve matching users up with agents whose risk profiles roughly match theirs. However, it might be the case that some users benefit most and are most satisfied when matched to risk profiles that are different from their own.\footnote{See Section 5.1 of \hyperref[Part1]{{Paper 1}} for a discussion.} In that case, we would need to develop ways of measuring user alignment for off-the-shelf tools that go beyond accuracy in capturing user risk preferences. More generally, further work should investigate how AI risk attitudes influence user trust in and satisfaction with agentic AI systems.

%% file: OpenAI_Appendix.tex
\subsection{Incorporating risk attitudes into expected utility}\label{incorporating-risk-attitudes-into-expected-utility}

We have characterized risk sensitivity as a necessary third factor when
describing an agent; we need their credences, their utilities,
\emph{and} their sensitivity to risk. However, some traditional
approaches to modeling risk aversion eschew this third factor by
building risk sensitivity into utilities themselves (von Neumann and
Morgenstern 1944, Savage 1972, Pettigrew 2015). The utility function
will be concave for a risk-averse actor, linear for a risk-neutral
actor, and convex for the risk-prone actor.

There are two ways to interpret the relationship between risk
sensitivity and utilities (which we will return to in more detail in
5.3). Consider Nate and Kate. We modeled them as assigning equally high
utility to eating at restaurant A and equally middling utility to B and
differing in their risk attitudes. Utilities are often thought of as
measures of the subjective value of an outcome for an agent. We assumed
that Nate and Kate value the taste, ambiance, etc. of the restaurants
the same way. However, we might think that the utilities involve more
than just these experiences. Kate does not like taking risks, so she
assigns a lower utility to the risky restaurant. We could take this to
reflect an aspect of her experience, such as the stress of uncertainty.
We might instead be somewhat more behaviorist. Someone's utilities
reflect their overall dispositions toward certain choices. Here,
utilities are attitudes about \emph{bets} rather than outcomes; since
Kate is risk averse, the utility of the risky bet is lower than the safe
bet.

We do not favor building risk aversion into utilities, for several
reasons. First, it might not be descriptively adequate. It (arguably)
cannot capture Allais preferences, where there is no consistent
assignment of utilities to amounts of money that captures agents'
preferences across bets, and it has systematic failures when used to
predict the economic behavior of actual agents (Abdellaoui, \emph{et
al.} 2011).

Suppose we dismiss this worry and grant that we can provide
mathematically equivalent descriptions of an agents' behavior by either:

\begin{enumerate}
\def\labelenumi{\arabic{enumi}.}
\item
  representing utilities and risk via separate variables, or
\item
  representing utilities and risk via a single variable that is a
  function of both.
\end{enumerate}

We think that there are methodological and theoretical reasons for
favoring the former. First, unpacking utilities and risk sensitivity
into two separate variables allows us to track the relative
contributions of each. For example, compare Kate to Tate. Like Nate,
Kate likes the experience of eating at restaurant A more than B but
disfavors it for reasons of risk aversion. Tate, on the other hand,
dislikes the food at restaurant A but is risk neutral. Kate and Tate
might assign equal utilities to A and B and thus behave the same way,
but those utilities stem from very different kinds of values. We want
our model to have the tools to represent what Nate and Kate have in
common with each other, but not with Tate. Lastly, in Section 5.3, we
will discuss a philosophical dispute about the nature of risk attitudes:
whether they are an intrinsic part of what is valued or an instrumental
means of getting what is valued. Keeping them notationally separate
allows us to remain agnostic on this front.

\subsection{{REU and WLU}}\label{reu-and-wlu}

To recall, EU maximization is a risk-neutral decision theory because it
doesn't allow for bad outcomes to be treated differently from good ones
or to treat low probabilities differently from high ones. The two most
prominent risk-sensitive decision theories among philosophers introduce
these abilities.

\subsubsection{Risk-weighted Expected Utility Theory}\label{risk-weighted-expected-utility-theory}

Building upon the rank-dependent risk theory of Quiggins (CITE), Buchak
(2013) develops and defends Risk-weighted Expected Utility (REU) theory
as a way of incorporating risk attitudes into expected value
comparisons. A risk-averse agent puts more weight on the worst-case
outcomes of a gamble than the best; that is, the worst case should
contribute more to their overall expected value calculation. In brief,
REU does this by: ranking the outcomes of a bet from worst to best,
diminishing the probabilities of better states, and reapportioning the rest of the
probability to better states --- all systematically done through a risk function $r$ that typically raises cumulative sums of those probabilities to some constant power. Her working example of $r$ squares the inner probability sums.

Suppose I offer you a bet on a fair coin: if it lands heads, I give you
200, and if it's tails, you lose 100. Instead of calculating EU by
taking the weighted average, it can be calculated by assuming a baseline
certainty of getting the utility of the worst case outcome
(x\textsubscript{1}), plus the probability that you get the additional
value of the second-worst outcome (x\textsubscript{2}) compared to the
worst case, and so on:

\begin{equation}
    EU(A) = \ \sum_{i = 1}^{n}\left\lbrack \left( \sum_{j = i}^{n}p\left( E_{j} \right) \right)(u\left( x_{i} \right) - u\left( x_{i - 1} \right)) \right\rbrack.
\end{equation}

In the bet I offered you above, you have a certainty of getting at least
$-100$, plus a $0.5$ chance of getting $300$ more than this, for a total
expected value of 50.

Now that we have an ordered list from worst- to best-case outcomes, we
can introduce the risk function, $r$, that places more decision weight on
those worst-case outcomes. REU does this by reducing (here, squaring inner sums)
the probabilities of jumping up to greater outcomes. In the bet I
offered you, you now have a certainty of getting at least $-100$ and now a
$0.5^2$ chance of getting $300$ more, so your risk-weighted
value is $-100 + 0.25(300) = -25$. This is worse, by your lights, than
refusing the bet.

More formally, take the list of outcomes of A from worst to best to be
\{E\textsubscript{1}, x\textsubscript{1}; \ldots; E\textsubscript{n},
x\textsubscript{n}\}, where x\textsubscript{i} is the consequence that
obtains in event E\textsubscript{i} (so E\textsubscript{1} is the event
of the worst case outcome obtaining, and x\textsubscript{1} is that
outcome's value). The REU of a bet A is:
\begin{equation}
    REU(A) = \ \sum_{i = 1}^{n}\left\lbrack r\left( \sum_{j = i}^{n}p\left( E_{j} \right) \right)(u\left( x_{i} \right) - u\left( x_{i - 1} \right)) \right\rbrack
\end{equation}

where $r$ is the risk function. If it raises inner sums by less than 1, the
agent will discount better outcomes and be risk averse. If does so by more
than 1, the agent will put more significance on better outcomes and be
risk seeking. It has EU as a special case when $r$ is the identity
function such that $r = 1$.

\subsubsection{Weighted-Linear Expected Utility
(WLU)}\label{weighted-linear-expected-utility-wlu}

Bottomley and Williamson (2023) defend Weighted-Linear Expected Utility
(WLU) as an alternative to REU. It departs from REU in one key way:
while REU introduces risk as a function of probabilities, WLU introduces
risk as a function of utilities (or values). It discounts (assigns less
decision weight) to better outcomes and amplifies worse outcomes.\footnote{Bottomley and Williamson view WLU as an improvement on REU, in particular because only the latter ``violate[s] the Betweenness axiom, which requires that you are indifferent to randomizing over two options between which you are already indifferent.'' (\emph{ibid.,} 697)}

WLU puts more weight on worst-case scenarios by adding a risk factor,
\emph{w}, that penalizes outcomes with higher utilities. Their working
example of \emph{w} for outcomes measured in money is $w(\$x) =
\frac{1}{1 + \ \sqrt[4]{x}}$. This risk weighting is applied to all
of the possible outcomes of an action. Then, you calculate the
\emph{relative weight} of each outcome, the outcome's risk-weighted
value divided by the risk-weighted value of all other outcomes, weighted
by their probability. Finally, the WLU of an action is the sum of the
utilities of all possible outcomes, weighted by their probabilities and
their \emph{relative weights}.

\begin{equation}
    WLU(A)\  = \ \sum_{i = 1}^{n}\left( \frac{w(x_{i})}{\Sigma_{j = 1}^{n}w(x_{j})p_{A}(x_{j})} \right)p_{A}(x_{i})u(x_{i})
\end{equation}

As desired, WLU puts more decision weight on worst-case outcomes,
displaying ``a high degree of responsiveness to bad outcomes coupled
with an almost risk-neutral attitude towards safe gambles''
(\emph{ibid.,} 14). It is stakes sensitive, tolerating a higher amount
of risk when the stakes are small (say, when gambling small amounts of
money) and less when the stakes are large (say, when gambling with one's
life savings).

\subsection{Prospect Theory}\label{prospect-theory}

EU and its risk-weighted extensions retain some
of the key assumptions of expected utility theory, and their goals are
largely normative rather than descriptive. Prospect Theory (Kahneman \&
Tversky 1979) departs from expected utility theory in both of these
respects. The theory is motivated by empirical lab results (some of
which we surveyed in Section 3) and attempts to predict the behavior of
actual agents. To do so, it introduces key assumptions about the kinds
of heuristics and biases that actual agents use when navigating
decisions under uncertainty. The theory has been extensively elaborated
upon, debated, and tested, and we cannot do service to all of these
developments (though we will return to some of these points in \hyperref[Part2]{{Paper 2}}).
Here, we will focus on the theory's key commitments and innovations.

First, standard decision theory assumes that people evaluate outcomes by
the overall amount of value that would result, the total amount of
assets that they would have in the final state of the bet. For example,
suppose you have $\$1$ million in existing assets and are considering
buying a lottery ticket with a $0.01$ chance of winning you $+\$1000$ and a
$0.99$ chance of losing you $-\$1$. Measured in terms of final assets, the
two possible outcomes are that you have $\$1,001,000$ or $\$999,999$.

However, for most people, ``the carriers of value or utility are changes
of wealth, rather than final asset positions that include current
wealth'' (Kahneman \& Tversky 1979, 273). People tend to evaluate
outcomes by their deviation from a reference point (typically, but not
always, the status quo). For example, if you evaluate the above bet in
terms of deviations from the status quo, the two relevant outcomes are
$+\$1000$ or $-\$1$. This has several implications. First, a person's
assessment of a bet can change depending on the choice of a reference
point, which can be influenced by framing effects. Therefore, the
assessment of a bet can depend on contextual factors. Second, bets will often be treated as having greater stakes and thus calling for different levels of risk sensitivity. We would predict that a millionaire would treat the
above bet as very low stakes and thus be risk tolerant. However, even a
millionaire may be risk averse when assessing the bet against a
reference point of $0$.

A second key psychological finding from Kahneman \& Tversky
(1979)\footnote{They also find interesting effects in subjects'
  reasoning about probabilities that show non-linearity in the
  significance of probabilities. For example, people treat certainties
  differently from other probabilities (e.g. being willing to pay more
  than reduce a chance of harm from $0.1$ to $0$ than from $0.2$ to $0.1$). We will
  not pay these much heed here in order to focus on other aspects of
  risk aversion.} is that people are very sensitive to whether an
outcome is framed as a loss or a gain and are much more loss avoidant
than gain seeking. For example, in an experiment from Williams (1966),
subjects were indifferent between a bet delivering 0 with certainty and
one delivering $100$ with probability $0.65$ and $-100$ with probability $0.35$
$(100, 0.65; -100, 0.35)$. This shows risk aversion, since the first bet has
an EU of 0 and the second bet has an EU of 30. However, they were also
indifferent between a bet delivering $-100$ with certainty and one
delivering $-200$ with probability 0.8 and 0 otherwise $(-200, 0.8; 0, 0.2)$.
This shows risk \emph{seeking}, since the first bet has an EU of -100
and the second bet has an EU of $-160$. In general, people are willing to
take risks to avoid losses and are risk averse when seeking gains.

This result manifests itself in economic behavior, as ``the minimal
compensation people demand to give up a good is often several times
larger than the maximum amount they are willing to pay for a
commensurate entitlement" (Levi 1992, 175). It also makes them sensitive
to framing effects, where the exact same bet is evaluated very
differently depending on whether it is described as loss avoidance or
gain seeking (Tversky and Kahneman 1981). For example, they presented
subjects with a choice between two programs for treating an epidemic
that would otherwise be expected to kill 600 people. These scenarios
were either described in terms of loss or gains:

Program A: with certainty 200 people will be saved (gain) / 400 people
will die (loss)

Program B: $1/3$ chance that 600 people will be saved and $2/3$ chance that 0
will be saved (gain) 
$1/3$ chance that 0 will die and $2/3$ chance that 600 will die (loss)

When the options were phrased in terms of gains (how many people could
be saved), most subjects (72\%) were risk averse, favoring A over B.
When the options were phrased in terms of losses (deaths), most (78\%)
were risk-seeking, favoring B over A.

Putting this together, Kahneman and Tversky (1979, 279) predict that a
typical agents' value function --- how much significance they place on
various outcomes --- is as follows:

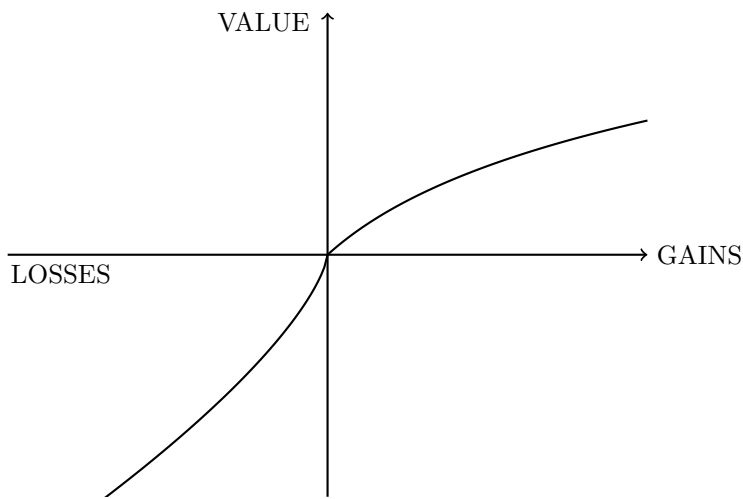
\begin{figure}[h!tb]
\centering
\begin{tikzpicture}
    \begin{axis}[
        axis lines = middle,
        xlabel = {GAINS},
        xtick = \empty,  
        ytick = \empty,  
        xmin = -3, xmax = 3,
        ymin = -2.5, ymax = 2.5,
        width=10cm,
        height=8cm,
        axis line style={->, thick}, 
        xlabel near ticks,
        ylabel near ticks,
        xlabel style={at={(current axis.right of origin)}, anchor=west},
        ylabel style={at={(current axis.above origin)}, anchor=south},
    ]
    \addplot[domain=0:3, samples=100, thick] {ln(x+1)};
    \addplot[domain=-3:0, samples=100, thick] {-1.5*abs(x)^(0.7)};
    \node at (axis cs:-2.5,-0.2) {LOSSES};  
    \node at (axis cs:-0.6,2.4) {VALUE};    
    \end{axis}
\end{tikzpicture}
\caption{Prospect Theory Value Function}
\end{figure}
The weighted value (V) that the agent will assign to a bet is given by:

\begin{equation}
V = \sum w(p_i) \, v(x_i)
\end{equation}

where \emph{p\textsubscript{i}} is perceived probability of outcome
\emph{x}, w(\emph{p}) is probability weighting function, and
v(\emph{x\textsubscript{i}}) is the value function assignment for
outcome \emph{x} (Levy 1992).

%% file: OpenAI_Bibliography.tex
\begin{hangparas}{2em}{1}

2012 Final Rule to Improve Transparency of Fees and Expenses to Workers
in 401(k)-Type Retirement Plans. (n.d.). U.S. Department of Labor.
\href{https://www.dol.gov/sites/dolgov/files/EBSA/about-ebsa/our-activities/resource-center/fact-sheets/final-rule-improve-transparency-of-fees-and-expenses.pdf}{[URL]}

Abdellaoui, M., Driouchi, A., \& l\textquotesingle Haridon, O. (2011).
Risk aversion elicitation: reconciling tractability and bias
minimization. Theory and Decision, 71, 63-80.

Aher, G. V., Arriaga, R. I., \& Kalai, A. T. (2023). Using large
language models to simulate multiple humans and replicate human subject
studies. In International Conference on Machine Learning (pp. 337-371).
PMLR.

Allais, M. (1953). Le comportement de l\textquotesingle homme rationnel
devant le risque: critique des postulats et axiomes de
l\textquotesingle école américaine. Econometrica: Journal of the
Econometric Society, 503-546.

Andersen, S., Harrison, G. W., Lau, M. I., \& Rutstrom, E. E. (2006).
Elicitation using multiple price lists. Experimental Economics, 9(4),
383--405.

Andreas, J. (2022). Language models as agent models. arXiv preprint
arXiv:2212.01681.

Argall, B. D., Chernova, S., Veloso, M., \& Browning, B. (2009). A
survey of robot learning from demonstration. Robotics and Autonomous
Systems, 57(5), 469-483.

Askell, A., Bai, Y., Chen, A., Drain, D., Ganguli, D., Henighan, T., ...
\& Kaplan, J. (2021). A general language assistant as a laboratory for
alignment. arXiv preprint arXiv:2112.00861.

Baier, A. (1986). Trust and antitrust. Ethics, 96(2), 231--260.

Baranowski, T. (1988). Validity and reliability of self report measures
of physical activity: an information-processing perspective. Research
Quarterly for Exercise and Sport, 59(4), 314-327.

Barocas, S., Hardt, M., \& Narayanan, A. (2023). Fairness and machine
learning: Limitations and opportunities. MIT press.

Barseghyan, L., Molinari, F., O\textquotesingle Donoghue, T., \&
Teitelbaum, J. C. (2013). Distinguishing probability weighting from risk
misperceptions in field data. American Economic Review, 103(3), 580--85.

Binswanger, H. P. (1980). Attitudes toward risk: Experimental
measurement in rural India. American Journal of Agricultural Economics,
62(3), 395-407.

{Bokern, M., Lohmann, E., Schürmann, O., \& Neumann, D. (2023). The
convergent and external validity of risk preference elicitation methods.
https://cris.maastrichtuniversity.nl/en/publications/the-convergent-and-external-validity-of-risk-preference-elicitati}

Borg, J. S., Sinnott-Armstrong, W., \& Conitzer, V. (2024). Moral AI:
And how we get there. Penguin Books Limited.

Bovens, L. (2019). The ethics of making risky decisions for others. In
M. D. White (Ed.), The Oxford Handbook of Ethics and Economics. Oxford
University Press.

Briggs, R. A. (2023). Normative theories of rational choice: Expected
utility. In E. N. Zalta \& U. Nodelman (Eds.), The Stanford Encyclopedia
of Philosophy (Winter 2023 ed.). Stanford University.
\href{https://plato.stanford.edu/archives/win2023/entries/rationality-normative-utility/}{[URL]}

Buchak, L. (2013). Risk and rationality. Oxford University Press.

Buchak, L. (2017). Taking risks behind the veil of ignorance. Ethics,
127(3), 610--44.

Buchak, L. (2019). Weighing the risks of climate change. The Monist,
102(1), 66-83.

Buchak, L. (2022). Normative theories of rational choice: Rivals to
expected utility. In E. N. Zalta (Ed.), The Stanford Encyclopedia of
Philosophy (Summer 2022 ed.). Stanford University.
\href{https://plato.stanford.edu/archives/sum2022/entries/rationality-normative-nonutility/}{[URL]}

Campo, S., Guerre, E., Perrigne, I., \& Vuong, Q. (2011). Semiparametric
estimation of first-price auctions with risk-averse bidders. The Review
of Economic Studies, 78(1), 112-147.

Castro, C., \& Pham, A. (2020). Is the attention economy noxious?
Philosophers, 20(17).

Chan, A., Salganik, R., Markelius, A., Pang, C., Rajkumar, N.,
Krasheninnikov, D., ... \& Maharaj, T. (2023). Harms from increasingly
agentic algorithmic systems. In Proceedings of the 2023 ACM Conference
on Fairness, Accountability, and Transparency (pp. 651-666).

Charles Schwab (2017). How to Determine Your Risk Tolerance Level.
\href{https://www.schwab.com/learn/story/how-to-determine-your-risk-tolerance-level}{[URL]}

Clark, J., \& Amodei, D. (2016, December 22). Faulty reward functions in
the wild. OpenAI.
\href{https://openai.com/blog/faulty-reward-functions/}{[URL]}

Cox, J. C., \& Oaxaca, R. L. (1996). Is bidding behavior consistent with
bidding theory for private value auctions?. Research in experimental
economics, 6, 131-148.

Dohmen, T. J., Falk, A., Huffman, D., Sunde, U., Schupp, J., \& Wagner,
G. G. (2005). Individual risk attitudes: New evidence from a large,
representative, experimentally-validated survey. IZA Discussion Paper
No. 1730.

Dohmen, T., Falk, A., Huffman, D., Sunde, U., Schupp, J., \& Wagner, G.
(2011). Individual risk attitudes: Measurement, determinants, and
behavioral consequences. Journal of the European Economic Association,
9(3), 522--550.

Donoghue v Stevenson AC 562 All ER (1932)
\href{https://www.scottishlawreports.org.uk/resources/donoghue-v-stevenson/case-report/}{[URL]}

Dworkin, G. (2020). Paternalism. In E. N. Zalta (Ed.), The Stanford
Encyclopedia of Philosophy (Fall 2020 ed.). Stanford University.
\href{https://plato.stanford.edu/archives/fall2020/entries/paternalism/}{[URL]}

Eckel, C. C., \& Grossman, P. J. (2002). Sex differences and statistical
stereotyping in attitudes toward financial risk. Evolution and Human
Behavior, 23(4), 281-295.

Eckel, C. C., \& Grossman, P. J. (2008). Men, women and risk aversion:
Experimental evidence. Handbook of Experimental Economics Results, 1,
1061-1073.

Esvelt, K. M. (2022). Delay, detect, defend: Preparing for a future in
which thousands can release new pandemics. Geneva Paper 29/22. Geneva
Centre for Security Policy.
\href{https://dam.gcsp.ch/files/doc/gcsp-geneva-paper-29-22}{[URL]}

Falk, A., Becker, A., Dohmen, T., Enke, B., Huffman, D., \& Sunde, U.
(2018). Global evidence on economic preferences. The quarterly journal
of economics, 133(4), 1645-1692.

Fisher, R. J., \& Katz, J. E. (2000). Social-desirability bias and the
validity of self-reported values. Psychology \& Marketing, 17(2),
105--120.

Forster, M., \& Sober, E. (1994). How to tell when simpler, more
unified, or less ad hoc theories will provide more accurate predictions.
The British Journal for the Philosophy of Science, 45(1), 1-35.

Gabriel, I. (2020). Artificial intelligence, values, and alignment.
Minds and Machines, 30(3), 411--437.
\href{https://doi.org/10.1007/s11023-020-09539-2}{[URL]}

Gabriel, I., Manzini, A., Keeling, G., Hendricks, L. A., Rieser, V.,
Iqbal, H., ... \& Manyika, J. (2024). The ethics of advanced AI
assistants. arXiv preprint arXiv:2404.16244.

Goetze, T. S. (2022). Mind the gap: Autonomous systems, the
responsibility gap, and moral entanglement. In 2022 ACM Conference on
Fairness, Accountability, and Transparency. ACM.

Greaves, H., Thomas, T., Mogensen, A., \& MacAskill, W. (2024). On the
desire to make a difference. Philosophical Studies, 1-28.

Hadfield-Menell, D., Milli, S., Abbeel, P., Russell, S. J., \& Dragan,
A. (2017). Inverse reward design. In Advances in Neural Information
Processing Systems (pp. 6765--6774).

Hajek, A. (2021). Risky business. Philosophical Perspectives, 35(1),
189-205.

Handa, K., Gal, Y., Pavlick, E., Goodman, N., Andreas, J., Tamkin, A.,
\& Li, B. Z. (2024). Bayesian preference elicitation with language
models. arXiv preprint arXiv:2403.05534.

Hansson, S. O. (2023). Risk. In E. N. Zalta \& U. Nodelman (Eds.), The
Stanford Encyclopedia of Philosophy (Summer 2023 ed.). Stanford
University.
\href{https://plato.stanford.edu/archives/sum2023/entries/risk/}{[URL]}

Harbaugh, W. T., Krause, K., \& Vesterlund, L. (2010). The fourfold
pattern of risk attitudes in choice and pricing tasks. The Economic
Journal, 120(545), 595--611.

Harrison, G. W. (2006). Hypothetical bias over uncertain outcomes. In J.
A. List (Ed.), Using Experimental Methods in Environmental and Resource
Economics (pp. 41--69). Edward Elgar.

Harrison, G. W. (2014). Real choices and hypothetical choices. In S.
Hess \& A. Daly (Eds.), Handbook of Choice Modelling (pp. 236--254).
Edward Elgar.

Harrison, G. W., \& Ross, D. (2017). The empirical adequacy of
cumulative prospect theory and its implications for normative
assessment. Journal of Economic Methodology, 24(2), 150--165.

Harrison, G. W., \& Rutström, E. E. (2008). Risk aversion in the
laboratory. In J. C. Cox \& G. W. Harrison (Eds.), Risk Aversion in
Experiments (Research in Experimental Economics, Vol. 12) (pp. 41-196).
Emerald Group Publishing Limited.

Harrison, G. W., \& Swarthout, J. T. (2016). Cumulative prospect theory
in the laboratory: A reconsideration (CEAR Working Paper No. 2016-05).
Center for Economic Analysis of Risk, Robinson College of Business,
Georgia State University.

Isaac, R. M., \& James, D. (2000). Just who are you calling risk averse?
Journal of Risk and Uncertainty, 20(2), 177--187.

Jones, K. (2012). Trustworthiness. Ethics, 123(1), 61--85.

Kahneman, D., \& Tversky, A. (1979). Prospect theory: An analysis of
decision under risk. Econometrica, 47(2), 263-292.

Kahneman, D., Knetsch, J. L., \& Thaler, R. H. (1991). The endowment
effect, loss aversion, and status quo bias. Journal of Economic
Perspectives, 5(1), 193-206.

Kirilenko, A., Kyle, A. S., Samadi, M., \& Tuzun, T. (2017). The flash
crash: High‐frequency trading in an electronic market. The Journal of
Finance, 72(3), 967-998.

Köchling, A., \& Wehner, M. C. (2020). Discriminated by an algorithm: A
systematic review of discrimination and fairness by algorithmic
decision-making in the context of HR recruitment and HR development.
Business Research, 13(3), 795-848.

Kosonen, P. (2022). How to discount small probabilities.
\href{https://www.petrakosonen.com/research/how-to-discount-small-probabilities.pdf}{[URL]}

Lazar, S. (2024). Legitimacy, authority, and democratic. In Oxford
Studies in Political Philosophy Volume 10 (p. 28). Oxford University
Press.

Lazer, D., Kennedy, R., King, G., \& Vespignani, A. (2014). The parable
of Google Flu: traps in big data analysis. science, 343(6176),
1203-1205.

Levy, J. S. (1992). An introduction to prospect theory. Political
Psychology, 13(2), 171-186.

Lin, C. C., Jaech, A., Li, X., Gormley, M. R., \& Eisner, J. (2020).
Limitations of autoregressive models and their alternatives. In
Proceedings of the Conference of the North American Chapter of the
Association for Computational Linguistics.

Lin, S., Hilton, J., \& Evans, O. (2021). TruthfulQA: Measuring how
models mimic human falsehoods. arXiv preprint arXiv:2109.07958.

List, J. A., \& Haigh, M. S. (2005). A simple test of expected utility
theory using professional traders. Proceedings of the National Academy
of Sciences, 102(3), 945-948.

Lum, K., \& Isaac, W. (2016). To predict and serve? Significance, 13(5),
14-19.

Machina, M. J., \& Siniscalchi, M. (2014). Ambiguity and ambiguity
aversion. In Handbook of the Economics of Risk and Uncertainty (Vol. 1,
pp. 729-807). North-Holland.

Mao, S., Zhang, N., Wang, X., Wang, M., Yao, Y., Jiang, Y., ... \& Chen,
H. (2024). Editing personality for large language models. arXiv preprint
arXiv:2310.02168v3.

Merrick, J. (2015). Immanuel Kant the, errrr, walker? Verso Magazine.
\href{https://www.versobooks.com/blogs/news/1963-immanuel-kant-the-errrr-walker}{[URL]}

Monton, B. (2019). How to avoid maximizing expected utility.
Philosophers\textquotesingle{} Imprint, 19.

Ng, A. Y., \& Russell, S. (2000). Algorithms for inverse reinforcement
learning. In Proceedings of the Seventeenth International Conference on
Machine Learning (Vol. 1, No. 2, p. 2).

Nguyen, C. T. (2022). Trust as an unquestioning attitude. Oxford Studies
in Epistemology, 7, 214-244.

Nisbett, R. E., \& Wilson, T. D. (1977). Telling more than we can know:
Verbal reports on mental processes. Psychological Review, 84(3), 231.

Oliver, A. (2003). A quantitative and qualitative test of the Allais
paradox using health outcomes. Journal of Economic Psychology, 24(1),
35-48.

O\textquotesingle Neil, C. (2017). Weapons of math destruction: How big
data increases inequality and threatens democracy. Crown.

Park, A. L. (2019). Injustice ex machina: Predictive algorithms in
criminal sentencing. UCLA Law Review, 19.

Paulhus, D. L. (1984). Two-component models of socially desirable
responding. Journal of Personality and Social Psychology, 46(3),
598--609.
\href{https://doi.org/10.1037/0022-3514.46.3.598}{[URL]}

Paulhus, D. L., \& Vazire, S. (2007). The self-report method. In R. W.
Robins, R. C. Fraley, \& R. F. Krueger (Eds.), Handbook of Research
Methods in Personality Psychology (pp. 224-239). Guilford Press.

Perez, E., Ringer, S., Lukoši ˙ ut¯ e, K., Nguyen, K., Chen, E., Heiner,
S., ... \& Kaplan, J. (2022). Discovering language model behaviors with
model-written evaluations. arXiv preprint arXiv:2212.09251.

Pettigrew, R. (2015). Risk, rationality and expected utility theory.
Canadian Journal of Philosophy, 45(5-6), 798-826.

Radford, A., Jozefowicz, R., \& Sutskever, I. (2017). Learning to
generate reviews and discovering sentiment. arXiv preprint
arXiv:1704.01444.

Ramsey, F. P. (1931). Truth and probability. In R. B. Braithwaite (Ed.),
The Foundations of Mathematics and Other Logical Essays (pp. 156-198).
Routledge.

Richman, W. L., Kiesler, S., Weisband, S., \& Drasgow, F. (1999). A
meta-analytic study of social desirability distortion in
computer-administered questionnaires, traditional questionnaires, and
interviews. Journal of Applied Psychology, 84(5), 754.

Roelofs, R., Shankar, V., Recht, B., Fridovich-Keil, S., Hardt, M.,
Miller, J., \& Schmidt, L. (2019). A meta-analysis of overfitting in
machine learning. Advances in Neural Information Processing Systems, 32.

Ross, S., Gordon, G., \& Bagnell, D. (2011). A reduction of imitation
learning and structured prediction to no-regret online learning.

{Sadigh, D., Landolfi, N., Sastry, S. S., Seshia, S. A., \& Dragan,
A. D. (2018). Planning for cars that coordinate with people: leveraging
effects on human actions for planning and active information gathering
over human internal state. Autonomous Robots, 42, 1405-1426.}

Schaal, S. (1996). Learning from demonstration. Advances in neural
information processing systems, 9

{Shavit, Y., Agarwal, S., Brundage, M., Adler, S.,
O\textquotesingle Keefe, C., Campbell, R., ... \& Robinson, D. G.
(2023). Practices for governing agentic AI systems. Research Paper,
OpenAI.}

{Silver, D., Huang, A., Maddison, C. J., Guez, A., Sifre, L., Van Den
Driessche, G., ... \& Hassabis, D. (2016). Mastering the game of Go with
deep neural networks and tree search. Nature, 529(7587), 484-489.}

{Smiley, L. (2022, March 8). "I\textquotesingle m the operator": The
aftermath of a self-driving tragedy. Wired.}

Snell, C., Klein, D., \& Zhong, R. (2022). Learning by distilling
context. arXiv preprint arXiv:2209.15189.

{Stark, F. (2016). Culpable carelessness: Recklessness and negligence
in the criminal law. Cambridge University Press.}

Stephan, M., Khazatsky, A., Mitchell, E., Chen, A. S., Hsu, S., Sharma,
A., \& Finn, C. (2024). Rlvf: Learning from verbal feedback without
overgeneralization. arXiv preprint arXiv:2402.10893.

Thoma, J. (2023). Taking risks on behalf of another. Philosophy Compass,
18(3), e12898.

Tversky, A., \& Kahneman, D. (1974). Judgment under Uncertainty:
Heuristics and Biases: Biases in judgments reveal some heuristics of
thinking under uncertainty. science, 185(4157), 1124-1131.

{Tversky, A., \& Kahneman, D. (1981). The framing of decisions and
the psychology of choice. Science, 211(4481), 453-458.}

{Tversky, A., \& Kahneman, D. (1992). Advances in prospect theory:
Cumulative representation of uncertainty. Journal of Risk and
Uncertainty, 5(4), 297-323.}

{Von Neumann, J. and Morgenstern, O. (1953) Theory of Games and
Economic Behavior. 3rd Edition, Princeton University Press, Princeton.}

Wakker, P., \& Deneffe, D. (1996). Eliciting von Neumann-Morgenstern
utilities when probabilities are distorted or unknown. Management
science, 42(8), 1131-1150.

Wakker, P., Erev, I., \& Weber, E. U. (1994). Comonotonic independence:
The critical test between classical and rank-dependent utility theories.
Journal of Risk and Uncertainty, 9, 195-230.

Weatherson, B. (2024). The End of Decision Theory.
https://philpapers.org/rec/WEATEO-9

{Williams, A. C. (1966). Attitudes toward speculative risks as an
indicator of attitudes toward pure risks. Journal of Risk and Insurance,
33(4), 577-586.}

\end{hangparas}